\documentclass{jfm}

\usepackage{subfiles}
\usepackage{graphicx}
\usepackage{epstopdf,epsfig,subfloat,subfig,floatrow}
\usepackage{newtxtext}
\usepackage{newtxmath}
\usepackage{natbib}
\usepackage{amsfonts}
\usepackage{amsmath}
\usepackage{amssymb,color,mathtools}
\usepackage{bm}
\usepackage{hyperref}
\usepackage{multirow,makecell,array,booktabs}

\usepackage{ulem}

\hypersetup{
    colorlinks = true,
    urlcolor   = blue,
    citecolor  = blue,
}

\newcommand{\RomanNumeralCaps}[1]
\linenumbers
\usepackage{soul,cancel}

\title{A kinetic model for rarefied flows of molecular gas with vibrational modes}

\author{      Qi Li
        \and Lei Wu
        \corresp{\email{wul@sustech.edu.cn}}
        }

\affiliation{
  Department of Mechanics and Aerospace Engineering, Southern University of Science and Technology, Shenzhen 518055, China
  }

\begin{document}
\maketitle

\begin{abstract}

A kinetic model is proposed for rarefied flows of molecular gas with rotational and temperature-dependent vibrational degrees of freedom. The model reduces to the Boltzmann equation for monatomic gas when the energy exchange between the translational and internal modes is absent, thus the influence of intermolecular potential can be captured. Moreover, not only the transport coefficients but also their fundamental relaxation processes are recovered. The accuracy of our kinetic model is validated by the direct simulation Monte Carlo method in several rarefied gas flows, including the  shock wave, Fourier flow, Couette flow, and the creep flow driven by Maxwell's demon. Then the kinetic model is adopted to investigate thermally-induced flows. By adjusting the viscosity index in the Boltzmann collision operator, we find that the intermolecular potential significantly influences the velocity and Knudsen force. Interestingly, in the transition flow regime, the Knudsen force exerting on a heated beam could reverse the direction when the viscosity index changes from 0.5 (hard-sphere gas) to 1 (Maxwell gas). This discovery is useful in the design of  micro-electromechanical systems for microstructure actuation and gas sensing.



\end{abstract}


\section{Introduction}

The non-equilibrium dynamics of molecular (diatomic/polyatomic) gas is commonly encountered in aerospace engineering. For example, at a high Mach number, the air surrounding an aircraft decelerates and heats up rapidly after compression by shock waves, which causes strong energy conversion from the translational energy into the internal energy. The temperature may reach thousands of degrees Kelvin and thus leads to significant changes in the physical and chemical properties of the gas \citep{Anderson2019, Ivano1998ARFM}. Under the assumption of thermodynamic equilibrium, the traditional Navier-Stokes-Fourier equations are used to predict the thermal environment and aerodynamic characteristics of the aircraft. And the influence of internal degrees of freedom (DoF) is taken into account by the variations of heat capacity and transport properties of molecular gas \citep{Malik1991}. On the other hand, when the thermodynamic nonequilibrium occurs, gases with different temperatures associated with various relaxation processes needs to be considered. And several sets of Navier-Stokes-type equations have been developed with multi-temperatures of different types of kinetic modes \citep{Colonna2006JTHT, Bruno2011PoF, Aoki2020PRE}.  

Since the macroscopic models are obtained at small Knudsen number, they are only applicable in the near-continuum flow regime. However, the gas could be in highly thermal nonequilibrium in many realistic situations, such as the reentry of aircraft into the atmosphere, where the gas flow changes from the continuum to the free-molecular regimes. Therefore, the treatment based on gas kinetic theory is inevitable, as the molecular dynamics simulation is limited to small spatial and temporal domains. The fundamental equation in gas kinetic theory is the Boltzmann equation, but it is only rigorously established for monatomic gas. For the molecular gas, its internal DoF pose difficulties in the modelling of rarefied gas dynamics. The heuristic way to describe the molecular gas dynamics in all flow regimes is the \cite{WangCS} (WCU) equation, which treats the internal DoF quantum mechanically and assigns each internal energy level an individual velocity distribution function. However, the complexity and excessive computational burden prevent the application of WCU equation.

The direct simulation Monte Carlo (DSMC) method \citep{Bird1994} is prevailing in simulating the rarefied gas dynamics \citep{Frezzotti2007EJMB, Pfeiffer2016PoF, Tantos2016IJHMT}. Although it is proven that DSMC is equivalent to the Boltzmann equation for monatomic gas~\citep{wagner_consist}, there are two drawbacks when applied to molecular gas flows. First, the bulk viscosity and the thermal conductivities cannot be recovered simultaneously. The reason lies in its phenomenological collision model of \cite{Borgnakke1975}, which realizes the correct exchange rate between the translational and internal energies to exactly recover the bulk viscosity \citep{Haas1994, Lavin2002PoF}. However, it cannot guarantee that the thermal conductivity, or its translational and internal components, is recovered at the same time \citep{Wu2020JFM, Li2021JFM}. Second, DSMC is not well suited to the simulation of low-speed flows due to its intrinsic stochastic nature. For instance, it has been found that the computational cost increases as $\text{Ma}^{-2}$ ($\text{Ma}$ is the Mach number) when the flow speed is approaching zero \citep{Hadjiconstantinou2003}. However, due to the rapid development of microelectromechanical techniques, the rarefied molecular gas conditions also exist in the flows at the microscale for a broad range of industrial applications \citep{Beskok_book}. And the speed of these small scale flows are usually much lower than the thermal velocity of gas molecules, thus making the DSMC time-consuming and even intractable in some cases.

Alternatively, kinetic models are proposed to imitate as closely as possible the behaviour of the WCU equation, and multiscale deterministic methods are developed to solve those kinetic models. The Bhatnagar-Gross-Krook (BGK) type kinetic models, which replace the Boltzmann collision operator with a single relaxation approximation~\citep{Bhatnagar1954}, are very popular. Notable success has been achieved by the BGK model in the modelling of the monatomic rarefied gas. However, the Prandtl number is incorrect in its standard model. To overcome this issue, the modified BGK models, such as the ellipsoidal-statistical BGK model \citep{Holway1966} and the Shakhov model \citep{Shakhov1968} have been proposed. These kinetic models have been extended to polyatomic rarefied gas by introducing additional internal  energy variables in the distribution function \citep{Morse1964, Rykov, Andries2000, Rahimi2016JFM, Wang2017JCP, Bernard2019JSC, Dauvois2020arXiv}, as well as the gas mixture of polyatomic molecules \citep{Klingenberg2018arXiv, Pirner2018JSP}. 
Besides, the Fokker–Planck models have been proposed \citep{Gorji2013, Mathiaud2020}, which take advantage of the continuous distribution functions in terms of stochastic velocity processes to speed up the stochastic particle methods. 

However, these models do not reduce to the Boltzmann equation for monatomic gases when the translational-internal energy exchange is absent. Therefore, these models cannot distinguish the influence of different intermolecular potentials. For example,  the uncertainties caused by different intermolecular potentials has been demonstrated in calculation of thermal creep slip on diffuse walls \citep{loyalka1990slip}, the thermal creep and Poiseuille flows \citep{Sharipov2009,Takata2011}, the viscous slip of the Couette flow \citep{Su2019JCP2}. On the other hand, all these kinetic model equations  concern only the transport coefficients, such as the thermal conductivity and bulk viscosity, while their fundamental relaxation processes are not captured, which are found to be important in rarefied molecular gas dynamics. For example, the relaxation rates of heat flux can significantly affect the creep flow driven by molecular velocity-dependent external force \citep{Li2021JFM}. Therefore, it is necessary to tackle the two difficulties when building a gas kinetic model for molecules with rational and vibrational DoF.




This rest of the paper is organized as follows. In \S\ref{sec:kinetic_model}, the transport coefficients and their intrinsic relation to relaxation rates are discussed, and the kinetic model is built based on the relaxation time approximation to reflect those relaxations. 
In \S\ref{FurtherModel}, the kinetic model is further developed to incorporate the Boltzmann collision operator to discern the influence of intermolecular potentials. 
In \S\ref{sec:validation}, the kinetic model is validated by DSMC in typical rarefied gas flows. Then, in \S\ref{sec:2D_thermal},  the kinetic model is applied to solve two-dimensional thermally induced microflow, and the influence of intermolecular potential on the thermal transpiration and the Knudsen force on micro-beam is investigated by varying the viscosity index. Finally, conclusions are presented in \S\ref{sec:conclusion}.

\section{Properties of molecular gas and relaxation-time approximation}\label{sec:kinetic_model}

A fundamental requirement in constructing a kinetic model is that all the transport coefficients are consistent with those obtained from the Boltzmann equation for monatomic gas or the WCU equation for molecular gas. Due to the excitation of internal DoF in molecular gas, additional relaxation processes occur between different type of energies, which lead to new transport coefficients such as  the bulk viscosity and internal thermal conductivity. The recovery of these new transport coefficients in kinetic model is crucial to accurately describe rarefied gas dynamics in many problems. For instances, the modelling of the shock wave requires correct bulk viscosity due to its high compressibility, while the modelling of thermal transpiration requires the recovery of translational thermal conductivity, rather than the total thermal conductivity~\citep{Mason1963JCP,porodnov1978thermal,Loyalka1979Polyatomic}. Therefore, in the following discussion, the transport coefficients, especially their intrinsic relaxation processes exclusively exist in molecular gas will be introduced, then the kinetic model will be established to recover these relaxation processes and transport coefficients.

\subsection{Kinetic description of molecular gas}

Both rotational and vibrational DoF of molecular gases are considered. In addition to the translational molecular velocity $\bm{v}$, the rotational energy $I_r$ and vibrational energy $I_v$ are introduced, and their corresponding numbers of DoF are $d_r$ and $d_v$. It is noted that the translational and rotational DoF are fully activated at relatively low temperature; for example, for nitrogen when the temperature is higher than $10~\text{K}$. Therefore, it is a common choice to use constant values of DoF for these modes. On the other hand, the vibrational DoF has not been significantly excited until $10^3~\text{K}$. Therefore, the vibrational DoF depends on the vibrational temperature $T_v$:
\begin{equation}\label{eq:temperature_dependent_dv}
	\begin{aligned}[b]
		d_v(T_v)=\frac{2T_{\text{ref}}/T_v}{\exp({T_{\text{ref}}/T_v})-1},
	\end{aligned}
\end{equation}
where $T_{\text{ref}}$ is the characteristic temperature of the active vibrational mode.

Thus, the distribution function of gases is denoted as $f(t,\bm{x},\bm{v},I_r,I_v)$, where $t$ is the time and $\bm{x}$ is the spatial coordinates. Macroscopic variables, such as the number density $n$, flow velocity $\bm{u}$, heat fluxes $\bm{q}_t,\bm{q}_r,\bm{q}_v$, pressure tensor $p_{ij}$, and temperatures $T_t,T_r,T_v$, are obtained by taking the moments of the distribution function:
\begin{equation}\label{eq:macroscopic_variables_f}
	\begin{aligned}[b]
		\left(n, n\bm{u},p_{ij}\right)=\int\left(1,\bm{v},mc_ic_j\right){f}\mathrm{d}\bm{v}\mathrm{d}I_r\mathrm{d}I_v, \\
		\left(\frac{3}{2}k_BT_t,\frac{d_r}{2}k_BT_r,\frac{{d_v(T_v)}}{2}k_BT_v\right)=\frac{1}{n}\int{\left(\frac{1}{2}mc^2,I_r,I_v\right)f}\mathrm{d}\bm{v}\mathrm{d}I_r\mathrm{d}I_v, \\
		\left(\bm{q}_t,\bm{q}_r,\bm{q}_v\right)=\int\bm{c}\left(\frac{1}{2}mc^2,I_r,I_v\right){f}\mathrm{d}\bm{v}\mathrm{d}I_r\mathrm{d}I_v,
	\end{aligned}
\end{equation}
where the subscripts $t,~r,~v$ indicate translational, rotational and vibrational components, respectively; $\bm{c}=\bm{v}-\bm{u}$ is the peculiar velocity, $m$ is the molecular mass, and $k_B$ is the Boltzmann constant.

We also define the temperature $T_{tr}$ to be the equilibrium temperature between the translational and rotational modes, $T_{tv}$ the equilibrium temperature between the translational and vibrational modes, and $T$ the equilibrium temperature over all DoF:
\begin{equation}\label{eq:Ttr_Ttv}
	\begin{aligned}[b]
		T_{tr}=\frac{3T_t+d_rT_r}{3+d_r}, \quad T_{tv}=\frac{3T_t+{{d_v(T_v)}}T_v}{3+{{d_v(T_{tv})}}}, \quad T=\frac{3T_t+d_rT_r+{{d_v(T_v)}}T_v}{3+d_r+{{d_v(T)}}},
	\end{aligned}
\end{equation}
and the corresponding pressures are $[p_t, p_r, p_v, p, p_{tr}, p_{tv}] = nk_B[T_t, T_r, T_v, T, T_{tr}, T_{tv}]$.  

\subsection{Relaxation processes in molecular gas}

In addition to the shear viscosity and translational heat conductivity in monatomic gas, the molecular gas possesses the bulk viscosity and internal thermal conductivities. The essence of these new transport coefficients are the relaxation of internal temperature and heat fluxes. This subsection is dedicated to the derivation of bulk viscosity and internal thermal conductivities, solely based on the relaxation processes.

\subsubsection{Bulk viscosity}

During the contraction or expansion of gas, the work done by pressure is converted to the translational energy immediately. However, in molecular gas, the molecules exhibit internal relaxation that exchanges the translational and internal energies in a finite time, which gives rise to the resistance that opposites to the volume change. This is known as the bulk viscosity.  

According to the Jeans-Landau-Teller equation, the rotational and vibrational relaxation at macroscopic level can be described as,
\begin{equation}\label{eq:Jeans_Landau_Teller}
	\frac{\mathrm{D}{T_r}}{\mathrm{D}{t}} = \frac{T_t-T_r}{\tau_r}, \quad \frac{\mathrm{D}{T_v}}{\mathrm{D}t} = \frac{T_t-T_v}{\tau_v},
\end{equation}
where ${\mathrm{D}}/{\mathrm{D}{t}}={\partial}/{\partial{t}}+\bm{u}\cdot{\partial}/{\partial{\bm{x}}}$ is the material derivative, $\tau_r$ and $\tau_v$ are the relaxation time  between the translational-rotational and translational-vibrational energy exchanges, respectively. Based on \eqref{eq:Ttr_Ttv} and \eqref{eq:Jeans_Landau_Teller}, the temperature change due to the effect of the relaxation alters $T$ to $T_t$ as follows:
\begin{equation}\label{eq:T_to_Tt_1}
	\begin{aligned}[b]
		T_t-T = \frac{1}{3+d_r+d_v(T_v)}\left(d_r\frac{\mathrm{D}{T_r}}{\mathrm{D}{t}}+d_v(T_v)\frac{\mathrm{D}{T_v}}{\mathrm{D}{t}}\right).
	\end{aligned}
\end{equation}
Considering the relaxation time $\tau_r$ and $\tau_v$ are much smaller than the timescale of gas volume change, where the deviation between equilibrium temperature $T$ and $T_t,~T_r,~T_v$ are small, the higher order terms of $T-T_r$ and $T-T_v$ can be ignored. Then, we have,
\begin{equation}\label{eq:T_to_Tt_2}
	\begin{aligned}[b]
		T_t - T = \frac{d_r\tau_r+d_v(T_v)\tau_v}{3+d_r+d_v(T_v)}\frac{\mathrm{D}{T}}{\mathrm{D}{t}}.
	\end{aligned}
\end{equation}

Ignoring the effect of shear viscosity and heat conduction, the energy conservation follows,
\begin{equation}\label{eq:conservation_pressure_work}
	\begin{aligned}[b]
		p_t\nabla\cdot\bm{u} + \frac{3+d_r+d_v(T_v)}{2}nk_B\frac{\mathrm{D}{T}}{\mathrm{D}{t}}=0.
	\end{aligned}
\end{equation}
Then, the pressure change due to the effect of the relaxation can be obtained by combining \eqref{eq:T_to_Tt_2} and \eqref{eq:conservation_pressure_work},
\begin{equation}\label{eq:p_to_pt_3}
	\begin{aligned}[b]
		p = p_t+2p_t\frac{d_r\tau_r+d_v(T_v)\tau_v}{[3+d_r+d_v(T_v)]^2}\nabla\cdot\bm{u}.
	\end{aligned}
\end{equation}
Thus, the bulk viscosity is obtained as
\begin{equation}\label{eq:mu_b}
	\begin{aligned}[b]
		\mu_b = 2p_t\frac{d_r\tau_r+d_v(T_v)\tau_v}{[3+d_r+d_v(T_v)]^2}.
	\end{aligned}
\end{equation}
It is shown that when the numbers of DoF are fixed, the bulk viscosity is determined by the translational pressure and relaxation times of internal modes.

\subsubsection{Thermal conductivity}

The rotational and vibrational modes in molecular gas carry the thermal energy and contribute  also  to the heat flux, while the conductance can be quite different from that of the translational one. In the continuum flow limit, the total thermal conductivity can determine the gas dynamics in addition to the viscosity and diffusivity. However, the thermal conductivity of a single type  mode may be important and even dominated when the gas is rarefied. For example, the mass flow rate in thermal transpiration is found to depend on the translational thermal conductivity of gas rather than the total thermal conductivity \citep{Mason1963JCP}.

In generally, the relaxation of translational and internal heat fluxes, $q_t$ and $q_{int}$, satisfies the following relation in spatially-homogeneous system~\citep{Mason1962}: 
\begin{equation}\label{eq:relaxation_qt_qint}
	\left[ 
      \begin{array}{cc} 
        \partial{\bm{q}_{t}}/{\partial{t}} \\ \partial{\bm{q}_{int}}/{\partial{t}}
      \end{array}
    \right]
    = -\frac{p_t}{\mu}
    \left[ 
      \begin{array}{cc} 
        A_{tt} & A_{ti} \\ A_{it} & A_{ii}
      \end{array}
    \right]
    \left[ 
      \begin{array}{cc} 
        \bm{q}_{t} \\ \bm{q}_{int} 
      \end{array}
    \right],
\end{equation}
where $\mu$ is the shear viscosity, the matrix $\bm{A}$ encapsulates the dimensionless thermal relaxation rates, and the subscripts $i$ represent the internal mode. From the Chapman-Enskog expansion, the thermal relaxation rates are related to the translational and internal thermal conductivities, $\kappa_t$ and $\kappa_{int}$, respectively, as
\begin{equation}\label{eq:relaxation_kt_kint}
	\left[ 
      \begin{array}{cc} 
        \kappa_{t} \\ \kappa_{int}
      \end{array}
    \right]
    = \frac{k_B\mu}{2m}
	\left[ 
      \begin{array}{cc} 
        A_{tt} & A_{ti} \\ A_{it} & A_{ii}
      \end{array}
    \right]^{-1}
    \left[ 
      \begin{array}{cc} 
        5 \\ d_{int} 
      \end{array}
    \right],
\end{equation}
where $d_{int}$ is the number of all internal DoF.

It will be convenient to use the following dimensionless \cite{Eucken1913} factor $f_{eu}$:
\begin{equation}\label{eq:feu}
	\begin{aligned}[b]
		c_vf_{eu}\equiv\frac{\kappa}{\mu}=\frac{\kappa_t+\kappa_{int}}{\mu},
	\end{aligned}
\end{equation}
where $\kappa$ is the total thermal conductivity,  and $c_v$ is the specific heat capacity at constant volume. Similarly, $f_t$ and $f_{int}$ represent the Eucken factors of the translational and internal modes, respectively,
\begin{equation}\label{eq:ft_fint}
	\begin{aligned}[b]
		f_t = \frac{2}{3}\frac{m\kappa_t}{k_B\mu}, \quad f_{int} = \frac{2}{d_{int}}\frac{m\kappa_r}{k_B\mu}.
	\end{aligned}
\end{equation}

The total Eucken factor $f_{eu}$ can be determined directly from the total thermal conductivity, which can be measured experimentally. However, those of the translational and internal parts are rather difficult to be obtained. Nevertheless, \cite{Mason1962} derived the approximate thermal relaxation rates $\bm{A}$,
\begin{equation}\label{eq:relaxation_q_Mason}
	\begin{aligned}[b]
		A_{tt}&=\frac{2}{3}+\frac{5d_{int}\tau}{18\tau_{int}}, \quad
		A_{ii}=\frac{\mu}{\rho D'}+\frac{3\tau}{6\tau_{int}}, \quad
		A_{ti}=-\frac{5\tau}{6\tau_{int}}, \quad
		A_{it}=-\frac{d_{int}\tau}{6\tau_{int}},
	\end{aligned}
\end{equation}
where $\rho=nm$ is the mass density, $\tau$ is the relaxation time of translational modes to reach equilibrium, $\tau_{int}$ is the relaxation time of translational-internal energy exchange and $D'$ is the average diffusion coefficient. Therefore, the translational and internal Eucken factors are determined,
\begin{equation}\label{eq:ft_fint_Mason}
	\begin{aligned}
		f_t&=\frac{5}{2}\left[1-\frac{5d_{int}\tau}{12\tau_{int}}\left(1-\frac{2}{5}\frac{\rho D'}{\mu}\right)\right], \quad
		f_{int}= \frac{\rho D'}{\mu}\left[1+\frac{5\tau}{4\tau_{int}}\left(1-\frac{2}{5}\frac{\rho D'}{\mu}\right)\right].
	\end{aligned}
\end{equation}
To match the experimental values of thermal conductivity, the internal relaxation time $\tau_{int}$ in the above equations has to be modified \citep{Mason1962}. However, from \eqref{eq:mu_b} it follows that the internal relaxation time determines the bulk viscosity, which means that the bulk viscosity and all thermal conductivities cannot be recovered simultaneously, if \eqref{eq:ft_fint_Mason} is used. To get rid of this problem, exact values of thermal relaxation rates $\bm{A}$ should be incorporated into the kinetic model.


\subsection{Kinetic model with relaxation time approximation}

It is well known that the evolution of the molecular gas distribution function is governed by the \cite{WangCS} equation, which is too complicated to be applied in realistic problems. Therefore, kinetic models are urgently needed to simplify the collision operator in the WCU equation. Well-known kinetic models are the stochastic~\cite{Borgnakke1975} model and the deterministic \cite{Rykov} and ellipsoidal-statistical BGK models~\citep{holway1966new,andries2000gaussian}, with the emphasis to recover the transport coefficients, rather than the essential relaxation process~\eqref{eq:relaxation_kt_kint}. To be specific, in both deterministic kinetic models, the cross-relaxation coefficients $A_{ti}$ and $A_{it}$ vanish. As a consequence, the ellipsoidal-statistical BGK model cannot recover $f_{tr}$ and $f_{int}$, although the total Eucken factor is correct; the Rykov model can recover $f_{tr}$ and $f_{int}$, and therefore has flexibility in the simulation of thermal transpiration, but in the rarefied flow driven by the Maxwell demon the velocity is incorrect~\citep{Li2021JFM}. 

We now try to build a kinetic model based on the Rykov model, due to its more freedom to reflect the relaxation process of heat fluxes. In this model, the elastic and inelastic collisions are considered separately with different relaxation time, which can be adjusted to give a correct bulk viscosity. And the reference distribution functions to which the distribution function relaxes contain the heat fluxes, so that the thermal conductivity can be recovered. Although the Rykov model is initially proposed for diatomic gas without vibrational modes, it has been extended to polyatomic gas~\citep{LeiJFM2015} and gases with vibrational modes~\citep{Titarev2018}.  By adjusting the heat fluxes in the reference distribution functions, \eqref{eq:relaxation_kt_kint} can be properly recovered.

For inelastic collisions, only the relaxation processes between translational-rotational and translational-vibrational DoF are considered, due to the weak rotational-vibrational relaxation. Thus, the evolution of the distribution function $f(\bm{x},\bm{v},I_r,I_v,t)$ under external body acceleration $\bm{a}$ is governed by
\begin{equation}\label{eq:governing_equation}
		\frac{\partial{f}}{\partial{t}}+\bm{v} \cdot \frac{\partial{f}}{\partial{\bm{x}}}+ \frac{\partial{(\bm{a}f)}}{\partial{\bm{v}}}
		= \underbrace{\frac{g_t-f}{\tau}}_{elastic} + \underbrace{\frac{g_r-g_t}{Z_r\tau} + \frac{g_v-g_t}{Z_v\tau}}_{inelastic} 
\end{equation}
where $Z_r$ and $Z_v$ are the rotational and vibrational collision number, respectively. Since the acceleration $\bm{a}$ could be velocity dependent under general consideration, it is kept inside the partial derivative with respect to $\bm{v}$. The reference distribution functions $g_t, g_r, g_v$ are expanded about the equilibrium distributions $E_t(T)\cdot E_r(T)\cdot E_v(T)$ in a series of orthogonal polynomials in variables peculiar velocity $\bm{c}$, rotational energy $I_r$, vibrational energy $I_v$ and corresponding moments $\bm{q_t}, \bm{q_r}, \bm{q_v}$:
\begin{equation}\label{eq:gt_gr_gv}
	\left.
	\begin{aligned}
		g_t =~&E_t(T_t) \cdot E_r(T_r) \cdot E_v(T_v) \cdot \left[{1 + \frac{2m\bm{q}_t\cdot{\bm{c}}}{15{k_B}{T_t}{p_t}} \left(\frac{mc^2}{2k_BT_t}-\frac{5}{2}\right)} \right. \\ 
		&\left. {+\frac{2m\bm{q}_r\cdot{\bm{c}}}{d_rk_BT_tp_r}\left(\frac{I_r}{k_BT_r}-\frac{d_r}{2}\right) + \frac{2m\bm{q}_v\cdot{\bm{c}}}{{{d_v(T_v)}}k_BT_tp_v}\left(\frac{I_v}{k_BT_v}-\frac{{d_v(T_v)}}{2}\right)}\right], \\
		g_r =~&E_t(T_{tr}) \cdot E_r(T_{tr}) \cdot E_v(T_v) \cdot \left[{1 + \frac{2m\bm{q}_0\cdot{\bm{c}}}{15{k_B}{T_{tr}}{p_{tr}}} \left(\frac{mc^2}{2k_BT_{tr}}-\frac{5}{2}\right)} \right. \\ 
		&\left. {+\frac{2m\bm{q}_1\cdot{\bm{c}}}{d_rk_BT_{tr}p_{tr}}\left(\frac{I_r}{k_BT_{tr}}-\frac{d_r}{2}\right) + \frac{2m\bm{q}_2\cdot{\bm{c}}}{{{d_v(T_v)}}k_BT_{tr}p_v}\left(\frac{I_v}{k_BT_v}-\frac{{d_v(T_v)}}{2}\right)}\right], \\
		g_v =~&E_t(T_{tv}) \cdot E_r(T_{r}) \cdot E_v(T_{tv}) \cdot \left[{1 + \frac{2m\bm{q}_0\cdot{\bm{c}}}{15{k_B}{T_{tv}}{p_{tv}}} \left(\frac{mc^2}{2k_BT_{tv}}-\frac{5}{2}\right)} \right. \\ 
		&\left. {+\frac{2m\bm{q}_1\cdot{\bm{c}}}{d_rk_BT_{tv}p_{r}}\left(\frac{I_r}{k_BT_{r}}-\frac{d_r}{2}\right) + \frac{2m\bm{q}_2\cdot{\bm{c}}}{{{d_v(T_{tv})}}k_BT_{tv}p_{tv}}\left(\frac{I_v}{k_BT_{tv}}-\frac{{d_v(T_{tv})}}{2}\right)}\right], 
	\end{aligned}
	\right\}
\end{equation}
with the equilibrium distribution functions,
\begin{equation}\label{eq:Et_Er_Ev}
	\left.
	\begin{aligned}
		E_t(T)&=n{\left(\frac{m}{2\pi k_BT}\right)}^{3/2}\exp{\left(-\frac{mc^2}{2k_BT}\right)}, \\
		E_r(T)&=\frac{I^{d_r/2-1}_{r}}{\Gamma(d_r/2)(k_BT)^{d_r/2}}\exp{\left(-\frac{I_r}{k_BT}\right)}, \\
		E_v(T)&=\frac{I^{{{d_v(T)}}/2-1}_{v}}{\Gamma({{d_v(T)}}/2)(k_BT)^{d_v/2}}\exp{\left(-\frac{I_v}{k_BT}\right)}.
	\end{aligned}
	\right\}
\end{equation}
where $\Gamma$ is the gamma function, ${\bm{q}_{0}}$, ${\bm{q}_{1}}$, and ${\bm{q}_{2}}$ are linear combinations of translational, rotational and vibrational heat fluxes.

\subsection{Determination of model parameters}

So far, the kinetic model equation \eqref{eq:governing_equation} with the reference distributions in \eqref{eq:gt_gr_gv} contain the free parameters $\bm{q}_0$, $\bm{q}_1$, $\bm{q}_2$, $Z_r$, $Z_v$, and $\tau$. They will be determined by the recovery of relaxation rates of shear stress, temperature, and heat fluxes, which corresponding to the recover of shear viscosity, bulk viscosity, and thermal conductivities, respectively.

\subsubsection{Relaxation of temperature}

For simplicity let us consider a spatial-homogeneous system without the external acceleration. Multiply the equation \eqref{eq:governing_equation} by $\frac{1}{2}mc^2$, $I_r$, $I_v$, and integrate them with respect to $\bm{v}$, $I_r$ and $I_v$, yielding
\begin{equation}\label{eq:dTt_dTr_dTv}
	\begin{aligned}[b]
		\frac{\partial T_t}{\partial t} =& ~\frac{T_{tr}-T_{t}}{Z_r\tau} + \frac{T_{tv}-T_{t}}{Z_v\tau}, \\
		\frac{\partial T_r}{\partial t} =& ~\frac{T_{tr}-T_{r}}{Z_r\tau}, \\
		\frac{\partial ({{d_v(T_v)}}T_v)}{\partial t} =& ~\frac{{{d_v(T_{tv})}}T_{tv}-{{d_v(T_v)}}T_{v}}{Z_v\tau}. 
	\end{aligned}
\end{equation}

Comparing to the Jeans–Landau-Teller equations \eqref{eq:Jeans_Landau_Teller}, the collision numbers relate to the relaxation time $\tau_r$ and $\tau_v$ are
\begin{equation}\label{eq:Zr_Zv}
	\begin{aligned}[b]
		Z_r = \frac{3\tau_r}{(3+d_r)\tau}, \quad Z_v = \frac{3\tau_v}{(3+d_v)\tau}.
	\end{aligned}
\end{equation}

Based on number density conservation and the definition of equilibrium temperature in equations \eqref{eq:Ttr_Ttv}, the conservation of total energy is guaranteed.


\subsubsection{Relaxation of heat flux}

In the original Rykov model, the relaxation of translational heat flux is independent of the rotational one, and vice versa. However, due to the energy exchange between different modes, it is necessary to consider the fact that the relaxations of heat fluxes are coupled within all the DoF. Thus, in analogy to~\eqref{eq:relaxation_qt_qint}, the relaxation of translational, rotational and vibrational heat fluxes are generalized to
\begin{equation}\label{eq:heat_flux_relaxation}
	\left[ 
      \begin{array}{ccc} 
        \partial{\bm{q}_{t}}/{\partial{t}} \\ \partial{\bm{q}_{r}}/{\partial{t}} \\ \partial{\bm{q}_{v}}/{\partial{t}}
      \end{array}
    \right]
    = -\frac{p_t}{\mu}
    \left[ 
      \begin{array}{ccc} 
        A_{tt} & A_{tr} & A_{tv} \\ A_{rt} & A_{rr} & A_{rv} \\ A_{vt} & A_{vr} & A_{vv}
      \end{array}
    \right]
    \left[ 
      \begin{array}{ccc} 
        \bm{q}_{t} \\ \bm{q}_{r} \\ \bm{q}_{v}
      \end{array}
    \right],
\end{equation}
where the dimensionless relaxation rates $\bm{A}$ is a $3\times3$ matrix including all three types of modes. Accordingly, $\bm{q}_0$, $\bm{q}_1$, $\bm{q}_2$ in reference distributions \eqref{eq:gt_gr_gv} can be determined in terms of $\bm{q}_t$, $\bm{q}_r$, $\bm{q}_v$ and the thermal relaxation rates $\bm{A}$. To be specific, the governing equation \eqref{eq:governing_equation} is multiplied by $\frac{1}{2}mc^2\bm{c}$, $I_r\bm{c}$ and $I_v\bm{c}$, respectively, and then are integrated with respect to $\bm{v}$, $I_r$ and $I_v$, yielding
\begin{equation}\label{eq:q0_q1_q2}
    \begin{bmatrix} 
      \bm{q}_{0} \\ \bm{q}_{1} \\ \bm{q}_{2}
	\end{bmatrix}
	= 
    \begin{bmatrix}		(2-3A_{tt})Z_{int}+1 & -3A_{tr}Z_{int} & -3A_{tv}Z_{int} \\		-A_{rt}Z_{int} & -A_{rr}Z_{int}+1 & -A_{rv}Z_{int} \\ 		-A_{vt}Z_{int} & -A_{vr}Z_{int} & -A_{vv}Z_{int}+1
    \end{bmatrix}
    \begin{bmatrix} 
      \bm{q}_{t} \\ \bm{q}_{r} \\ \bm{q}_{v}
    \end{bmatrix},
\end{equation}
where $Z_{int}=\left({1}/{Z_r}+{1}/{Z_v}\right)^{-1}$.

\subsubsection{Shear viscosity and bulk viscosity}

As it is discussed above, other than the shear viscosity, the bulk viscosity arises from the resistance of contraction or expansion in molecular gas, due to the energy exchange between translational and internal motions. And both of them can be derived based on the Chapman-Enskog expansion \citep{CE}, when the system is close to equilibrium. To the second approximation of the distribution, it is assumed $f=f^{(0)}+f^{(1)}$, where $f^{(0)}=E_t(T)E_r(T)E_v(T)$ is the equilibrium distribution at temperature $T$. Let $\mathcal{D}f\equiv{\partial{f}}/{\partial{t}}+\bm{v} \cdot {\partial{f}}/{\partial{\bm{x}}}+\bm{a} \cdot {\partial{f}}/{\partial{\bm{v}}}$, and consider $\mathcal{D}^{(0)}f=0$, according to Chapman-Enskog expansion and the governing equation \eqref{eq:governing_equation}, we have
\begin{equation}\label{eq:Df(1)_J}
	\begin{aligned}[b]
		f^{(1)} = g_t - f^{(0)} +\frac{1}{Z_r}\left(g_r-g_t\right) +\frac{1}{Z_v}\left(g_v-g_t\right)-\tau\mathcal{D}^{(1)}f,
	\end{aligned}
\end{equation}
where
\begin{equation}\label{eq:Df(1)}
	\begin{aligned}[b]
		\mathcal{D}^{(1)}f=~&\frac{\partial{f^{(0)}}}{\partial{t}}+\bm{v} \cdot \frac{\partial{f^{(0)}}}{\partial{\bm{x}}}+\bm{a} \cdot \frac{\partial{f^{(0)}}}{\partial{\bm{v}}} \\
		=~&f^{(0)}\left[\left( \left(\frac{mc^{2}}{2k_BT}-\frac{5}{2}\right) +\left(\frac{Ir}{k_BT}-\frac{d_r}{2}\right) +\left(\frac{Iv}{k_BT} -\frac{d_v}{2} \right)\right)\bm{c}\cdot\nabla\ln{T} \right. \\
		&\left. +\frac{2}{(3+d_r+d_v)}\left(\frac{d_r+d_v}{3}\left(\frac{mc^{2}}{2k_BT}-\frac{3}{2}\right) -\left(\frac{Ir}{k_BT}-\frac{d_r}{2}\right) -\left(\frac{Iv}{k_BT}-\frac{d_v}{2}\right)\right)\frac{\partial{u_i}}{\partial{x_i}} \right. \\
		&\left. +\frac{m}{k_BT}c_{<i}c_{j>}\frac{\partial{u_{i}}}{\partial{x_{j}}} \right],
	\end{aligned}
\end{equation}
$c_{<i}c_{j>}=c_ic_j-c^2\delta_{ij}$, and  $\delta_{ij}$ is the Kronecker delta function.

The pressure tensor $p_{ij}$ is calculated as
\begin{equation}
	\begin{aligned}[b]
		p_{ij}&=\int{mc_ic_j(f^{(0)}+f^{(1)})}\mathrm{d}\bm{v}\mathrm{d}I_r\mathrm{d}I_v \\
		&= \left(p_t+\frac{1}{Z_r}(p_{tr}-p_t)+\frac{1}{Z_v}(p_{tv}-p_t)\right)\delta_{ij}-p\tau\frac{\partial{u_{<i}}}{\partial{x_{j>}}}-p\tau\frac{2(d_r+d_v)}{3(3+d_r+d_v)}\frac{\partial{u_k}}{\partial{x_k}}\delta_{ij} \\
		&= p\delta_{ij}-p\tau\frac{\partial{u_{<i}}}{\partial{x_{j>}}}-2p\tau\frac{(3+d_r)d_rZ_r+(3+d_v)d_vZ_v}{3\left(3+d_r+d_v\right)^2}\frac{\partial{u_i}}{\partial{x_i}}\delta_{ij},
	\end{aligned}
\end{equation}
where ${\partial{u_{<i}}}/{\partial{x_{j>}}}={\partial{u_{i}}}/{\partial{x_{j}}}+{\partial{u_{j}}}/{\partial{x_{i}}}-\frac{2}{3}({\partial{u_{k}}}/{\partial{x_{k}}})\delta_{ij}$. The shear viscosity $\mu$ and bulk viscosity $\mu_b$ are then obtained:
\begin{equation}
\begin{aligned}[b]
	\mu(T_t)&=p_t\tau, \\ \mu_b(T_t)&=2p_t\tau\frac{(3+d_r)d_rZ_r+(3+d_v)d_vZ_v}{3\left(3+d_r+d_v\right)^2}.
	\end{aligned}
\end{equation}
Therefore, it is shown that the ratio $\mu_b/\mu$ depends only on the numbers of internal DoF and corresponding collision numbers. Larger $Z_r$ or $Z_v$ makes the energy exchange between translational and internal motions more difficult, thus lead to higher bulk viscosity.

\subsubsection{Thermal conductivity and Eucken factors}

Consider a homogeneous system of molecular gas at rest, where the spatial derivatives of flow velocity vanish in \eqref{eq:Df(1)}, the  translational, rotational and vibrational heat fluxes can be calculated based on \eqref{eq:macroscopic_variables_f} and \eqref{eq:heat_flux_relaxation}. Eventually we have
\begin{equation}\label{eq:dq_dT}
	\begin{aligned}[b]
	\left[ 
      \begin{array}{ccc} 
        \bm{q}_t \\ \bm{q}_r \\ \bm{q}_v
      \end{array}
    \right]
	&= \int{\bm{c}\left[ 
      \begin{array}{ccc} 
        \frac{1}{2}mc^2 \\ I_r \\ I_v
      \end{array}
    \right] \left(f^{(0)}+f^{(1)}\right)}\mathrm{d}\bm{v}\mathrm{d}I_r\mathrm{d}I_v \\
	&= \tau\left[ 
      \begin{array}{ccc} 
        \partial{\bm{q}_{t}}/{\partial{t}} \\ \partial{\bm{q}_{r}}/{\partial{t}} \\ \partial{\bm{q}_{v}}/{\partial{t}}
      \end{array}
    \right] +\left[ 
		\begin{array}{ccc} 
		  \bm{q}_t \\ \bm{q}_r \\ \bm{q}_v
		\end{array}
	\right] - \frac{k_B\mu}{2m}
    \left[ 
      \begin{array}{ccc} 
        5 \\ d_r \\ d_v(T_v)
      \end{array}
    \right]
	\nabla{T}.
	\end{aligned}
\end{equation}
Consider $(\bm{q}_t, \bm{q}_r, \bm{q}_v)=-(\kappa_t, \kappa_r, \kappa_v)\nabla T$, then the thermal conductivities are 
\begin{equation}\label{eq:kappa_A}
	\left[ 
      \begin{array}{ccc} 
        \kappa_t \\ \kappa_r \\ \kappa_v
      \end{array}
    \right]
	= \frac{k_B\mu}{2m}
	\left[ 
      \begin{array}{ccc} 
        A_{tt} & A_{tr} & A_{tv} \\ A_{rt} & A_{rr} & A_{rv} \\ A_{vt} & A_{vr} & A_{vv}
      \end{array}
    \right]^{-1}
    \left[ 
      \begin{array}{ccc} 
        5 \\ d_r \\ d_v(T_v)
      \end{array}
    \right],
\end{equation}
And the dimensionless parameters Eucken factors are calculated based on  \eqref{eq:ft_fint}:
\begin{equation}\label{eq:EuckenFactor_A}
	\left[ 
      \begin{array}{ccc} 
        f_t \\ f_r \\ f_v
      \end{array}
    \right]
	= 
	\left[ 
      \begin{array}{ccc} 
        3A_{tt} & d_rA_{tr} & d_v(T_v)A_{tv} \\ 3A_{rt} & d_rA_{rr} & d_v(T_v)A_{rv} \\ 3A_{vt} & d_rA_{vr} & d_v(T_v)A_{vv}
      \end{array}
    \right]^{-1}
    \left[ 
      \begin{array}{ccc} 
        5 \\ d_r \\ d_v(T_v)
      \end{array}
    \right].
\end{equation}

Clearly, the elements in matrix $\bm{A}$ cannot be fully determined even though all the Eucken factors $f_t, f_r, f_v$ (thermal conductivities $\kappa_t, \kappa_r, \kappa_v$ equivalently) are fixed. In other words, in molecular gas, having all the transport coefficients is not enough to exactly describe the relaxation of heat flux, which may lead to uncertainty in predicting macroscopic gas dynamics~\citep{Li2021JFM}. Therefore, it is necessary to recovery the thermal relaxation rates in the kinetic model correctly.

\section{Kinetic model with Boltzmann collision operator}\label{FurtherModel}

In practical numerical simulations, it is better to eliminate the internal energy variables ${I_r,~I_v}$, by introducing the following reduced velocity distribution functions ${f_0,~f_1,~f_2}$:
\begin{equation}\label{eq:f0_f1_f2}
\left(f_0,f_1,f_2\right)=\iint_{0}^{\infty}\left(1,I_r,I_v\right)f\left(t,\bm{x},\bm{v},I_r,I_v\right)\mathrm{d}{I_r}\mathrm{d}{I_v}.
\end{equation}
Then, the governing equation~\eqref{eq:governing_equation} can be transferred to three coupled equations:
\begin{equation}\label{eq:reduced_governing_equation}
\frac{\partial{f_l}}{\partial{t}}+\bm{v} \cdot \frac{\partial{f_l}}{\partial{\bm{x}}}+\frac{\partial{(\bm{a}f_l)}}{\partial{\bm{v}}}
= \frac{g_{lt}-f_l}{\tau} + \frac{g_{lr}-g_{lt}}{Z_r\tau} + \frac{g_{lv}-g_{lt}}{Z_v\tau}, \quad l=0,1,2,
\end{equation}
where the reduced reference velocity distribution functions are
\begin{equation}\label{eq:g0t_g0r_g0v}
\begin{aligned}[b]
g_{0t}&=E_t(T_t)\left[1+\frac{2m\bm{q}_t\cdot{\bm{c}}}{15{k_B}{T_t}{p_t}} \left(\frac{mc^2}{2k_BT_t}-\frac{5}{2}\right)\right], \\
g_{0r}&=E_t(T_{tr})\left[1+\frac{2m\bm{q}_0\cdot{\bm{c}}}{15{k_B}{T_{tr}}{p_{tr}}} \left(\frac{mc^2}{2k_BT_{tr}}-\frac{5}{2}\right)\right], \\
g_{0v}&=E_t(T_{tv})\left[1+\frac{2m\bm{q}_0\cdot{\bm{c}}}{15{k_B}{T_{tv}}{p_{tv}}} \left(\frac{mc^2}{2k_BT_{tv}}-\frac{5}{2}\right)\right], 
\end{aligned}
\end{equation}
and
\begin{equation}\label{eq:g1t_g1r_g1v}
\begin{aligned}[b]
g_{1t}&=\frac{d_r}{2}k_BT_rg_{0t}+\frac{m\bm{q}_r\cdot{\bm{c}}}{p_t}E_t(T_t), \\
g_{1r}&=\frac{d_r}{2}k_BT_{tr}g_{0r}+\frac{m\bm{q}_1\cdot{\bm{c}}}{p_{tr}}E_t(T_{tr}), \\
g_{1v}&=\frac{d_r}{2}k_BT_{r}g_{0v}+\frac{m\bm{q}_1\cdot{\bm{c}}}{p_{tv}}E_t(T_{tv}), \\
g_{2t}&=\frac{{d_v(T_v)}}{2}k_BT_vg_{0t}+\frac{m\bm{q}_v\cdot{\bm{c}}}{p_t}E_t(T_t), \\
g_{2r}&=\frac{{d_v(T_v)}}{2}k_BT_{v}g_{0r}+\frac{m\bm{q}_2\cdot{\bm{c}}}{p_{tr}}E_t(T_{tr}), \\
g_{2v}&=\frac{{d_v(T_{tv})}}{2}k_BT_{tv}g_{0v}+\frac{m\bm{q}_2\cdot{\bm{c}}}{p_{tv}}E_t(T_{tv}). 
\end{aligned}
\end{equation}

The macroscopic quantities defined in \eqref{eq:macroscopic_variables_f} can be calculated based on the reduced velocity distribution functions:
\begin{equation}\label{eq:macroscopic_variables_f0_f1_f2}
	\begin{aligned}[b]
		\left(n, n\bm{u},p_{ij}\right)=\int\left(1,\bm{v},mc_ic_j\right){f_0}\mathrm{d}\bm{v}, \\
		\left(\frac{3}{2}k_BT_t,\frac{d_r}{2}k_BT_r,\frac{{d_v(T_v)}}{2}k_BT_v\right)=\frac{1}{n}\int{\left(\frac{1}{2}mc^2f_0,f_1,f_2\right)}\mathrm{d}\bm{v}, \\
		\left(\bm{q}_t,\bm{q}_r,\bm{q}_v\right)=\int{\bm{c}\left(\frac{1}{2}mc^2f_0,f_1,f_2\right)}\mathrm{d}\bm{v}.
	\end{aligned}
\end{equation}

It is noted that although the kinetic model~\eqref{eq:governing_equation} is proposed as per classical mechanics, i.e., the vibrational energy levels are continuous. From the perspective of quantum mechanics, the discrete levels of vibrational energy need to be involved \citep{Anderson2019}, and this large number of DoF due to the internal modes makes the trace of distribution function  time-consuming. Fortunately, this is not necessary since the fundamental task is to obtain the evolution of macroscopic measurable quantities. It is shown that the complexity arising from the discrete vibrational energy can be eliminated with the reduced distribution technique \citep{Mathiaud2020}. Therefore, by summation over all vibrational DoF and energy levels, the kinetic model proposed in this work is not restricted by the classical mechanics treatment of internal DoF. 

Obviously, all molecules relax with the same speed in the relaxation-time approximation~\eqref{eq:governing_equation}, which is not very physical, since in general molecules with larger peculiar velocity has larger collision probability and hence smaller relaxation time; in fact, when \eqref{eq:governing_equation} is used, the temperature of normal shock wave will be overpredicted~\citep{LeiJFM2015}.  To circumvent this problem, by observing that the elastic collision term in \eqref{eq:reduced_governing_equation} with $l=0$ is just the Shakhov-type approximation of the Boltzmann collision operator for monatomic gas~\citep{Shakhov1968,Shakhov_S}, we replace the elastic collision term $(g_{0t}-f_0)/\tau$ with the Boltzmann collision operator $Q(f_0)$ in monatomic gas:
\begin{equation}\label{eq:Boltzmann_collision_operator}
	Q(f_0) = \int_{\mathbb{R}^3}{\int_{\mathbb{S}^2}{B(\cos{\theta},{\left| \bm{v}-\bm{v}_* \right|})[f_0(\bm{v}'_*)f_0(\bm{v}')-f_0(\bm{v}_*)f_0(\bm{v})]\mathrm{d}{\Omega}}\mathrm{d}{\bm{v}_*}},
\end{equation} 
so that the relaxation time depends on the molecular velocity. Meanwhile, $g_{1t}$ and $g_{2t}$ are modified correspondingly~\citep{LeiJFM2015}, resulting in the following kinetic model for molecular gas:
\begin{equation}\label{eq:kinetic_model_equation}
	\begin{aligned}[b]
		\frac{\partial{f_0}}{\partial{t}}+\bm{v} \cdot \frac{\partial{f_0}}{\partial{\bm{x}}}+  \frac{\partial{(\bm{a}f_0)}}{\partial{\bm{v}}} &= Q(f_0) + \frac{g_{0r}-g_{0t}}{Z_r\tau} + \frac{g_{0v}-g_{0t}}{Z_v\tau}, \\
		\frac{\partial{f_1}}{\partial{t}}+\bm{v} \cdot \frac{\partial{f_1}}{\partial{\bm{x}}}+ \frac{\partial{(\bm{a}f_1)}}{\partial{\bm{v}}} &= \frac{g_{1t}'-f_1}{\tau} + \frac{g_{1r}-g_{1t}}{Z_r\tau} + \frac{g_{1v}-g_{1t}}{Z_v\tau}, \\
		\frac{\partial{f_2}}{\partial{t}}+\bm{v} \cdot \frac{\partial{f_2}}{\partial{\bm{x}}}+ \frac{\partial{(\bm{a}f_2)}}{\partial{\bm{v}}} &= \frac{g_{2t}'-f_2}{\tau} + \frac{g_{2r}-g_{2t}}{Z_r\tau} + \frac{g_{2v}-g_{2t}}{Z_v\tau},
	\end{aligned}
\end{equation}
with
\begin{equation}\label{eq:glt_v}
	\begin{aligned}[b]
		g_{1t}'&=\frac{d_r}{2}k_BT_r[\tau Q(f_0)+f_0]+\frac{m\bm{q}_r\cdot{\bm{c}}}{p_t}E_t(T_t), \\
		g_{2t}'&=\frac{{d_v(T_v)}}{2}k_BT_v[\tau Q(f_0)+f_0]+\frac{m\bm{q}_v\cdot{\bm{c}}}{p_t}E_t(T_t). 
	\end{aligned}
\end{equation}

Since the Shakhov model and the Boltzmann equation have the same shear viscosity and translational thermal conductivity, it can be shown that the new model~\eqref{eq:kinetic_model_equation} has the same transport coefficients with the model~\eqref{eq:reduced_governing_equation}.

Note that in~\eqref{eq:Boltzmann_collision_operator}, $\theta$ is the deflection angle of collision,  $\bm{v}$ and $\bm{v}_*$ are the velocities of the two molecules before collision, while $\bm{v}'$ and $\bm{v}'_*$ are the velocities of the two molecules after collision, and $\Omega$ is the solid angle. $B(\cos{\theta},{\left| \bm{v}-\bm{v}_* \right|})$ is the collision kernel, which incorporates the role of intermolecular potential. When the inverse power-law potential is considered, the collision kernel is modelled as~\citep{Lei2013,lei_Jfm}
\begin{equation}
	\begin{aligned}[b]
		B&=\frac{5\sqrt{\pi{m}k_BT_0}(4k_BT_0/m)^{(2\omega-1)/2}}{64\pi\mu(T_0)\Gamma^2(9/4-\omega/2)}\sin^{(1-2\omega)/2}\left(\frac{\theta}{2}\right)
		\cos^{(1-2\omega)/2}\left(\frac{\theta}{2}\right)
		|\bm{v}_r|^{2(1-\omega)},
	\end{aligned}
\end{equation}
where $\omega$ is the viscosity index, that is,
\begin{equation}\label{eq:viscosity_temperature}
	\mu(T)=\mu(T_0)\left(\frac{T}{T_0}\right)^\omega.
\end{equation}
Therefore, this kinetic model is able to distinguish the role of intermolecular potentials~\citep{Sharipov2009,Takata2011,lei_Jfm,wuPoF2015}, while the models based on the relaxation-time approximation do not have this capability. 



\section{Validation of the kinetic model}\label{sec:validation}


To evaluate the accuracy of the kinetic model~\eqref{eq:kinetic_model_equation}, numerical solutions of one-dimensional Fourier flow, Couette flow, thermal creep flow and normal shock wave in nitrogen with constant vibrational DoF are compared with DSMC solutions. The kinetic model equations are solved by the discretized velocity method with the fast spectral method for the Boltzmann collision operator~\citep{Lei2013,lei_Jfm}, while DSMC simulations are conducted using the open source code SPARTA~\citep{SPARTA}.

In the following paper, dimensionless variables will be presented. The density, velocity, temperature, stress, and heat flux are normalized by the reference number density $n_0$, the most probable speed $v_m=\sqrt{2k_BT_0/m}$, the reference temperature $T_0$, $n_0k_BT_0$, and $n_0k_BT_0v_m$, respectively. The spatial variable is normalized by the characteristic flow length $L_0$, and the Knudsen number is defined as
\begin{equation}\label{eq:Kn}
	\text{Kn}=\frac{\mu(T_0)}{n_0L_0}\sqrt{\frac{\pi}{2mk_BT_0}}.
\end{equation}

\subsection{Relaxation rates extracted from DSMC}\label{subsec:A_from_DSMC}

Since the bulk viscosity and thermal conductivity cannot be adjusted independently in DSMC, we extract the thermal relaxation rates from the DSMC and apply to our kinetic model, to make a fair comparison. With the fixed shear viscosity and self-diffusion coefficient, the collision number $Z_r$ and $Z_v$ are the only parameters that affect the thermal relaxation rates in DSMC. Here we take $Z_r=2.667$ and $Z_v=10Z_r$.


\begin{figure}[t]
	\centering
	\subfloat[]{\includegraphics[scale=0.22,clip=true]{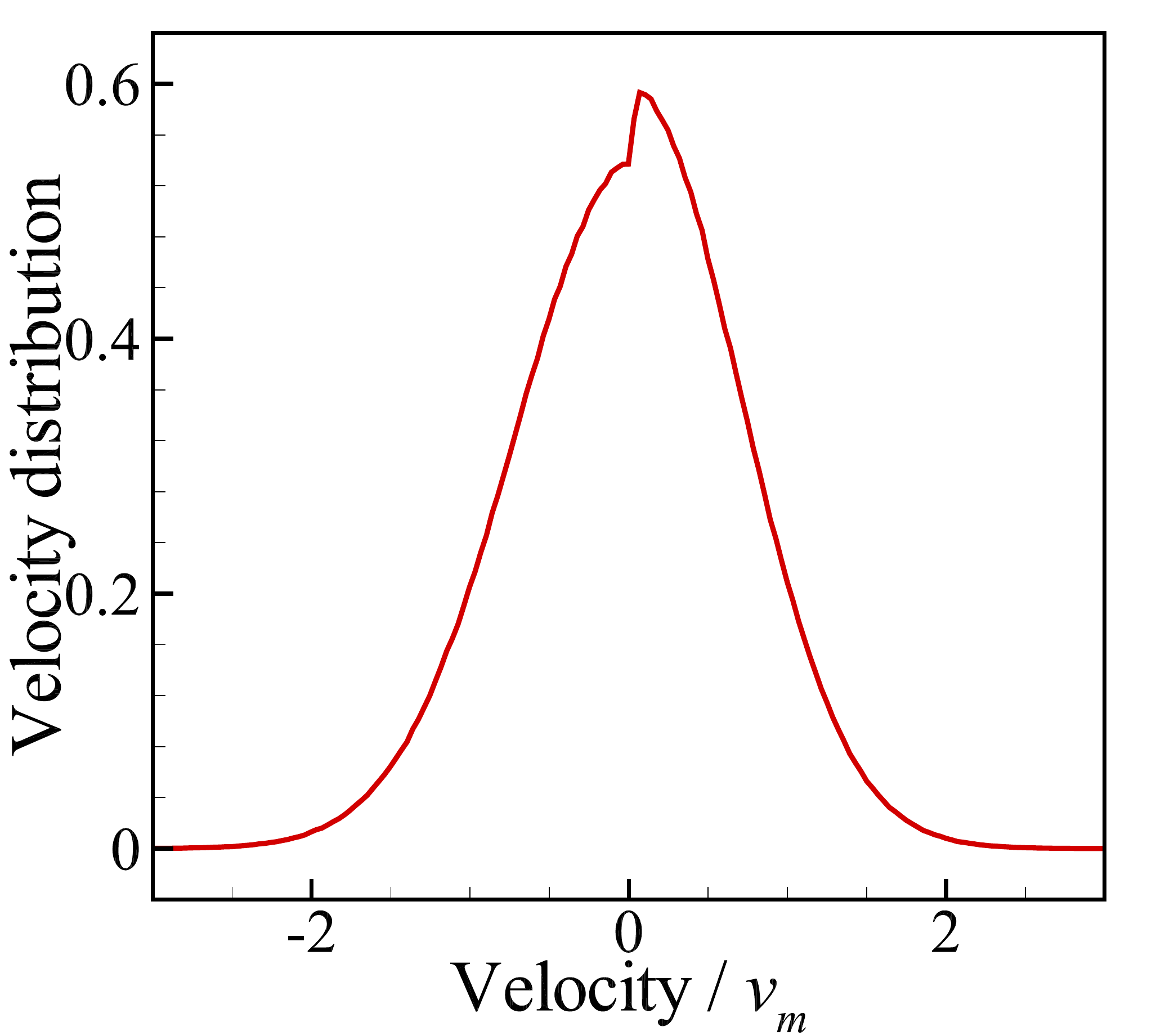}\label{fig:A_DSMC:a}}  \quad
	\subfloat[]{\includegraphics[scale=0.22,clip=true]{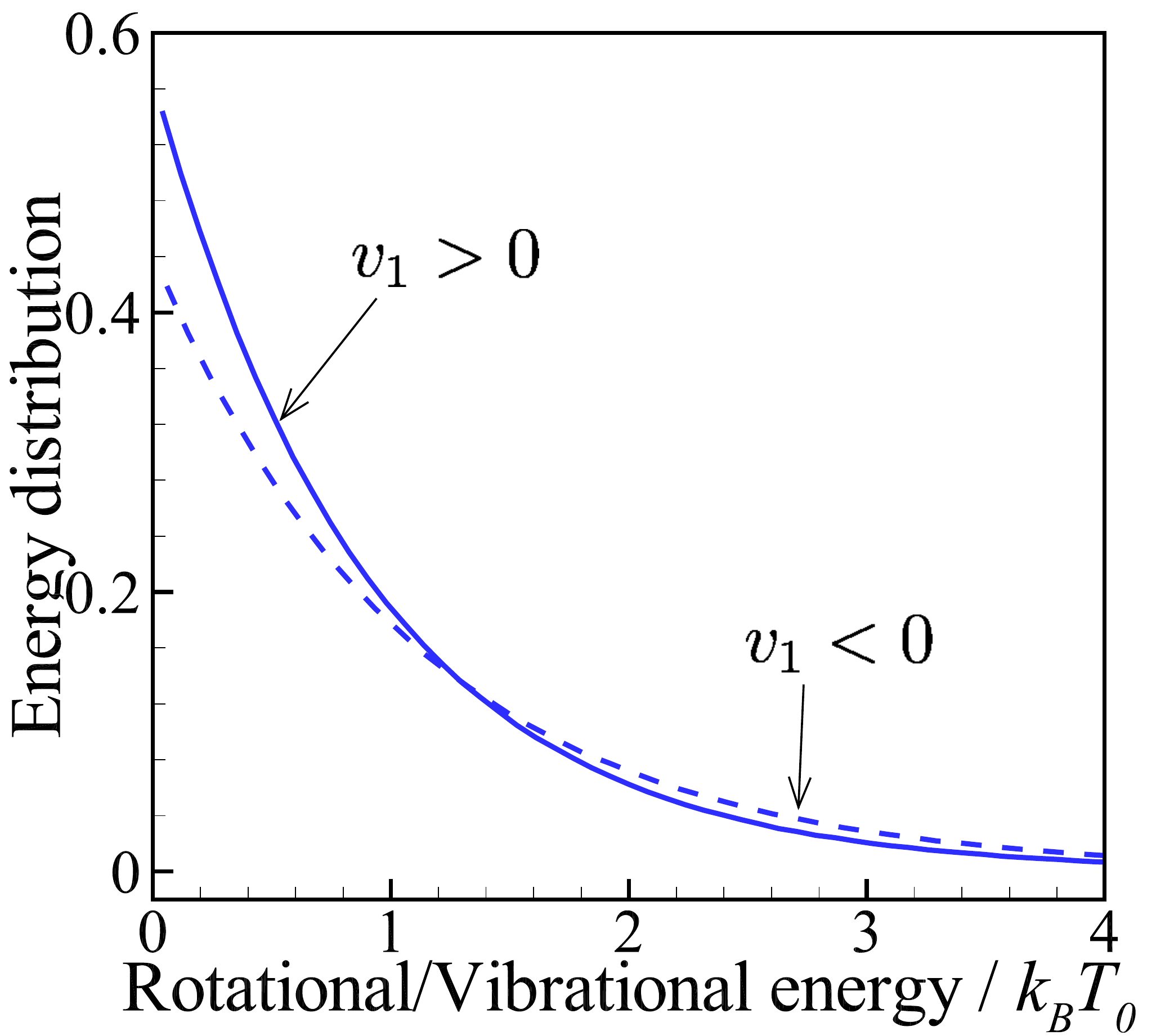}\label{fig:A_DSMC:b}} \\
	\vskip 0.5cm
	\subfloat[]{\includegraphics[scale=0.235,clip=true]{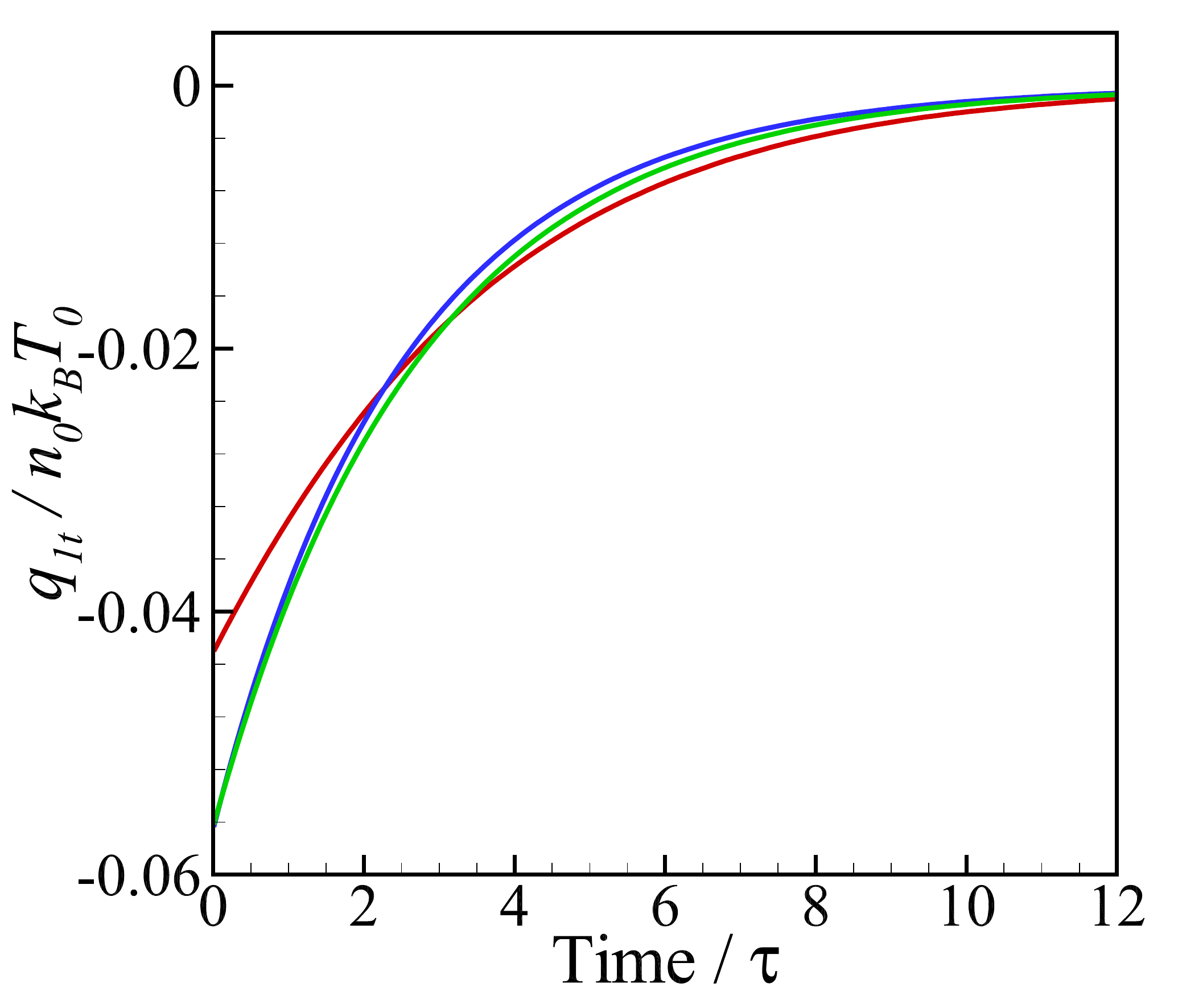}\label{fig:A_DSMC:c}}  \quad
	\subfloat[]{\includegraphics[scale=0.23,clip=true]{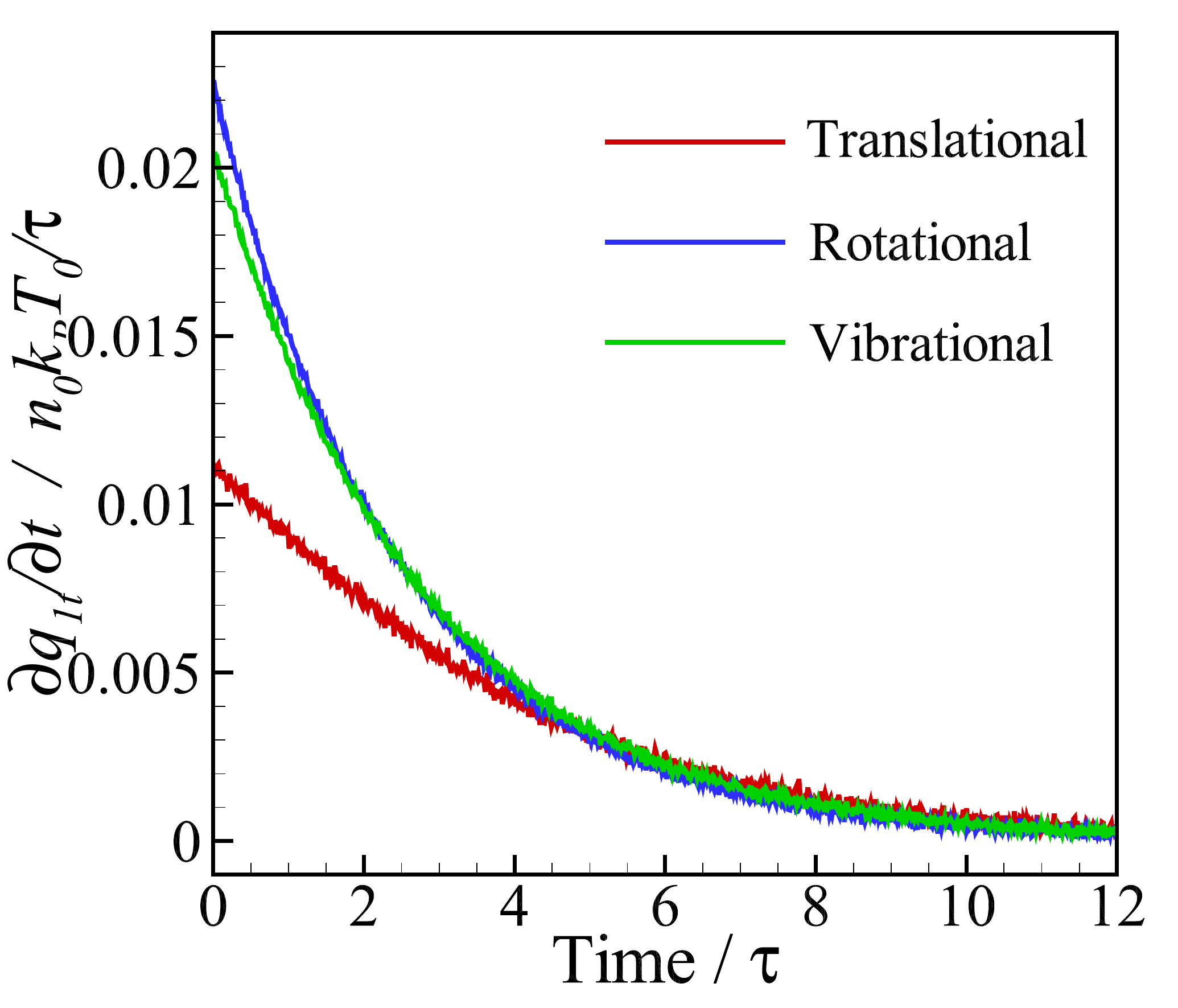}\label{fig:A_DSMC:d}}
	\caption{Extraction of the thermal relaxation rates $\bm{A}$ in~\eqref{eq:heat_flux_relaxation} from the DSMC simulation. Special distributions of (a) the molecular velocity and (b) rotational/vibrational energy are designed to generate initial heat flux. (c) The evolution of heat fluxes and (d) their time derivatives are monitored until the system reaches thermal equilibrium.}
	\label{fig:A_DSMC}
\end{figure}

Similar to the procedure of extracting thermal relaxation rates for the translational and rotational DoF from DSMC~\citep{Li2021JFM}, here a homogeneous system of nitrogen is simulated, which consists of $10^6$ simulation particles in a cubic cell of the size $(10~\text{nm})^3$. The periodic condition is applied at all boundaries. Binary collisions are described by the variable soft sphere model, and the system parameters and properties of nitrogen used in the simulations are: $d_r=d_v=2$, $n_0=2.69\times10^{25}~\text{m}^{-3}$,  $T_0=5000~\text{K}$,  $m=4.65\times10^{-26}~\text{kg}$, the molecular diameter is $d=4.11\times10^{-10}~\text{m}$, the viscosity index is $\omega=0.74$,  the angular scattering parameter is $\alpha=1.36$, and the Schmidt number is $Sc=1/1.34$~\citep{Bird1994}. Initially,  simulation particles with positive velocity in the $x_1$ direction follow the equilibrium distribution at $4500$~K, while those moving in the opposite direction follow the equilibrium distribution at $5500$~K, see  figure \ref{fig:A_DSMC:a} and \ref{fig:A_DSMC:b}, so that initial heat fluxes in all DoF are generated. Then the evolution of heat flux is monitored until the entire system reaches thermal equilibrium, see figure \ref{fig:A_DSMC:c}. Ensemble averaged is taken from 3000 independent runs to get the time derivative of heat flux in figure \ref{fig:A_DSMC:d}. Finally, the following relaxation rates are extracted by solving the linear regression problem \eqref{eq:heat_flux_relaxation} with the least squares method:
\begin{equation}\label{eq:relaxation_rates}
    \left[ 
      \begin{array}{ccc} 
        A_{tt} & A_{tr} & A_{tv} \\ A_{rt} & A_{rr} & A_{rv} \\ A_{vt} & A_{vr} & A_{vv}
      \end{array}
    \right]
    =
	\left[ 
      \begin{array}{ccc} 
        ~0.786 & -0.208 & ~0.003 \\ -0.047 & ~0.883 & -0.049 \\ -0.004 & -0.038 & ~0.772
      \end{array}
    \right].
\end{equation}
Hence, according to~\eqref{eq:EuckenFactor_A}, we have $f_t=2.3635$, $f_r=1.3979$, $f_v=1.3825$, and $f_{eu}=1.807$. With these parameters, our kinetic model is uniquely determined.

\subsection{Fourier flow}\label{subsec:validation_FourierFlow}

The heat transfer in the nitrogen gas between two parallel plates located at $x_2=0$ and $L_0$ are considered, where the temperature of the lower and upper plates are $T_l=0.8T_0$ and $T_u=1.2T_0$, respectively. The averaged number density of nitrogen is set to be $n_0$, and the characteristic length $L_0$ is chosen to be the distance between two plates. The Knudsen numbers considered are  $\text{Kn}=0.1$ and 1. The diffuse boundary conditions are adopted, so that the reflected distributions are
\begin{equation}\label{eq:BC_FourierFlow}
	\begin{aligned}[b]
		&x_2=0,~v_2\ge0: \quad f_0=\frac{n_{in}(x_2=0)}{n_0}E_t(T_l), \quad
		f_1=\frac{d_r}{2}k_BT_lf_0, \quad
		f_2=\frac{d_v}{2}k_BT_lf_0, \\
		&x_2=L_0,~v_2\le0: \quad f_0=\frac{n_{in}(x_2=L_0)}{n_0}E_t(T_u), \quad
		f_1=\frac{d_r}{2}k_BT_uf_0, \quad
		f_2=\frac{d_v}{2}k_BT_uf_0,
	\end{aligned}
\end{equation}
where $n_{in}$ is determined by the flux of incident number density of gas at the plates:
\begin{equation}\label{eq:BC_FourierFlow_n_in}
	\begin{aligned}[b]
		n_{in}(x_2=0)&= -\left(\frac{2m\pi}{k_BT_l}\right)^{1/2}\int_{v_2<0}v_2f_0\mathrm{d}\bm{v}, \\
		n_{in}(x_2=L_0)&= \left(\frac{2m\pi}{k_BT_u}\right)^{1/2}\int_{v_2>0}v_2f_0\mathrm{d}\bm{v}.
	\end{aligned}
\end{equation}

\begin{figure}[t]
	\centering
	\subfloat[]{\includegraphics[scale=0.18,clip=true]{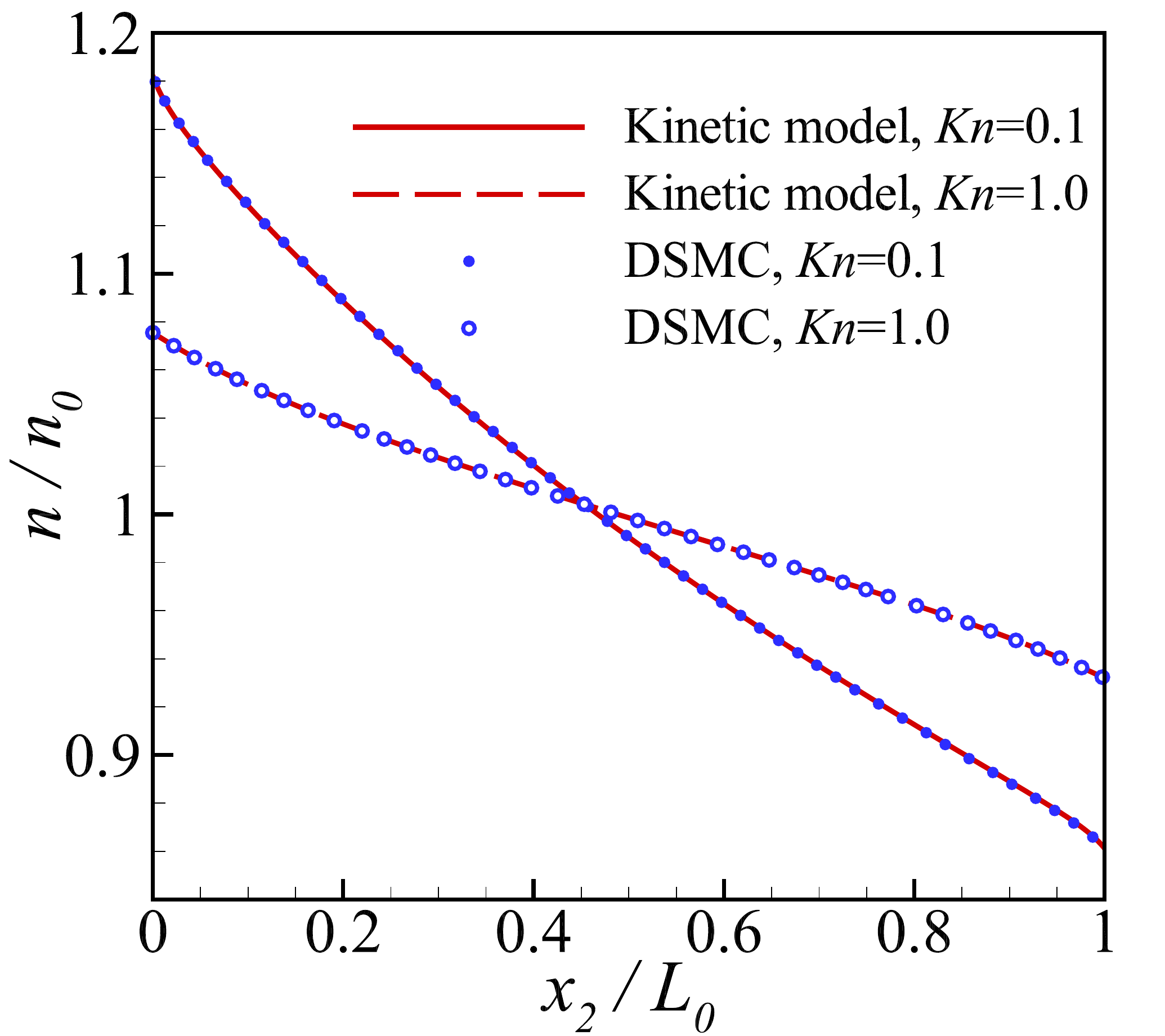}\label{fig:1DFourierFlow:a}}
	\subfloat[]{\includegraphics[scale=0.18,clip=true]{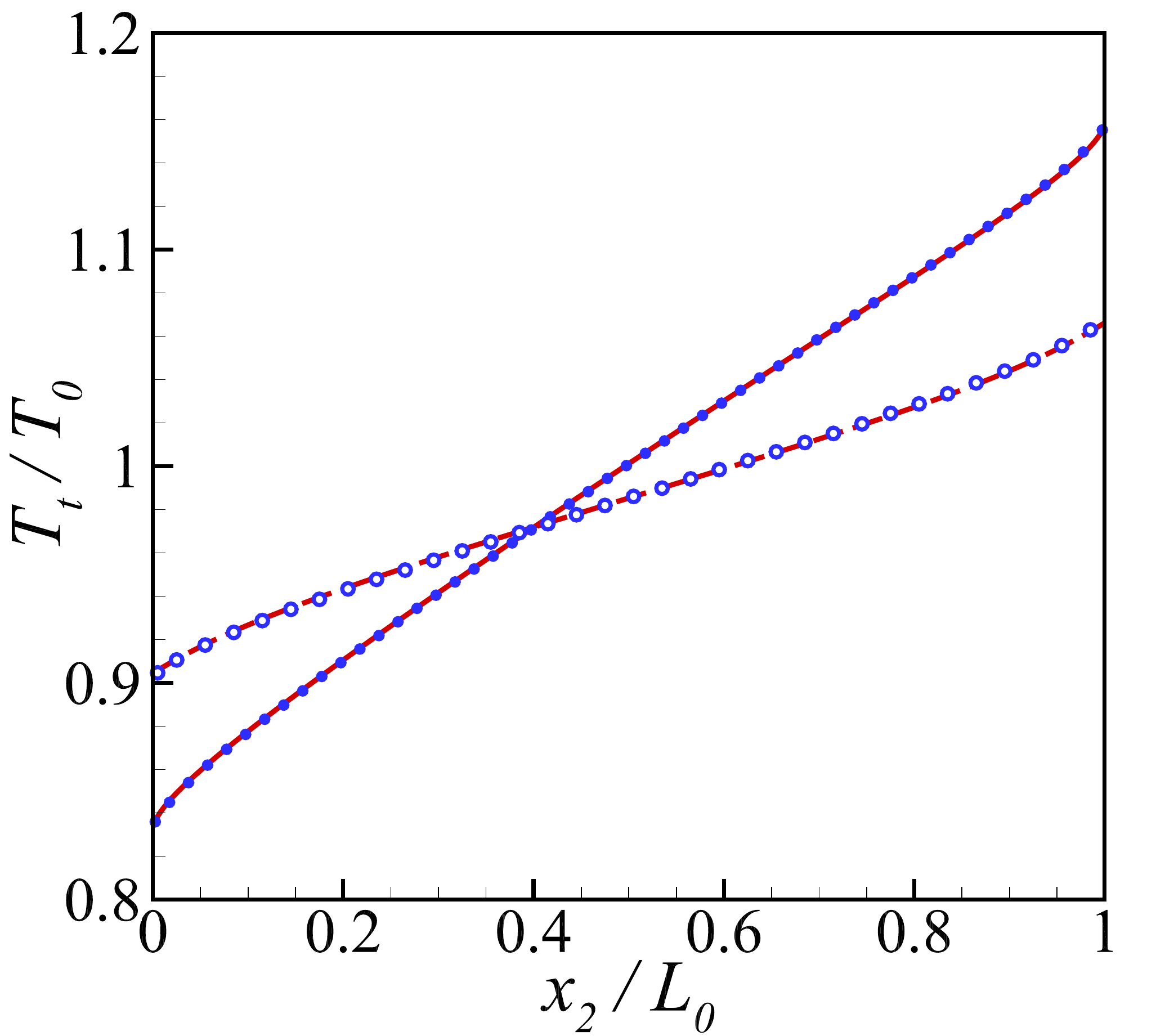}\label{fig:1DFourierFlow:b}}
	\subfloat[]{\includegraphics[scale=0.18,clip=true]{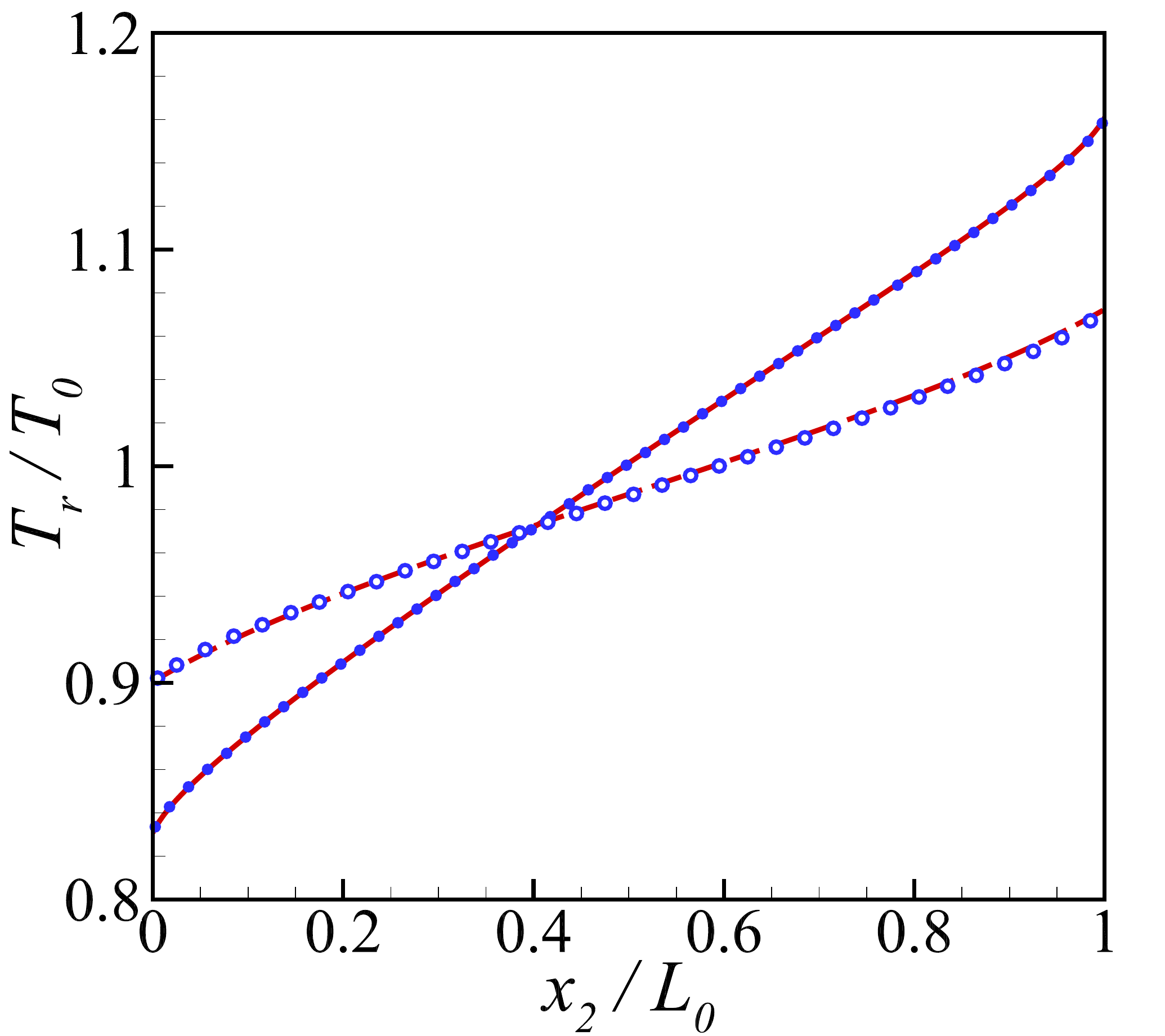}\label{fig:1DFourierFlow:c}} \\
	\subfloat[]{\includegraphics[scale=0.18,clip=true]{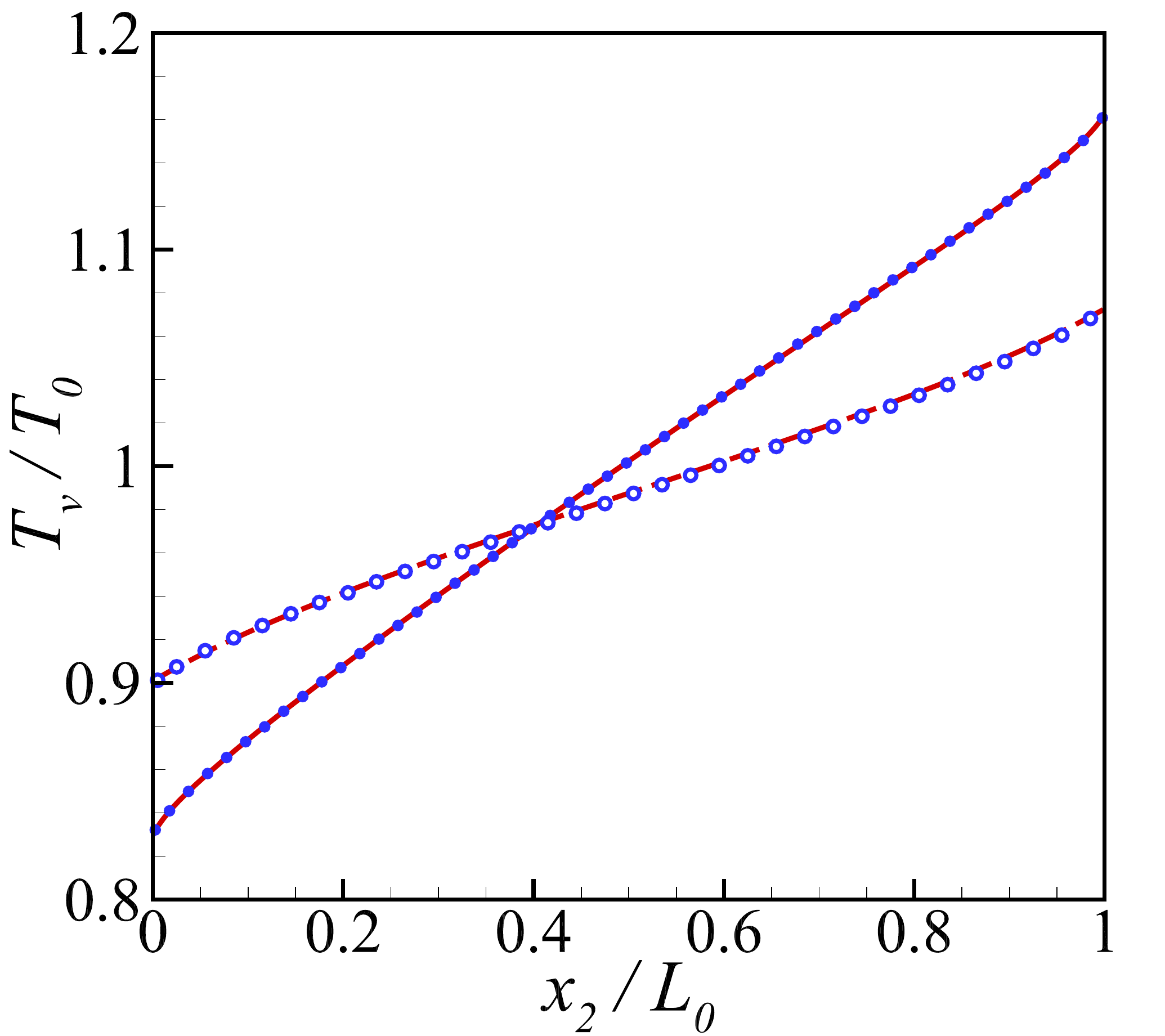}\label{fig:1DFourierFlow:d}}
	\subfloat[]{\includegraphics[scale=0.18,clip=true]{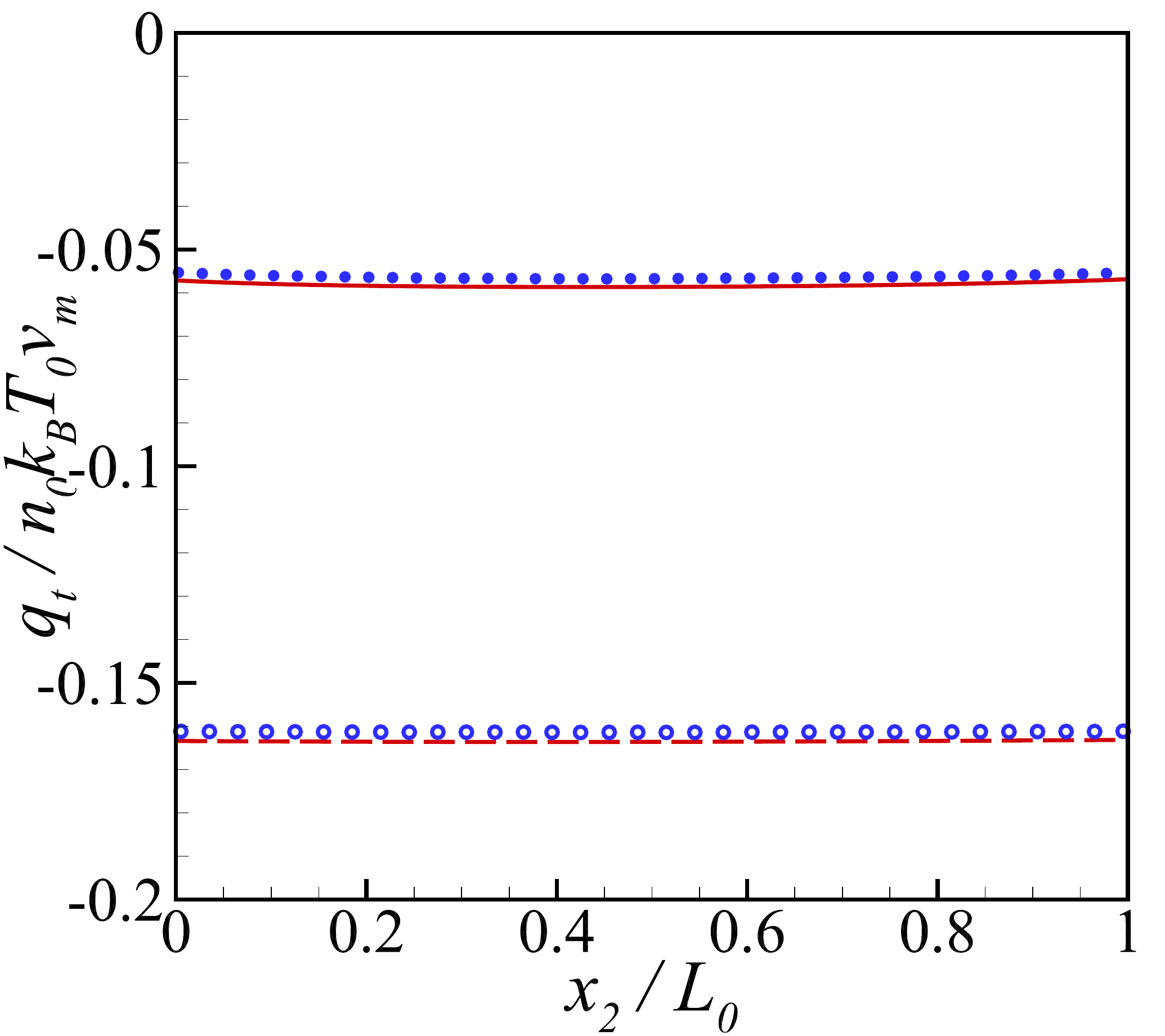}\label{fig:1DFourierFlow:e}}
	\subfloat[]{\includegraphics[scale=0.18,clip=true]{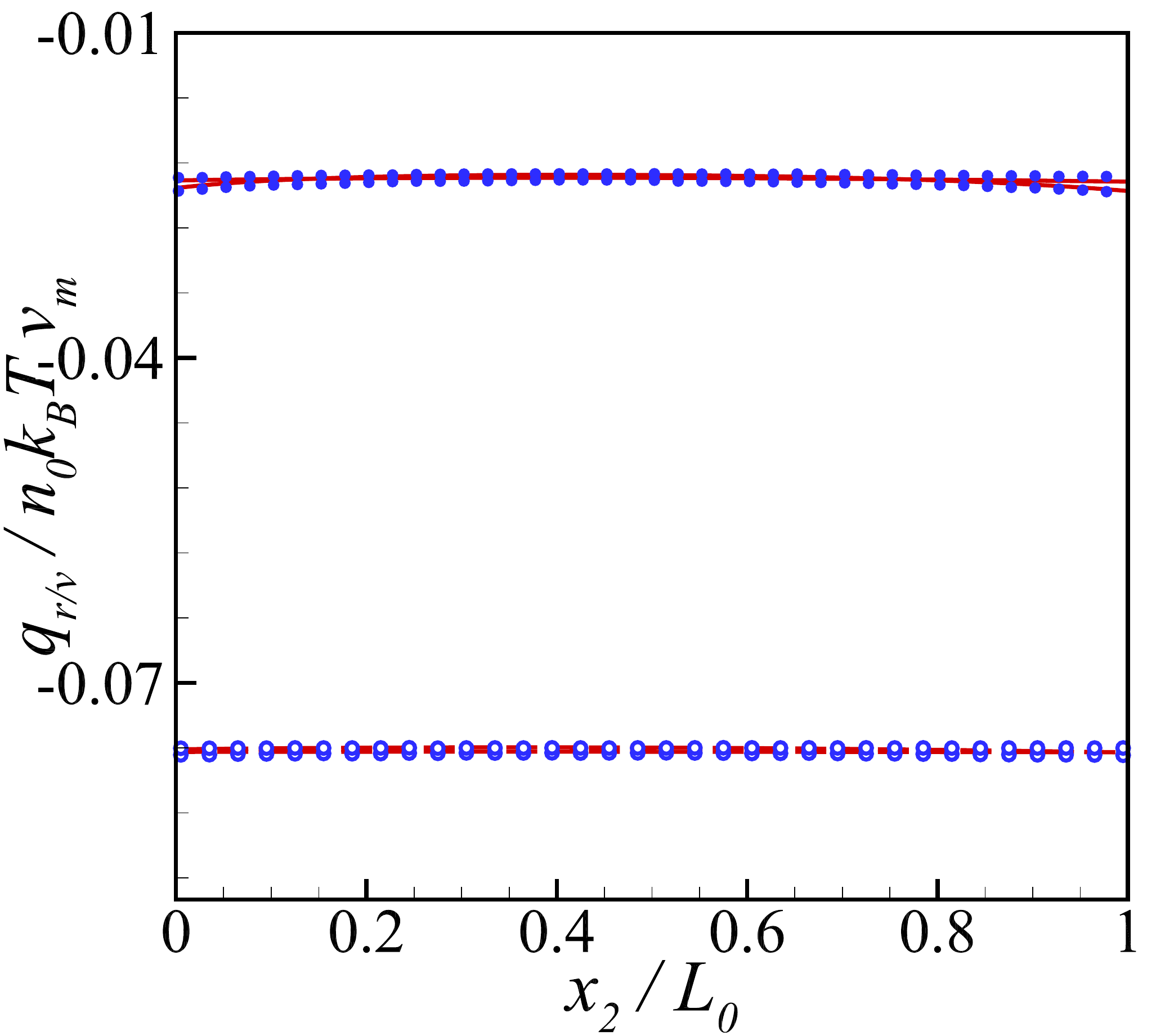}\label{fig:1DFourierFlow:f}}
	\caption{
		Comparisons of the (a) density, (b) translational temperature, (c) rotational temperature, (d) vibrational temperature, (e) translational heat flux and (f) rotational/vibrational heat flux of nitrogen between our kinetic model (lines) and DSMC  (circles) for the Fourier flows.
	}
	\label{fig:1DFourierFlow}
\end{figure}

Numerical results from the kinetic model~\eqref{eq:kinetic_model_equation} and  DSMC are shown in figure \ref{fig:1DFourierFlow}. For both  $\text{Kn}=0.1$ and $\text{Kn}=1$, excellent agreement in the density and temperature are observed, where the maximum relative error in the translational heat flux is less than $3\%$. Meanwhile,  profiles of translational, rotational and vibrational temperatures nearly overlap, although the relaxation times for different DoF are different. Additionally, the rotational and vibrational heat flux are almost the same (figure~\ref{fig:1DFourierFlow:f}), due to the close values of the rotational and vibrational thermal conductivities. Thus, it is clearly seen that the values of collision number $Z_r$ and $Z_v$ do not have influence on the distribution of macroscopic quantities for the steady-state planar Fourier flow. 

We define an effective thermal conductivity of the system by 
\begin{equation}
\kappa_{e}=-q\frac{L_0}{\Delta T}.
\end{equation}
With the increase of Knudsen number, $\kappa_e$ decreases due to the wall confinement that effectively increases the thermal resistance. For instances, the ratio between the translational and rotational/vibrational thermal conductivities in the continuum limit is around $\kappa_t/\kappa_{r,v}=2.55$, which decreases to 2.42 and 2.21 when $\text{Kn}=0.1$ and $\text{Kn}=1$, respectively. 

\begin{figure}[t]
	\centering
	\subfloat[]{\includegraphics[scale=0.24,clip=true]{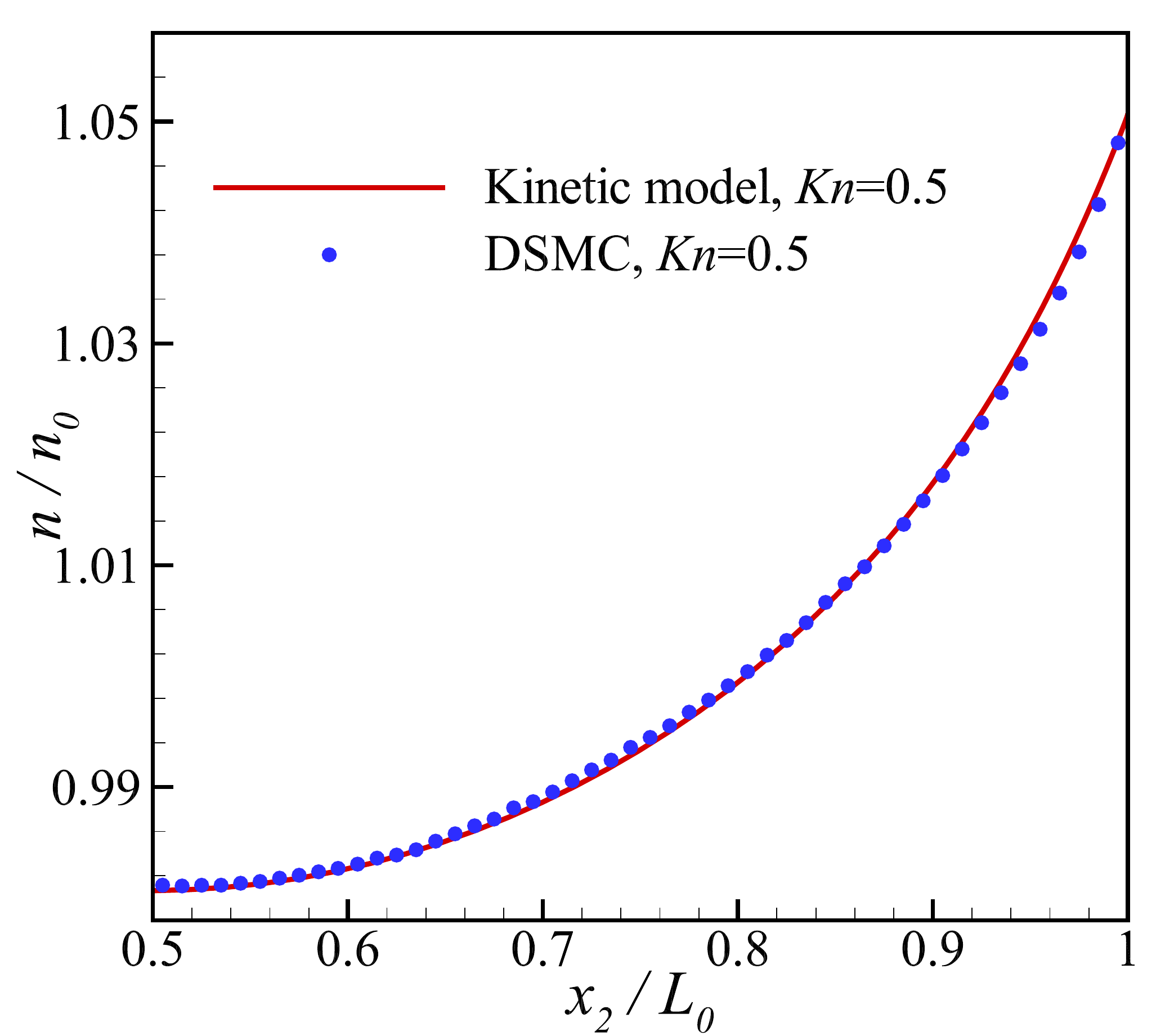}} \quad
	\subfloat[]{\includegraphics[scale=0.24,clip=true]{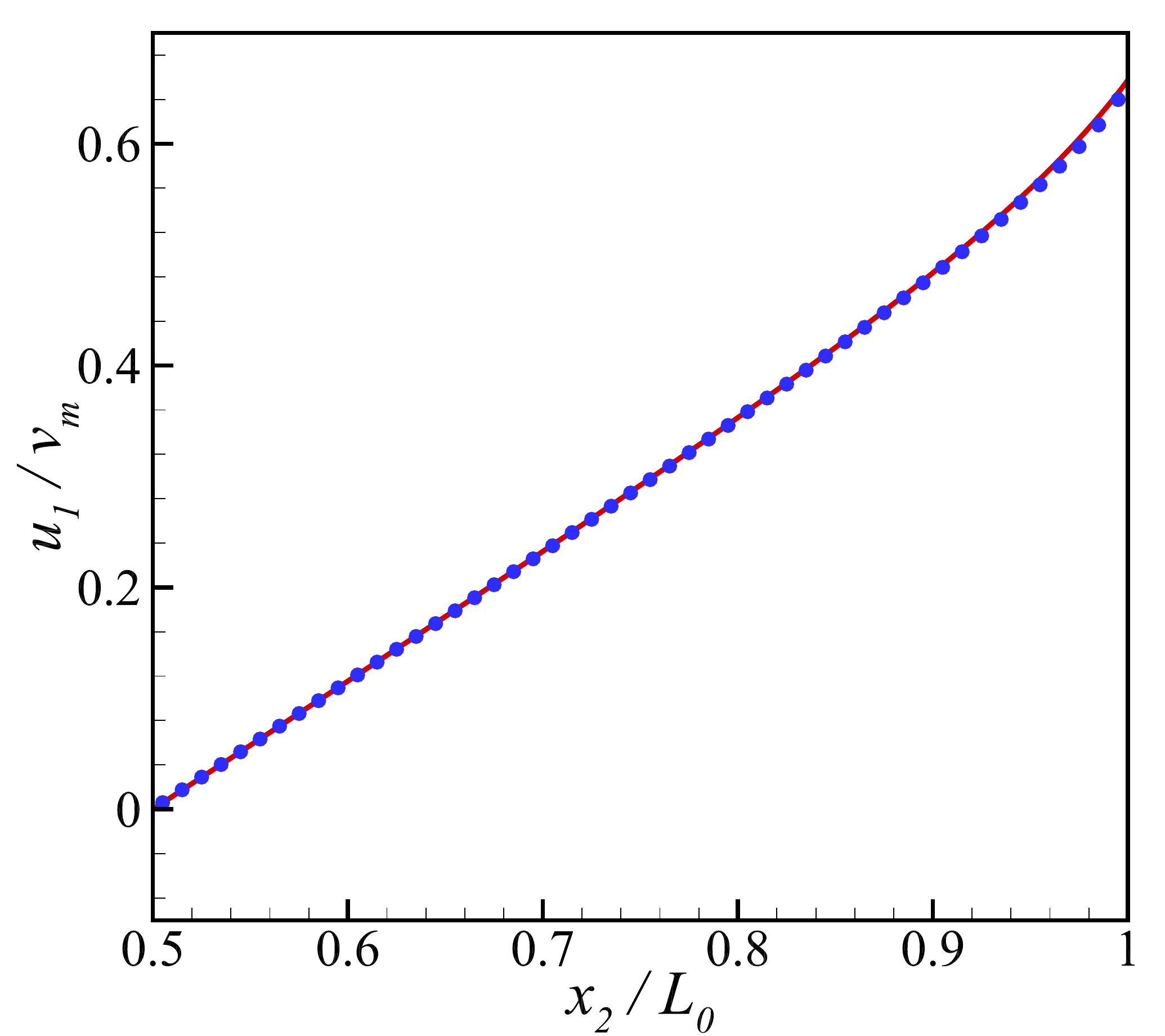}} \\
	\subfloat[]{\includegraphics[scale=0.24,clip=true]{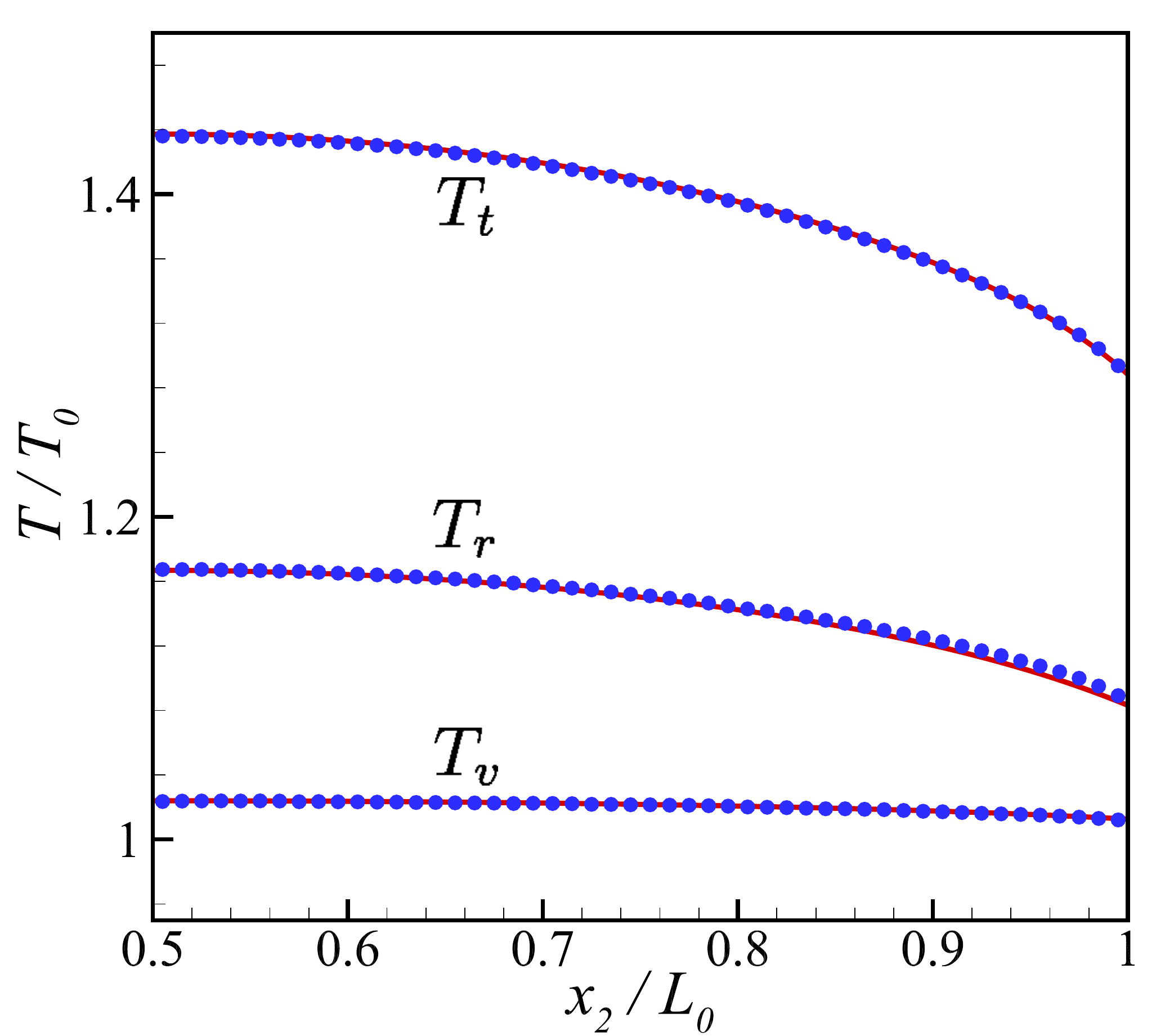}} \quad
	\subfloat[]{\includegraphics[scale=0.24,clip=true]{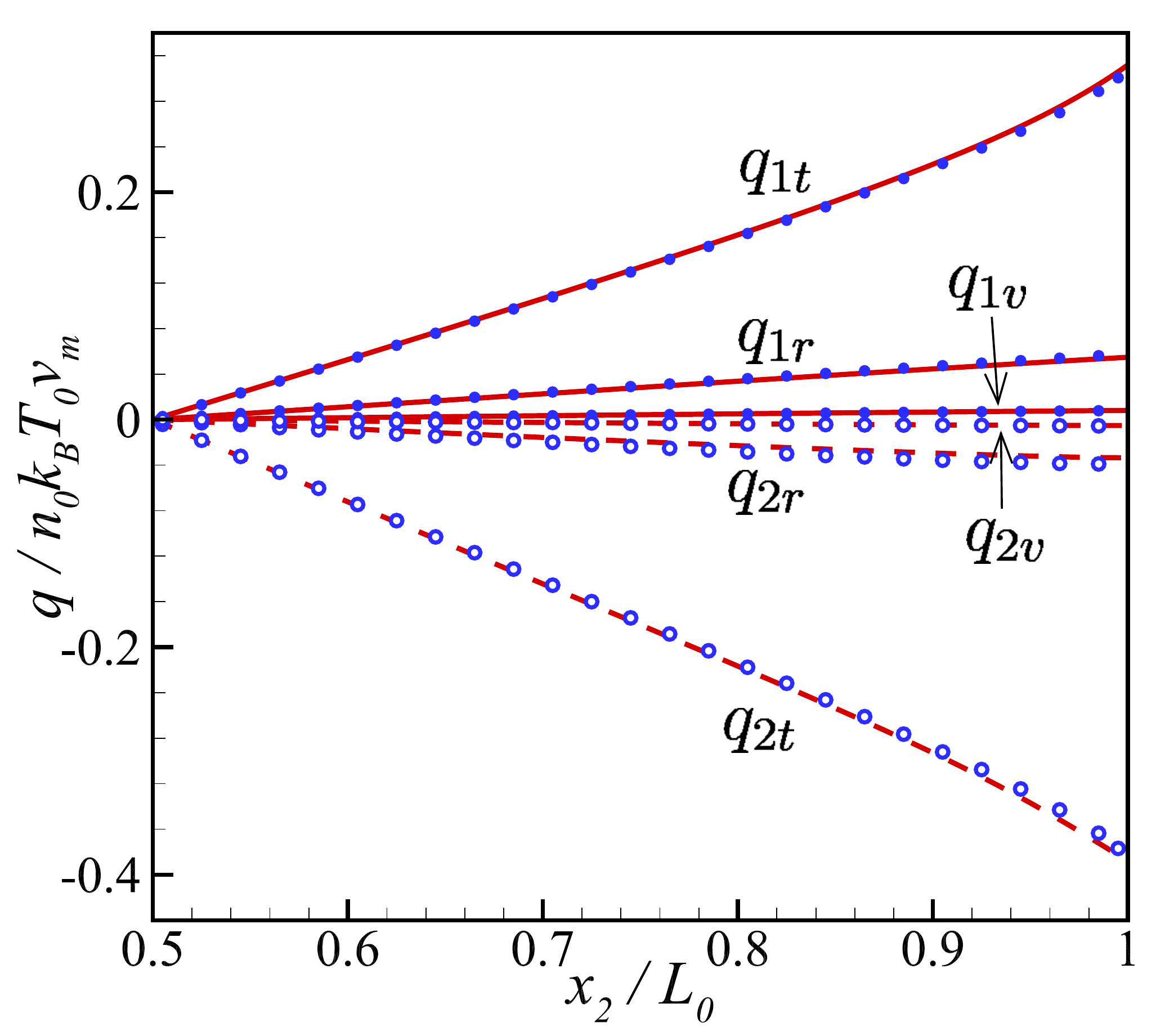}}
	\caption{Comparisons of the (a) density, (b) flow velocity, (c) temperature and (d) heat flux $q_1$ in the flow direction and $q_2$ perpendicular to flow direction of nitrogen, between our kinetic model (red lines) and DSMC simulations (blue circles) for the one-dimensional Couette flow at $\text{Kn}=0.5$.}
	\label{fig:1DCouetteFlow}
\end{figure}

\subsection{Couette flow}

The configuration of the Couette flow is the same as the Fourier flow, while the temperature of both plates are kept the same at $T_0$, and the velocity of lower and upper plates are $v_1=-v_m$ and $v_1=v_m$, respectively. Due to the symmetry, only half of the  domain $(L_0/2\le{}x_2\le{}L_0)$ is simulated. The diffuse boundary condition at $x_2=L_0$ yields:
\begin{equation}\label{eq:BC_CouetteFlow}
	v_2\le{}0: \quad 
	f_0=\frac{n_{in}(x_2=L_0)}{n_0}E_t(T_u), \quad
	f_1=\frac{d_r}{2}k_BT_0f_0, \quad
	f_2=\frac{d_v}{2}k_BT_0f_0,
\end{equation}
where $n_{in}(x_2=L_0)$ is determined as the same way as~\eqref{eq:BC_FourierFlow_n_in}, while the symmetrical condition at $x_2=L_0/2$ reads
\begin{equation}
	v_2\ge{}0: \quad 
	f_0=f_0(-v_1,-v_2,v_3), \quad
	f_1=\frac{d_r}{2}k_BTf_0, \quad
	f_2=\frac{d_v}{2}k_BTf_0, 
\end{equation}

The results from our kinetic model and the DSMC simulation at $\text{Kn}=0.5$ are shown in figure~\ref{fig:1DCouetteFlow}, which demonstrates the accuracy of our model. The vibrational temperature is much lower than the rotational one, since in this problem the energy increase in internal DoF only comes from the exchange with translational ones. Thus, larger collision number leads to less increase in internal temperature at the same distance from the wall (due to the infrequent relaxation with the translational mode), and also contributes less to the heat flux.

\subsection{Creep flow driven by the Maxwell demon}

\begin{figure}[t]
	\centering
	\subfloat[]{\includegraphics[scale=0.24,clip=true]{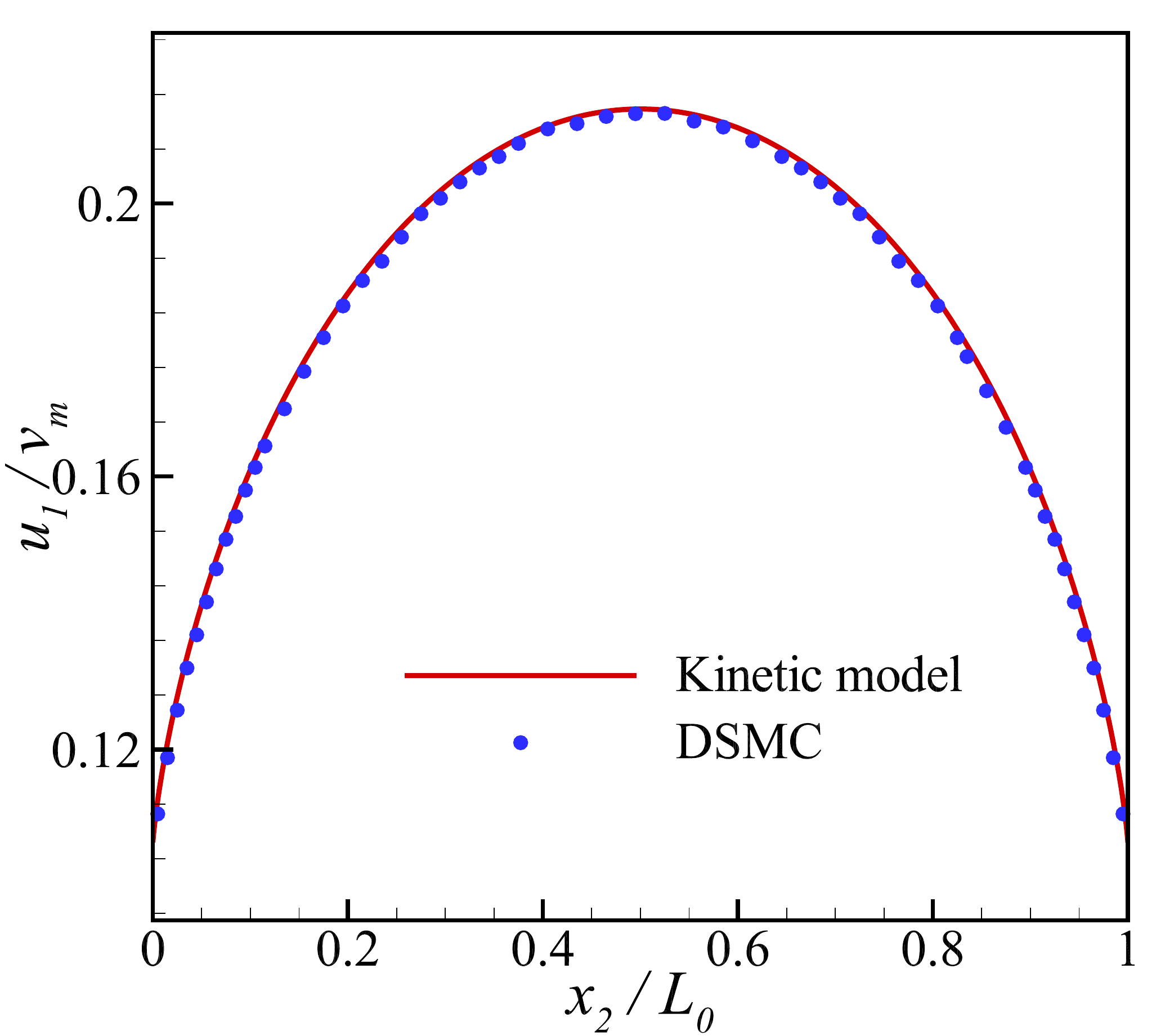}\label{fig:1DCreepFlow:a}} \quad
	\subfloat[]{\includegraphics[scale=0.24,clip=true]{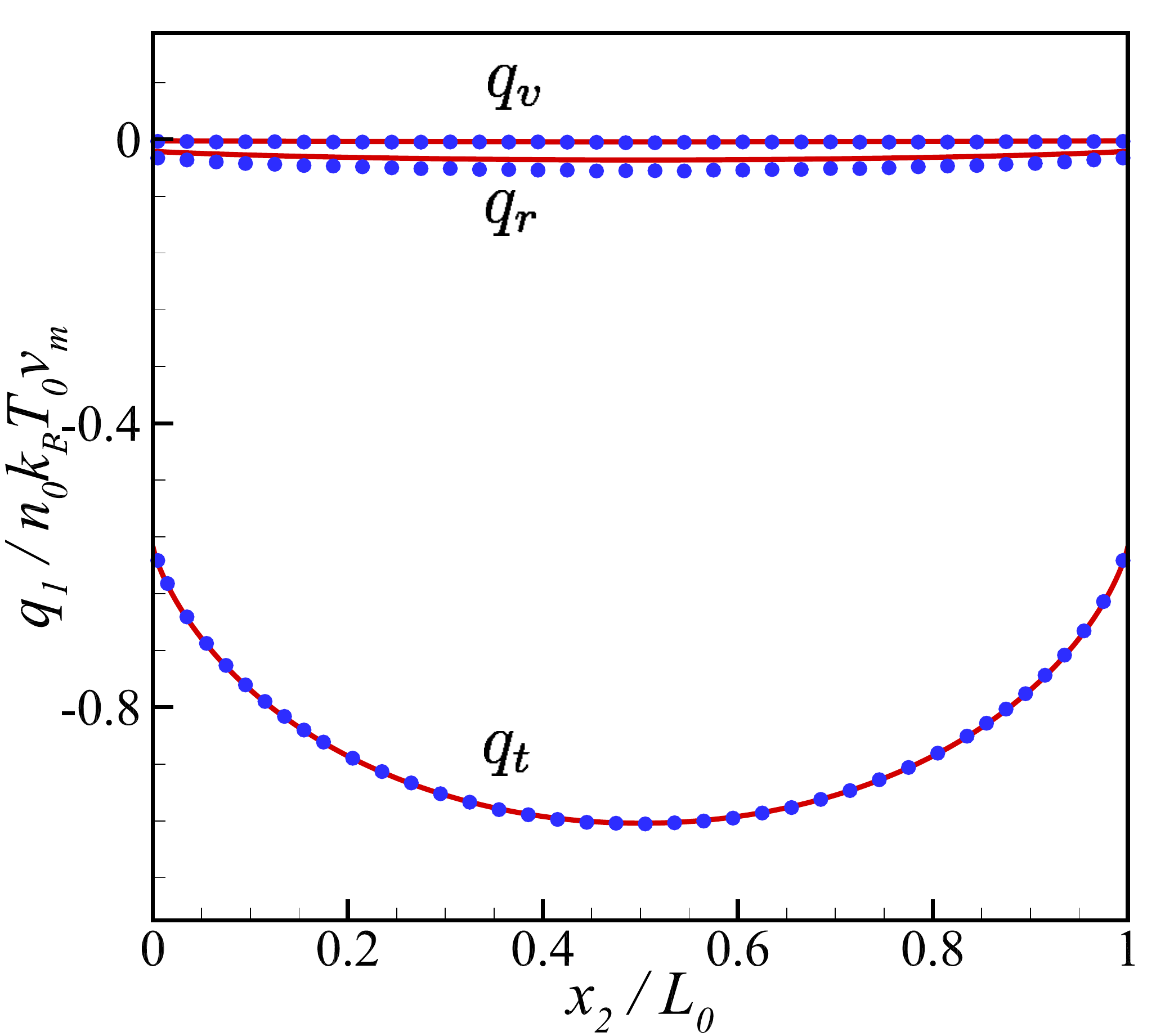}\label{fig:1DCreepFlow:b}} 
	\caption{Comparisons of the (a) velocity and (b) heat flux in flow direction of nitrogen between kinetic model (lines) and DSMC simulations (circles) for one-dimensional creep flow driven by the Maxwell demon at $\text{Kn}=1$. Both the flow velocity and the heat flux have been further normalized by ${2a_0L_0}/{v_m^2}$.}
	\label{fig:1DCreepFlow}
\end{figure}

The creep flow driven by the Maxwell demon is a thought test~\citep{Li2021JFM}, where each gas molecule is subjected to an external acceleration based on its kinetic energy:
\begin{equation}\label{eq:CreepFlow_acceleration}
	a_1 = a_0\left(\frac{v_1^2}{v_m^2}-\frac{3}{2}\right).
\end{equation}
That is, fast molecules are forced towards the positive direction, while the slow molecules move in opposite direction. 

Consider the nitrogen flow driven by the Maxwell demon confined between two parallel plates with distance $L_0$ apart. To solve the force-driven flow, we choose small values of $a_0$ so that the gas flow deviates only slightly from the global equilibrium; the acceleration acting on the molecules is linearised, which results in the source terms at right-hand side of model equations \eqref{eq:kinetic_model_equation}:
\begin{equation}\label{eq:CreepFlow_equation}
	\begin{aligned}[b]
		\frac{\partial{f_0}}{\partial{t}}+\bm{v} \cdot \frac{\partial{f_0}}{\partial{\bm{x}}} &= Q(f_0) + \frac{g_{0r}-g_{0t}}{Z_r\tau} + \frac{g_{0v}-g_{0t}}{Z_v\tau}-\frac{2a_0L_0}{v_m^2}v_1E_t(T_0)\left(\frac{v_1^2}{v_m^2}-\frac{5}{2}\right), \\
		\frac{\partial{f_1}}{\partial{t}}+\bm{v} \cdot \frac{\partial{f_1}}{\partial{\bm{x}}} &= \frac{g_{1t}'-f_1}{\tau} + \frac{g_{1r}-g_{1t}}{Z_r\tau} + \frac{g_{1v}-g_{1t}}{Z_v\tau}-\frac{d_ra_0L_0}{v_m^2}v_1k_BT_0E_t(T_0)\left(\frac{v_1^2}{v_m^2}-\frac{5}{2}\right), \\
		\frac{\partial{f_2}}{\partial{t}}+\bm{v} \cdot \frac{\partial{f_2}}{\partial{\bm{x}}} &= \frac{g_{2t}'-f_2}{\tau} + \frac{g_{2r}-g_{2t}}{Z_r\tau} + \frac{g_{2v}-g_{2t}}{Z_v\tau}-\frac{d_va_0L_0}{v_m^2}v_1k_BT_0E_t(T_0)\left(\frac{v_1^2}{v_m^2}-\frac{5}{2}\right).
	\end{aligned}
\end{equation}
The plates at rest are fully diffuse, then the boundary conditions are simply given by \eqref{eq:BC_FourierFlow} and \eqref{eq:BC_FourierFlow_n_in}, but with the wall temperature replaced by $T_0$. 

\begin{figure}[t]
	\centering
	\subfloat[]{\includegraphics[scale=0.24,clip=true]{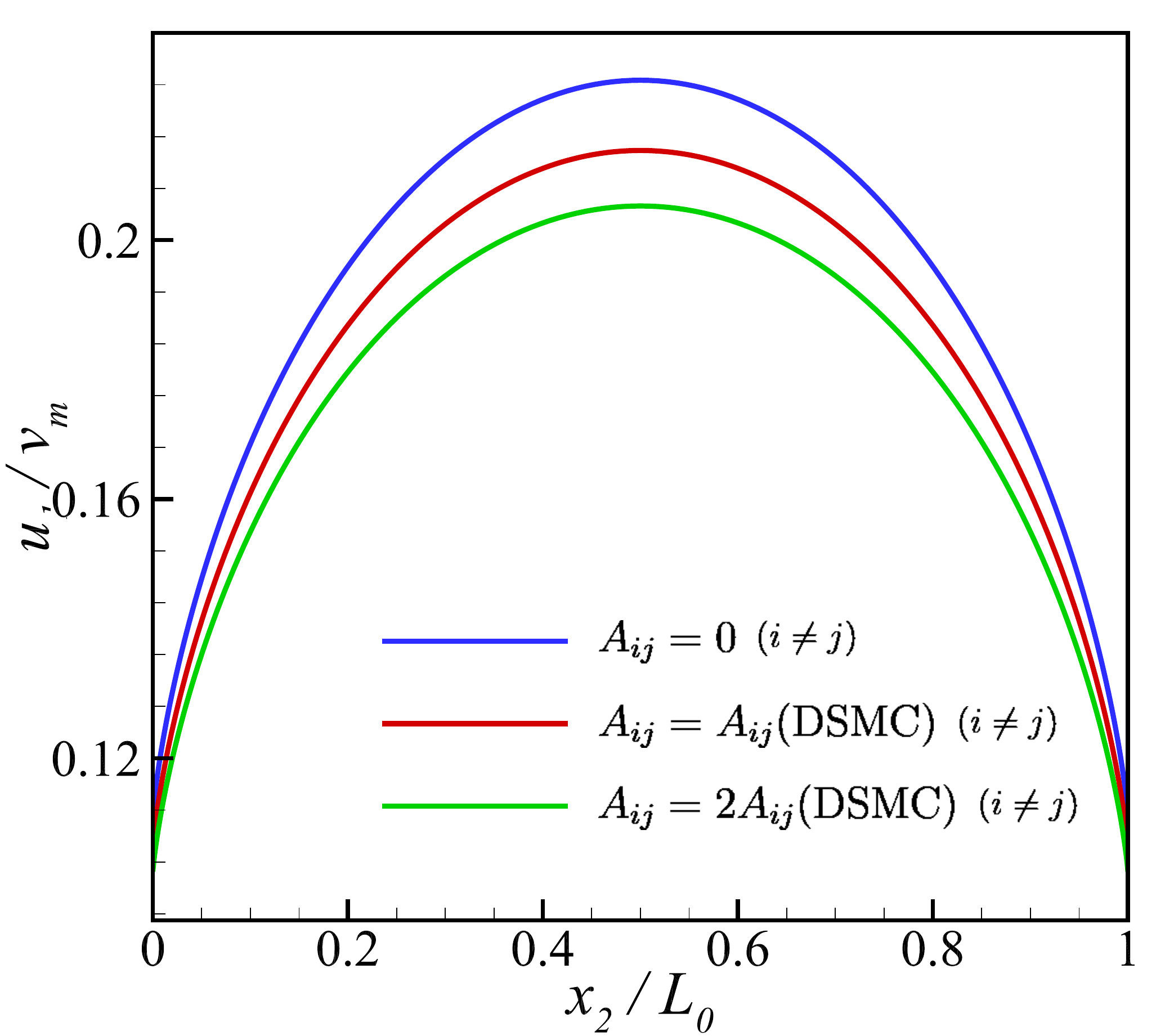}\label{fig:1DCreepFlow_Aij:a}} \quad
	\subfloat[]{\includegraphics[scale=0.24,clip=true]{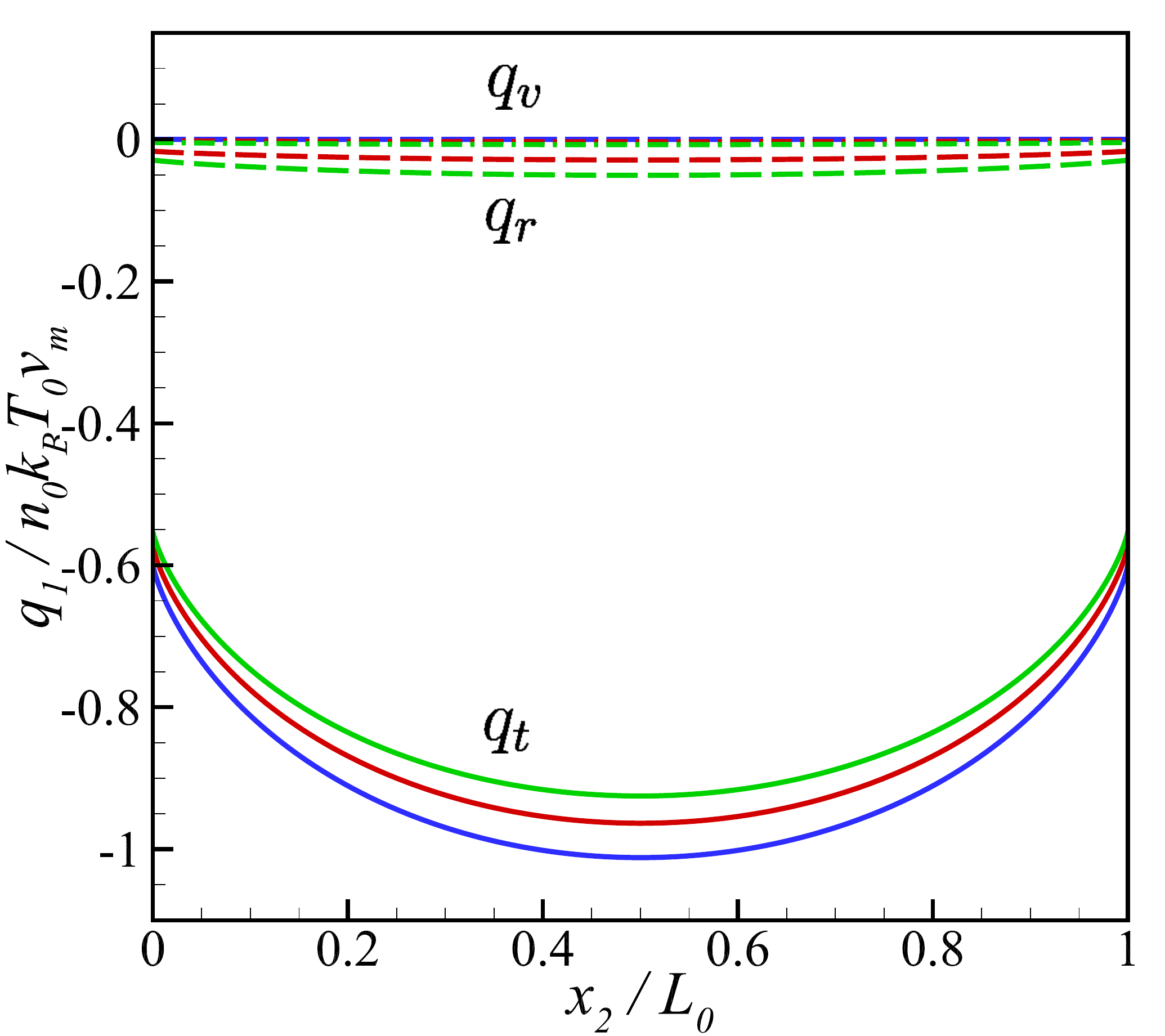}\label{fig:1DCreepFlow_Aij:b}} 
	\caption{Same as figure~\ref{fig:1DCreepFlow}, except that the off-diagonal elements in $A$ are set to be zero (blue), the values from DSMC (red) and double of those from DSMC (green), respectively.}
	\label{fig:1DCreepFlow_Aij}
\end{figure}

Figure \ref{fig:1DCreepFlow} shows the good agreement between the solution of our kinetic model and  DSMC at $\text{Kn}=1$. The rotational/vibrational heat flux is one/two order of magnitude smaller than the translational heat flux, which shows negligible contribution to the total heat transfer in this problem. 

To assess the influence of the thermal relaxation rates on the creep flow, two more cases are conducted by varying the values of the matrix $A$ but keeping the Eucken factors fixed. More specifically, the off-diagonal elements in $A$ in the two cases are set to be zero and double of those given by DSMC, respectively. The values of diagonal elements are calculated based on \eqref{eq:EuckenFactor_A} using the fixed Eucken factors. Figure \ref{fig:1DCreepFlow_Aij} shows that these relaxation rates affect the  flow velocity and heat fluxes, despite that the thermal conductivities are fixed. In particular, when the off-diagonal elements in $A$ are zero, the heat fluxes of different types of DoF are decoupled, so that the internal heat fluxes are exactly zero. This situations occur in  many traditional kinetic models, such as the Rykov model and the ellipsoidal-statistical BGK model. This example demonstrates the importance of recovering the fundamental thermal relaxation process rather than the apparent thermal conductivities in rarefied gas flow simulations. 

\subsection{Normal shock wave}\label{subsec:validation_ShockWave}

\begin{figure}[t]
	\centering
	\subfloat[]{\includegraphics[scale=0.24,clip=true]{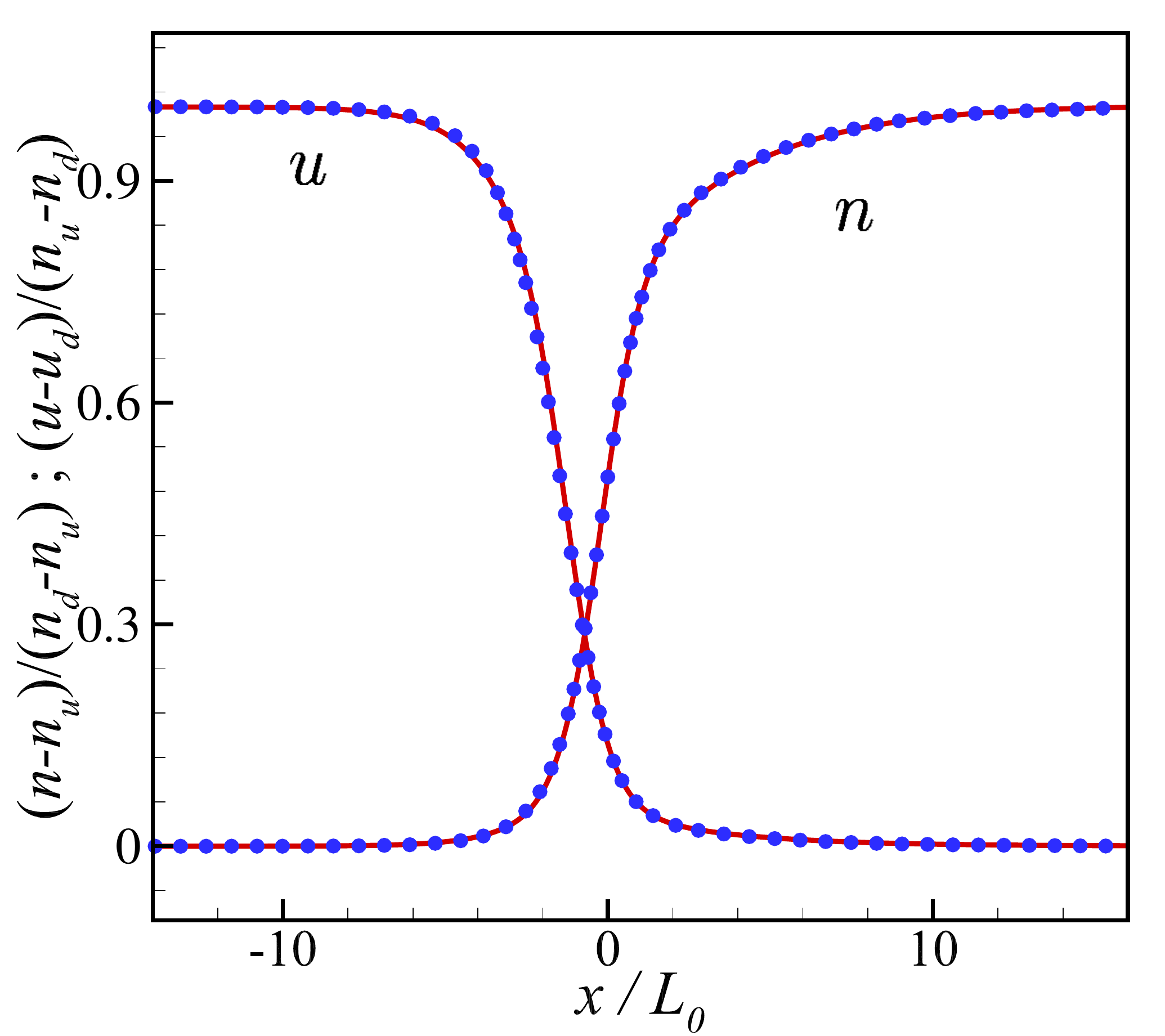}\label{fig:1DShockWave:a}} \quad
	\subfloat[]{\includegraphics[scale=0.24,clip=true]{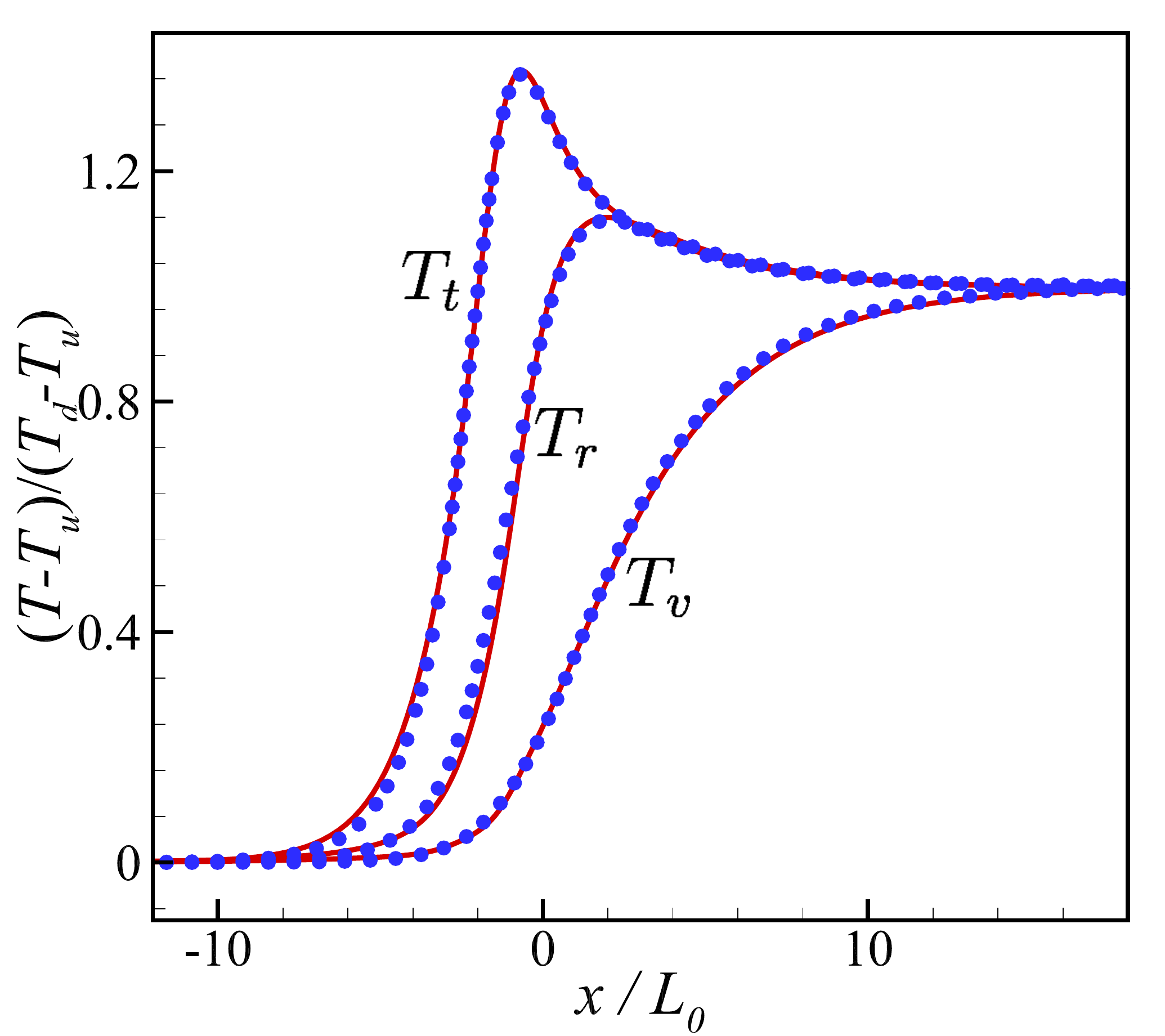}\label{fig:1DShockWave:b}} \\
	\subfloat[]{\includegraphics[scale=0.24,clip=true]{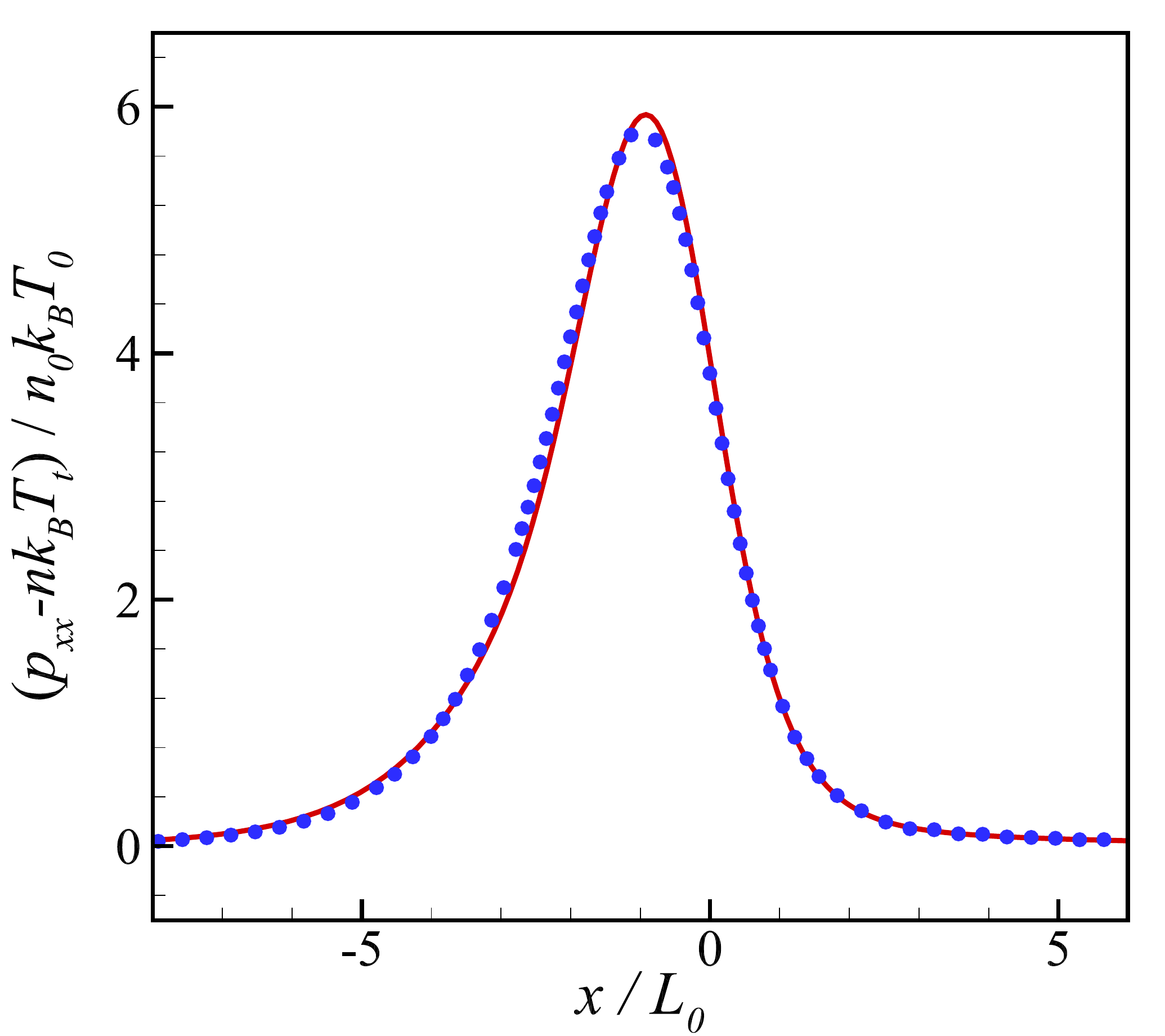}\label{fig:1DShockWave:c}} \quad
	\subfloat[]{\includegraphics[scale=0.24,clip=true]{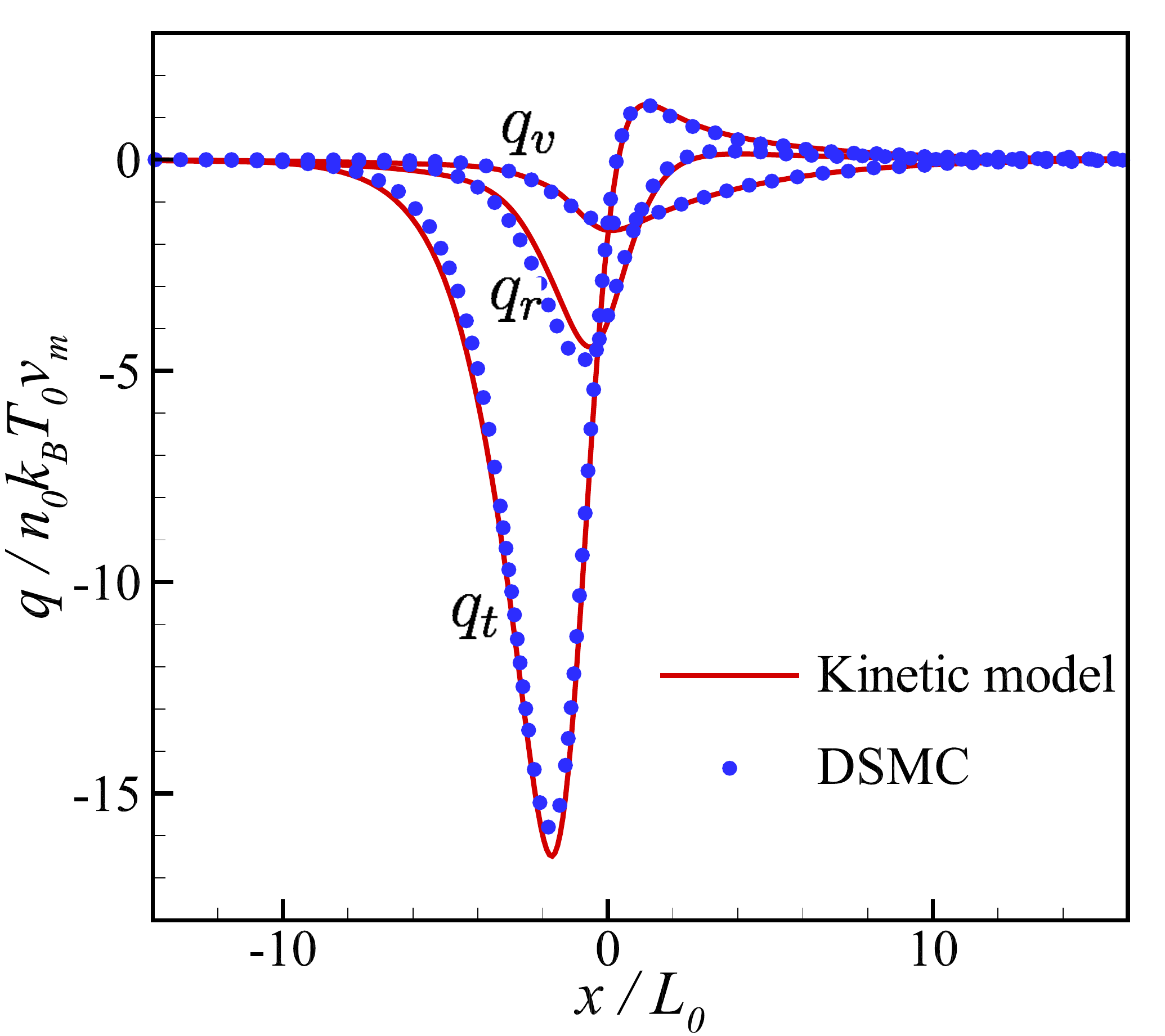}\label{fig:1DShockWave:d}}
	\caption{Comparisons of the (a) density and velocity, (b) temperature, (c) deviated pressure and (d) heat flux of nitrogen between our kinetic model (lines) and DSMC (circles) for normal shock wave at $\text{Ma}=5$.}
	\label{fig:1DShockWave}
\end{figure}

In the simulations of normal shock wave of nitrogen, the upstream number density $n_u=n_0=2.69\times10^{25}\text{m}^{-3}$ and temperature ${T_u=T_0=3993.8\text{K}}$ are chosen to be the reference values, which also determine the characteristic length to be $L_0={16\mu(T_0)}/({5n_0\sqrt{2\pi mk_BT_0}})$ and hence $\text{Kn}={5\pi}/{16}$ in this problem. The total length of the simulation domain is $90L_0$, so that the boundary conditions at both ends can be approximated by equilibrium states (the wave front is initially located at $x=0$):
\begin{equation}\label{eq:BC_ShockWave}
	\begin{aligned}[b]
		&x=-30L_0,~v\ge{}0: \quad f_0=\frac{n_{u}}{n_0}E_t(T_u), ~f_1=\frac{d_r}{2}k_BT_uf_0, ~f_2=\frac{d_v}{2}k_BT_uf_0, \\
		&x=60L_0,~v\le{}0: \quad ~~f_0=\frac{n_{d}}{n_0}E_t(T_d), ~f_1=\frac{d_r}{2}k_BT_df_0, ~f_2=\frac{d_v}{2}k_BT_df_0,
	\end{aligned}
\end{equation}
where the subscripts $u, d$ represent the upstream and downstream end, respectively. Given the Mach number, macroscopic quantities at the downstream end are determined by the Rankine–Hugoniot relation.

Numerical results of both the kinetic model~\eqref{eq:kinetic_model_equation} and DSMC are compared in figure \ref{fig:1DShockWave}, when the Mach number is $\text{Ma}=5$. As expected, the model equations reproduce the structure of normal shock wave with high accuracy. The rotational and vibrational collision numbers, $Z_r$ and $Z_v$, which affect energy exchange rate between internal and translational modes, play roles in the  difference of rotational and vibrational temperatures. That is, the distance for vibrational temperature to reach equilibrium are much longer than that for rotational modes. This is consistent with the fact that we set $Z_v=10Z_r$.

\section{Application to two-dimensional thermally induced microflow}\label{sec:2D_thermal}

Have validated the kinetic model~\eqref{eq:kinetic_model_equation}, we investigate the thermal transpiration of molecular gas in a cavity and the Knudsen force on a micro-beam. The viscosity index is varied to examine the effect of intermolecular potential on thermally induced flows. The deterministic numerical method is suited in this case since the flow speed is usually very small.

\subsection{Thermal transpiration in cavity}\label{subsec:transpiration}

Consider a two-dimensional rectangular cavity with aspect ratio of 5, and the length of the short side is set to be the characteristic length $L_0$. The temperature of the two ends are maintained at $T_w(x_1=0)=0.8T_0$ and $T_w(x_1=5L_0)=1.2T_0$, respectively, and that of the side walls  are linearly distributed from $0.8T_0$ to $1.2T_0$. Only the lower half of the cavity $(0\le{}x_1\le5L_0,~0\le{}x_2\le{}L_0/2)$ is simulated owing to the symmetry, and all walls scatter gas molecules diffusely, so that the boundary conditions at the solid walls are
\begin{equation}\label{eq:BC_transpiration}
f_0=\frac{n_{in}(x_1)}{n_0}E_t(T_w(x_1)), \quad ~f_1=\frac{d_r}{2}k_BT_w(x_1)f_0, \quad ~f_2=\frac{d_v}{2}k_BT_w(x_1)f_0,
\end{equation}
while that at the symmetry line ($x_2=L_0/2, v_2\le{}0$) is 
\begin{equation}
\begin{aligned}[b]
f_0=f_0(v_1,-v_2,v_3), \quad
f_1=\frac{d_r}{2}k_BTf_0, \quad
f_2=\frac{d_v}{2}k_BTf_0,
\end{aligned}
\end{equation}
where $n_{in}(x_1)$ is determined similar to \eqref{eq:BC_FourierFlow_n_in}.

\subsubsection{Flow filed and its mechanism}

\begin{figure}[t]
	\centering
	{\includegraphics[scale=0.5,clip=true]{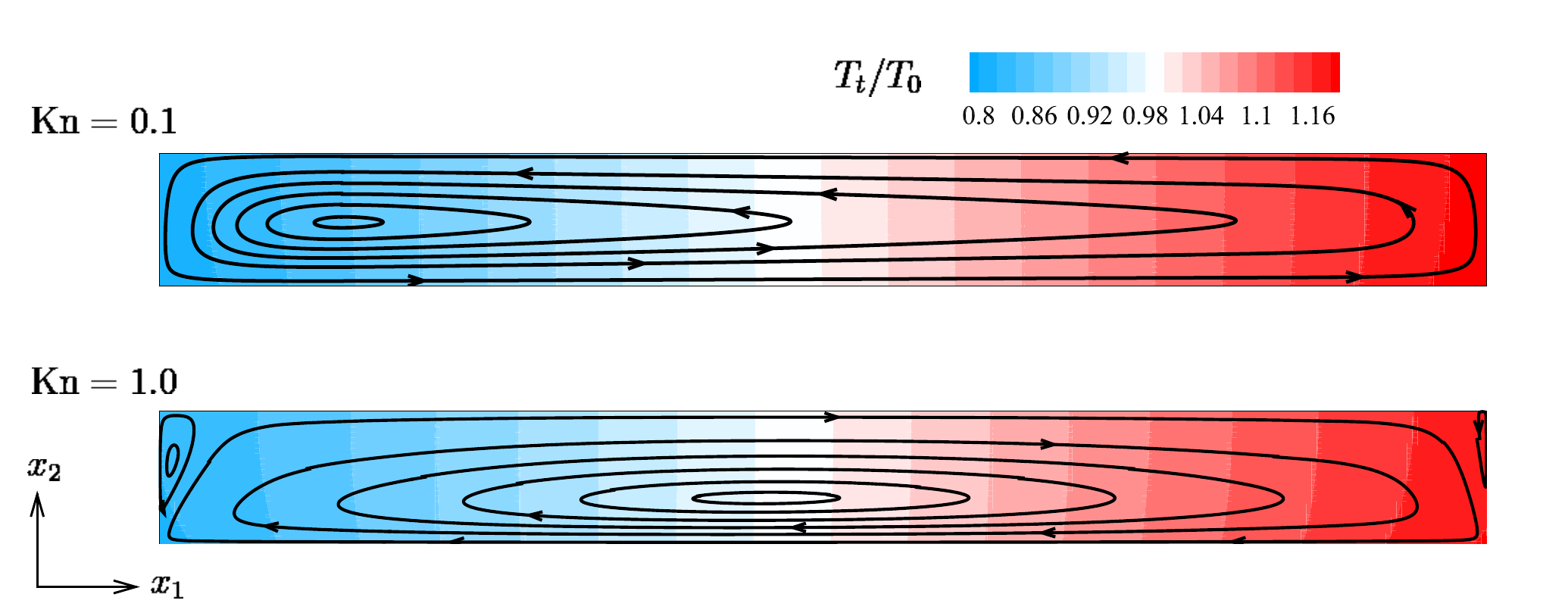}}
	\caption{The flow field of thermal transpiration of nitrogen in the cavity solved by kinetic model equations at $\text{Kn}=0.1$ and 1, and viscosity index $\omega=0.74$.}
	\label{fig:2DTranspiration_flow}
\end{figure}

The flow field is show in figure \ref{fig:2DTranspiration_flow}, when $\text{Kn}=0.1$ and $1$. A large vortex occupies almost the entire simulation domain, which is formed by two competing mechanisms. The diffuse wall generates a steady thermal transpiration, which pushes the gas from the cold end to the hot end. As a result, gas molecules accumulate at the hot end and increase the pressure there, which induces the Poiseuille flow from the hot end to the cold end. Consequently, the thermal transpiration and the Poiseuille flow that are in the opposite directions form the entire vortex. Since in the steady state the horizontal mass flow rate along any vertical line is zero, the rotational direction of the vortex depends on the relative strength of the thermal transpiration and Poiseuille flow. For example, when the Knudsen number is small, the parabolic velocity profile of the Poiseuille flow between two parallel plates is rather steep (the velocity at the channel centre is much larger than that near the solid wall), while that of thermal is very flat. Therefore, the major vortex rotates anticlockwise. On the contrary, when the Knudsen number is large, the velocity profile in thermal transpiration is steeper than that in the Poiseuille flow, so the major vortex rotates clockwise. 

\subsubsection{Influence of intermolecular potential}

\begin{figure}[t]
	\centering
	\subfloat[]{\includegraphics[scale=0.19,clip=true]{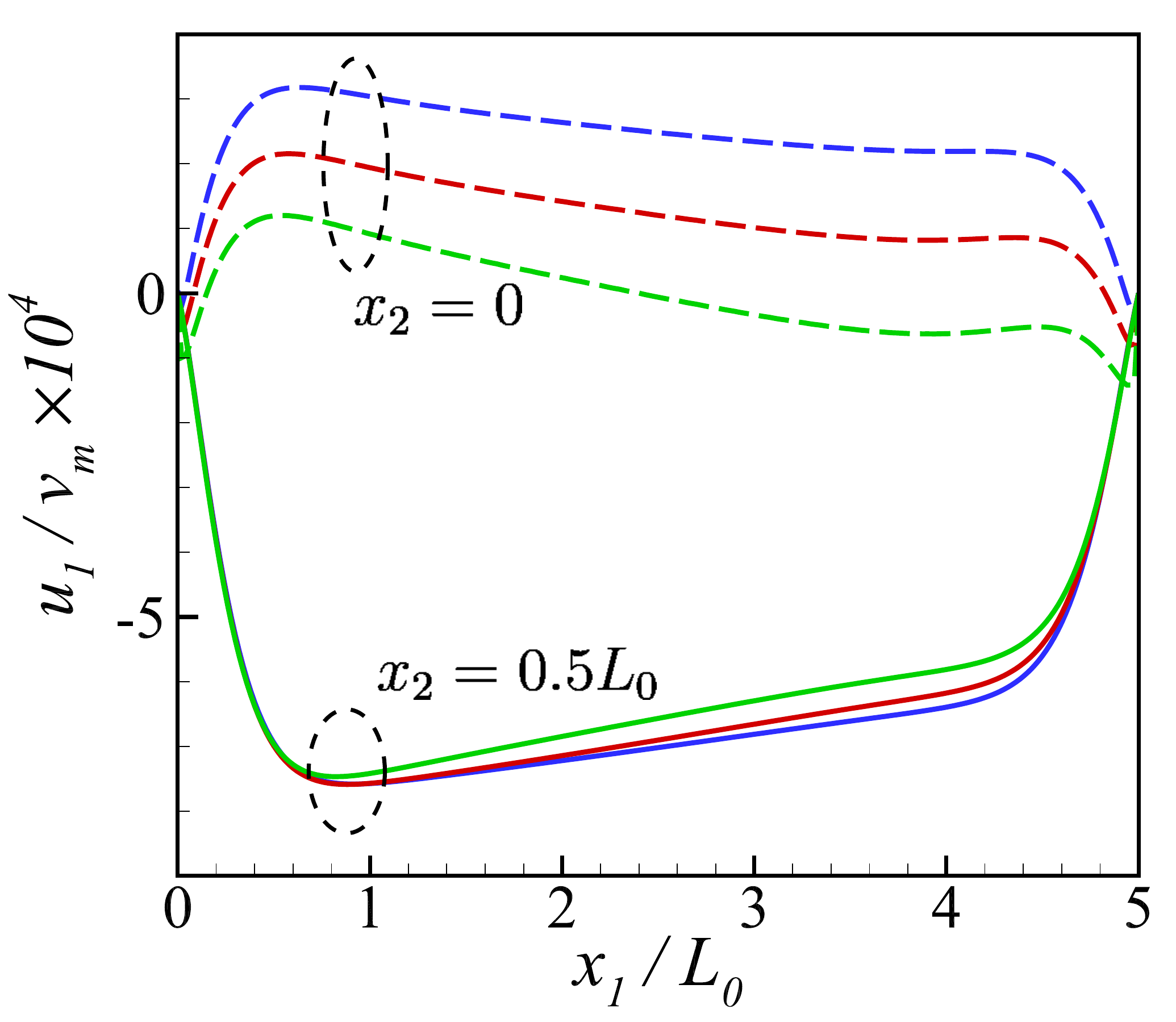}\label{fig:2DTranspiration_w:a}}
	\subfloat[]{\includegraphics[scale=0.19,clip=true]{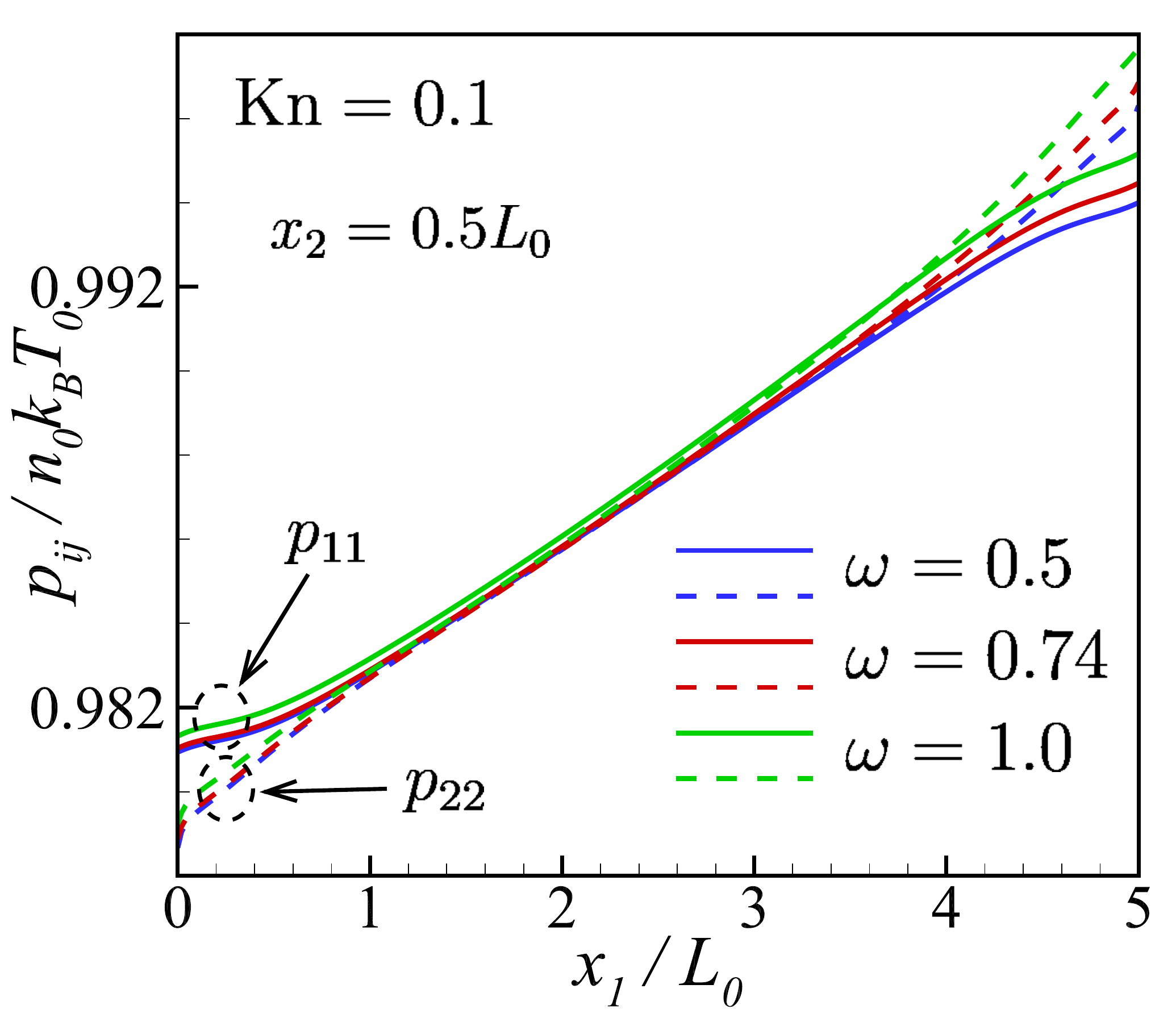}\label{fig:2DTranspiration_w:b}}
	\subfloat[]{\includegraphics[scale=0.19,clip=true]{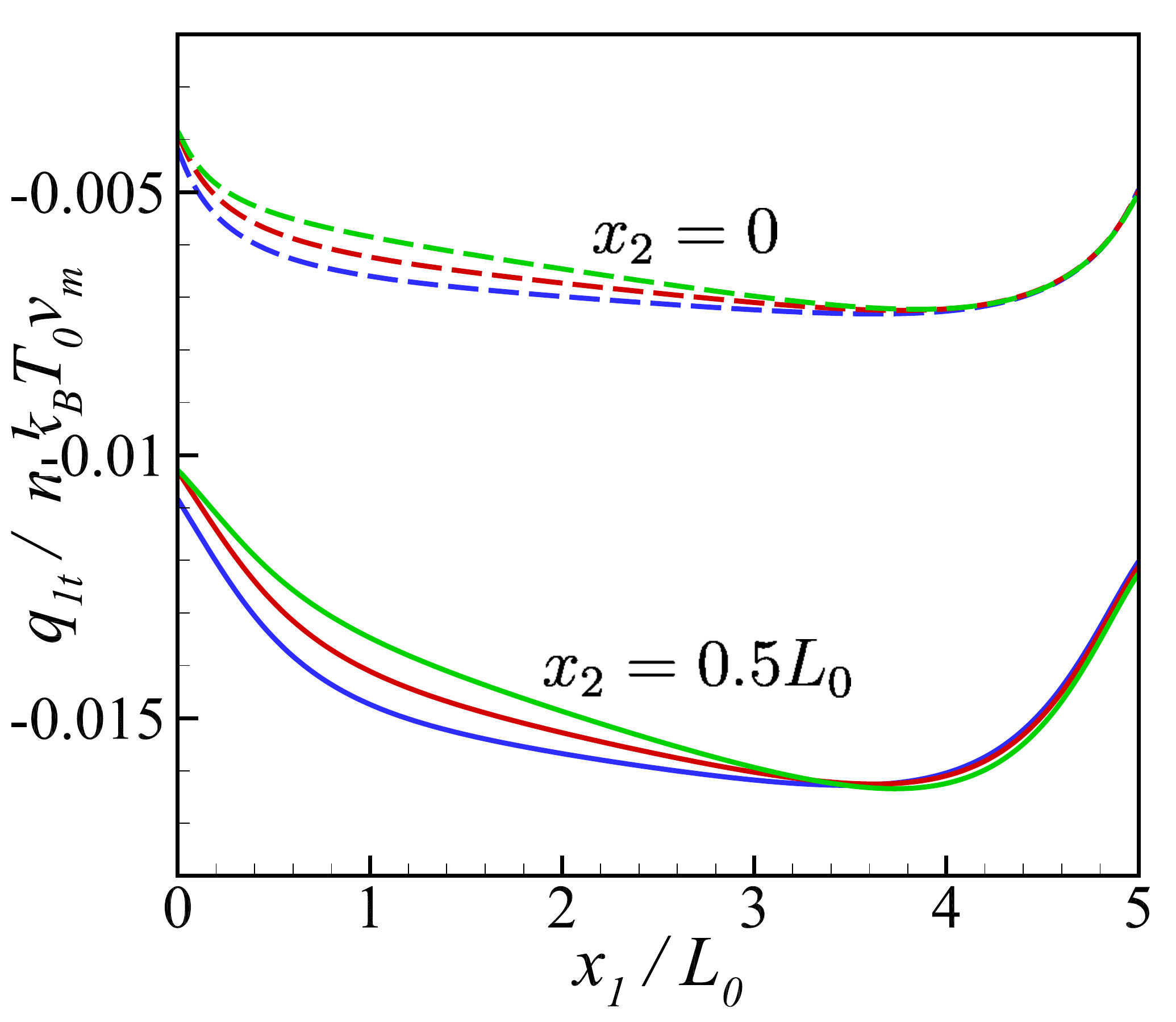}\label{fig:2DTranspiration_w:c}} \\
	\subfloat[]{\includegraphics[scale=0.19,clip=true]{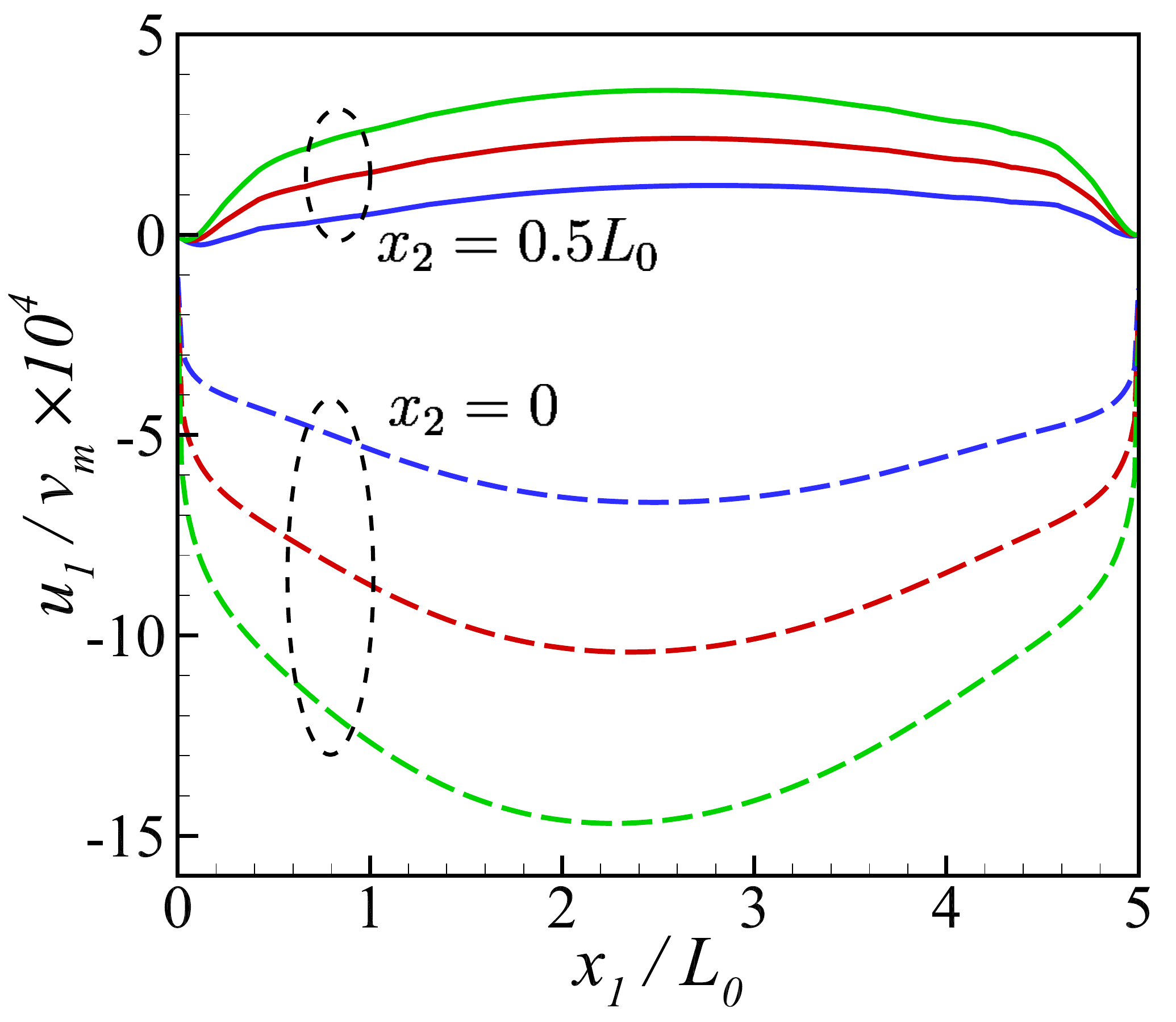}\label{fig:2DTranspiration_w:d}}
	\subfloat[]{\includegraphics[scale=0.19,clip=true]{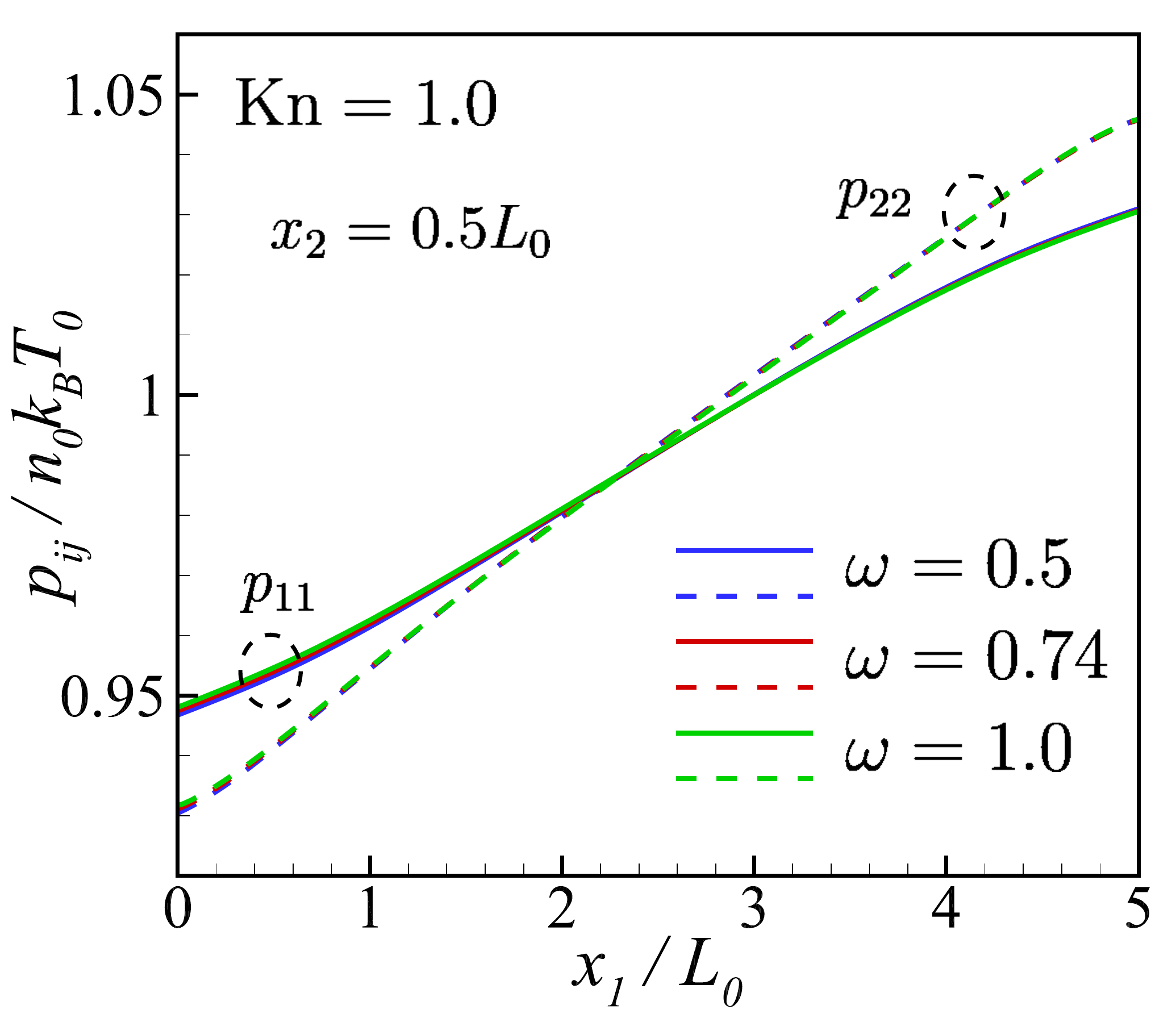}\label{fig:2DTranspiration_w:e}}
	\subfloat[]{\includegraphics[scale=0.19,clip=true]{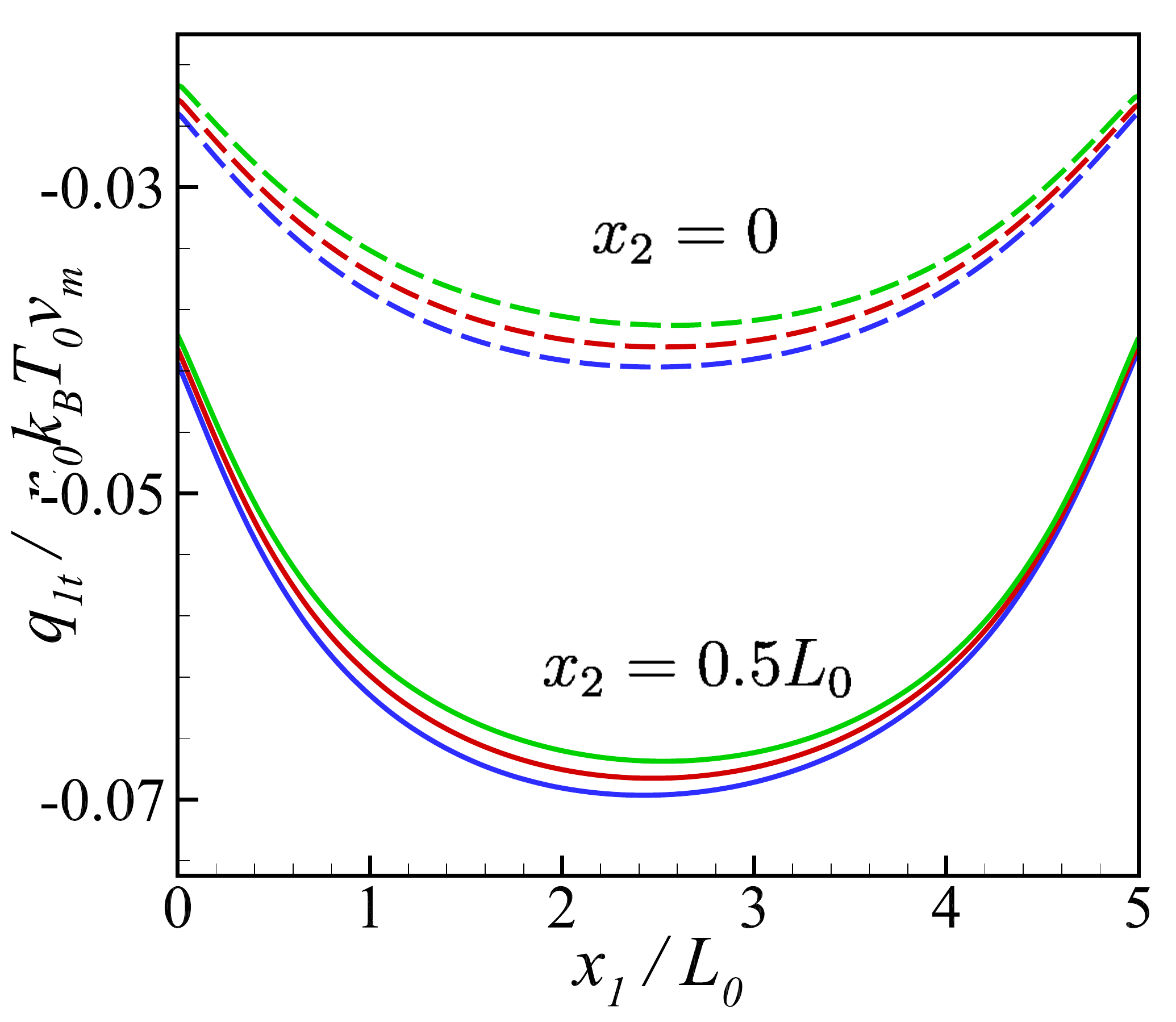}\label{fig:2DTranspiration_w:f}}
	\caption{The thermal transpiration of nitrogen in a cavity. The  viscosity index is $\omega=0.5$ (blue), 0.74 (red) and 1.0 (green). (a, d) velocity in $x_1$ direction, (b, e) normal pressure $p_{11}$ and $p_{22}$, (c, f) translational heat flux in $x_1$ direction. Macroscopic quantities are plotted along the central line $x_2=0.5L_0$ (solid lines) and the wall $x_2=0$ (dashed lines). The Knudsen number is $\text{Kn}=0.1$ and 1 in the first and second rows, respectively.}
	\label{fig:2DTranspiration_w}
\end{figure}

The influence of the intermolecular potential is investigated by varying the viscosity index $\omega$, while fixing the relaxation rates and all transport coefficients at the reference temperature. The values of viscosity index are chosen to represent the hard sphere molecules ($\omega=0.5$), the nitrogen molecules ($\omega=0.74$) and Maxwell molecules ($\omega=1$).  
Figure \ref{fig:2DTranspiration_w} compares the horizontal velocity, normal pressure and translational heat flux distribution with different $\omega$, when $\text{Kn}=0.1$ and 1. Clearly, the most significant impact of the intermolecular potential is the change of the flow velocity. When $\text{Kn}=0.1$, the slip velocity on the wall ($x_2=0$) varies dramatically with $\omega$, while that along the central line ($x_2=0.5L_0$) does not change that much. When $\text{Kn}=1$, significant variation of the flow velocity occurs over the entire domain, and the maximum speed located at the wall increases by 2.2 times, when $\omega$ is changed from 0.5 to 1. 

The pressure difference generated at the two ends of the cavity is shown in figure \ref{fig:2DTranspiration_w:b} and \ref{fig:2DTranspiration_w:e}. It is seen that, the change in normal pressure in both $x_1$ and $x_2$ directions is slight at $\text{Kn}=0.1$, and even negligible at $\text{Kn}=1$. Besides, the variation of translational heat flux is also not that significant. In general, the heat flux in the low temperature region decreases with the increase of $\omega$, since the effective shear viscosity and hence the thermal conductivity is lower in this region based on \eqref{eq:viscosity_temperature}. The situation in the high temperature region is reversed, i.e., the heat flux increases with $\omega$. Since the temperature varies from $0.8T_0$ to $1.2T_0$ in the system, the difference in viscosity and thermal conductivity does not exceed 10\% when $\omega$ changing from 0.5 to 1, and thus the difference in heat flux is also within this range.

\subsection{Knudsen force on micro-beam}

The Knudsen force acting on a heated micro-beam adjacent to a cold substrate is a mechanical force created by the surrounding thermally nonequilibrium rarefied gas. As one type of the thermal forces, which emerges mainly in micro/nano devices with integrated heaters due to the advent of microfabrication techniques nowadays, the Knudsen force has been investigated numerically and experimentally \citep{Passian2003PRL, Li2013PRE, Pikus2019Vacuum}. Both the magnitude and direction of the force is important, since it may significantly affect the performance of many micro/nano devices, say, the accuracy of atomic force microscopy~\citep{Passian2003Um}. The Knudsen force induced by molecular gas has not been systematically studied, especially the underlying mechanism and the influence of intermolecular potential. 

The system considered here is a heated micro-beam ($L_0\times2L_0$) placed inside a cold chamber ($5L_0\times10L_0$), and the centre of the micro-beam shift towards negative $x_2$ direction by a distance $L_0$ with respect to the centre of the chamber. The temperature of the micro-beam and chamber are maintained at $1.2T_0$ and $0.8T_0$, respectively, and all the surface are fully diffuse. Due to the symmetry, only the right half of the system ($0\le{}x_1\le{}5L_0,~0\le{}x_2\le{}5L_0$) is simulated.

The implicit discontinuous Galerkin (DG) method is employed in the numerical simulations \citep{Su2020JSC}, and the fourth-order approximating polynomials are used in the DG scheme. The computational domain is partitioned by unstructured triangles with refinement in the vicinity of the beam surfaces. To be specific, the total number of elements is 1790 when $\text{Kn}=0.1$ and 1738 when $\text{Kn}=1$. The molecular velocity space is truncated by $[-7v_m,7v_m]^3$, and 64 non-uniform velocity points are used to discretize $v_1$ and $v_2$, while 32 uniform points are used for $v_3$. And in the velocity space the fast spectral method is incorporated into the DG discretization to evaluate the collision operator, where $32\times32\times32$ equidistant frequencies are employed.

\subsubsection{Flow filed and its mechanism}

\begin{figure}[t]
	\centering
	\subfloat[]{\includegraphics[scale=0.3,clip=true]{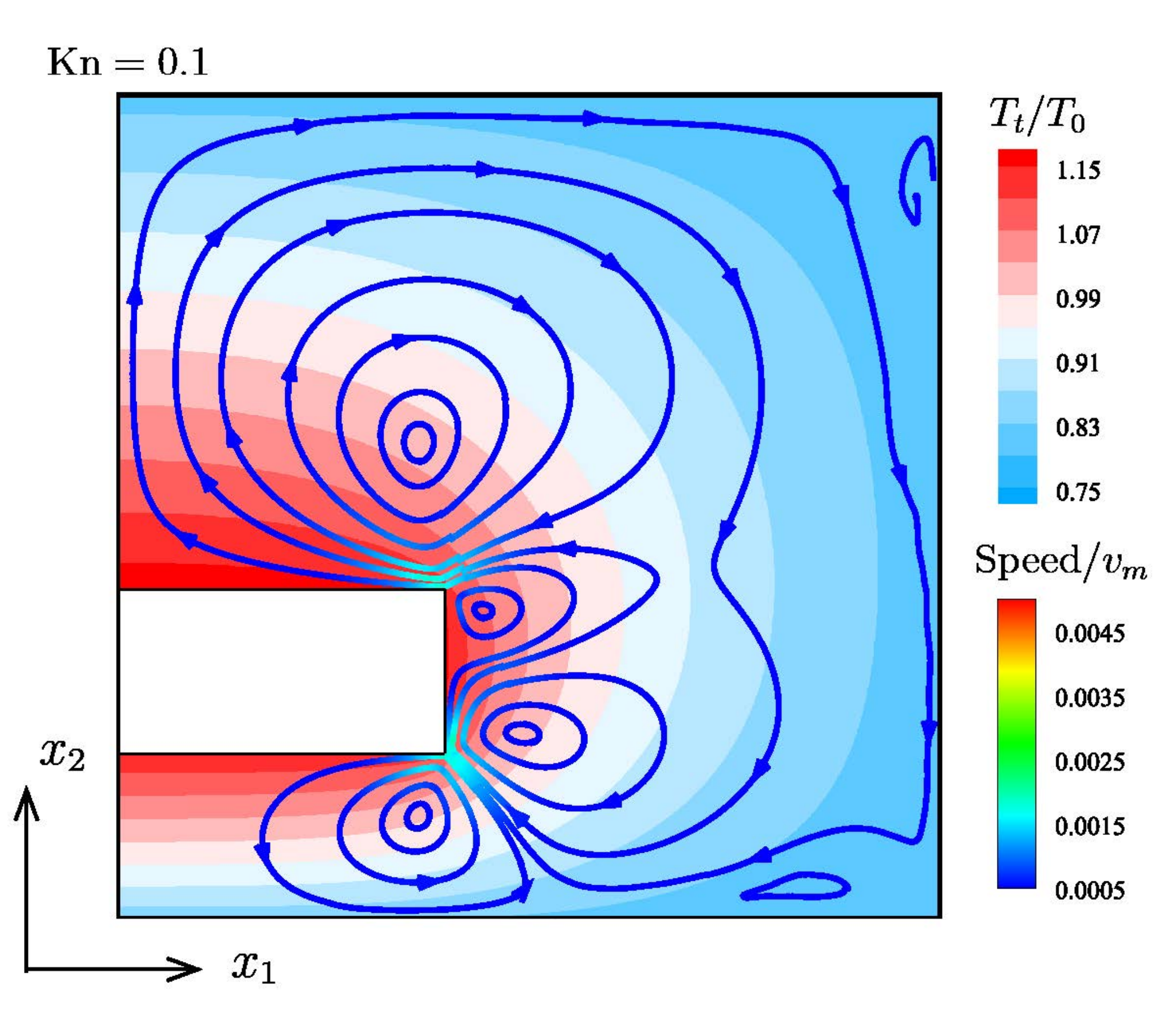}}
	\subfloat[]{\includegraphics[scale=0.3,clip=true]{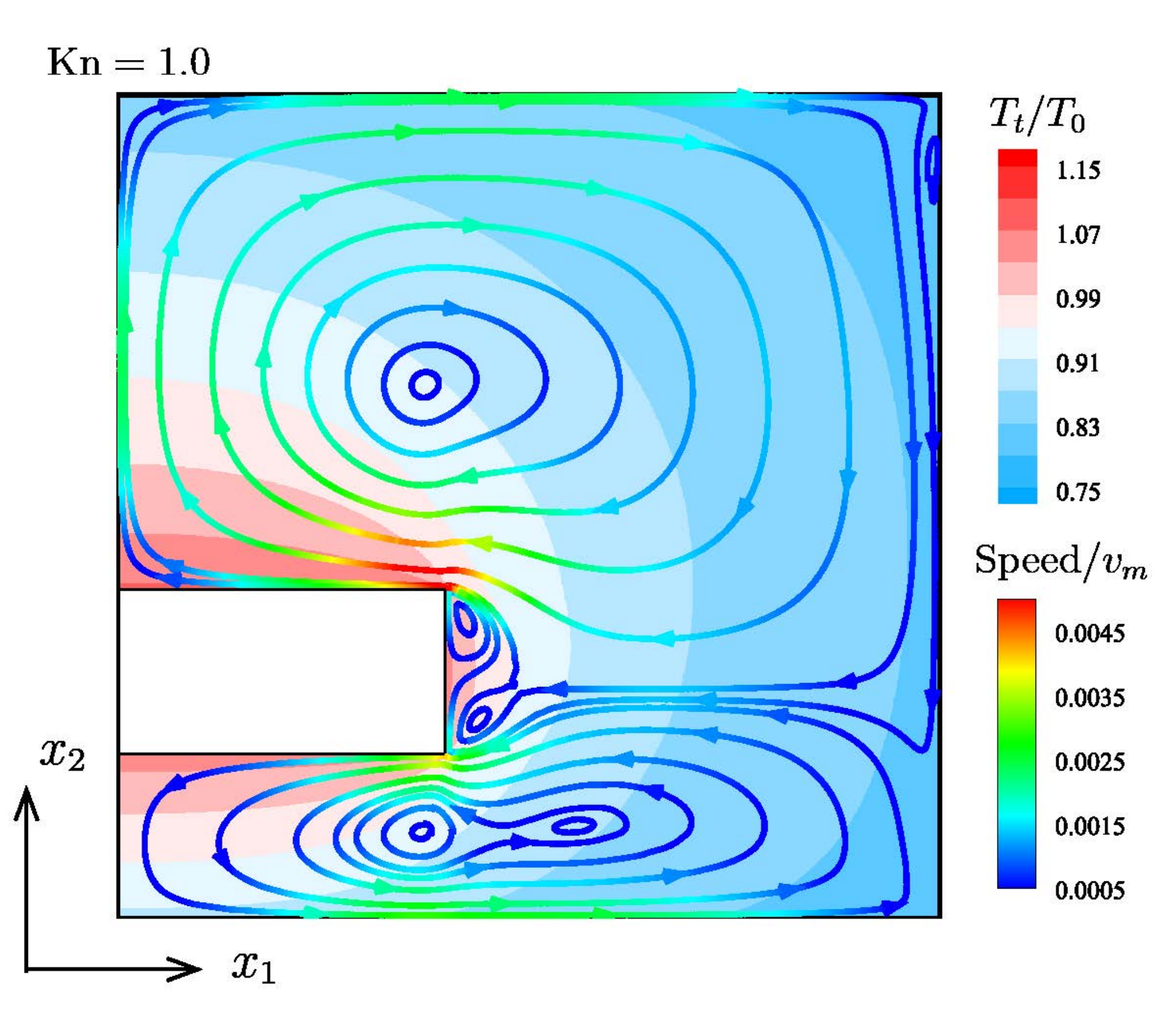}}
	\caption{Flow field of nitrogen surrounding the heated micro-beam in cavity, which is solved by the kinetic model~\eqref{eq:kinetic_model_equation} when (a) $\text{Kn}=0.1$ and (b) $\text{Kn}=1$, and viscosity index $\omega=0.74$. The background contour is the distribution of translational temperature.}
	\label{fig:2DKnudsenForce_flow}
\end{figure}

The translational temperature contour and flow field are shown in figure \ref{fig:2DKnudsenForce_flow}. The flow structures are similar for the two cases at $\text{Kn}=0.1$ and $1$, although the relative strength varies significantly. The temperature of gas changes rapidly around the sharp corners, and forms large temperature gradient normal to the surface of the micro-beam. In analogy to the asymptotic analysis of Boltzmann equation for monatomic gas \citep{Sone2002Book}, three types of thermally induced flow exist around the micro-beam in the molecular gas, namely, the thermal stress slip flow, the nonlinear thermal stress flows, and the thermal edge flow. The first two are caused by the normal temperature gradient along the wall, and the flow speed reaches the maximum values at the sharp corners. However, they are in the direction opposite to that shown in figure \ref{fig:2DKnudsenForce_flow}. On the contrary, the mechanism of the thermal edge flow is similar to that of the thermal transpiration, and the flow is in the same direction as that observed in figure \ref{fig:2DKnudsenForce_flow}. The thermal edge flow is fairly strong within a wide range of Knudsen number, especially becomes strongest in the transition regime. As demonstrated in figure~\ref{fig:2DKnudsenForce_flow}, the flow speed is much larger when $\text{Kn}=1$ than that when $\text{Kn}=0.1$.

Figure \ref{fig:2DKnudsenForce_edge_Kn1_w:b} shows the magnitude of heat flux along the surfaces of the heated micro-beam at $\text{Kn}=1$, where the result of nitrogen gas ($\omega=0.74$) is indicated by red line. And its direction is normal to the surface, due to the isothermal walls. The strongest heat flux occurs at the corners of the micro-beam, where the normal temperature gradient is largest. Despite the different collision numbers of rotational and vibrational modes, the rotational and vibrational heat fluxes are almost the same. Meanwhile, it is found that the total internal heat flux is approximately equal to the translational one. Therefore, although the relaxation time of internal relaxation is usually much longer than that of the translational mode, the heat fluxes carried by rotational and vibrational modes could be considerable. In particular, when a gas molecule consists of more atoms, it could have more number of internal DoF and thus contributes more to the total heat flux.

\subsubsection{The Knudsen force on the micro-beam}

\begin{figure}[t]
	\centering
	\subfloat[]{\includegraphics[scale=0.3,clip=true]{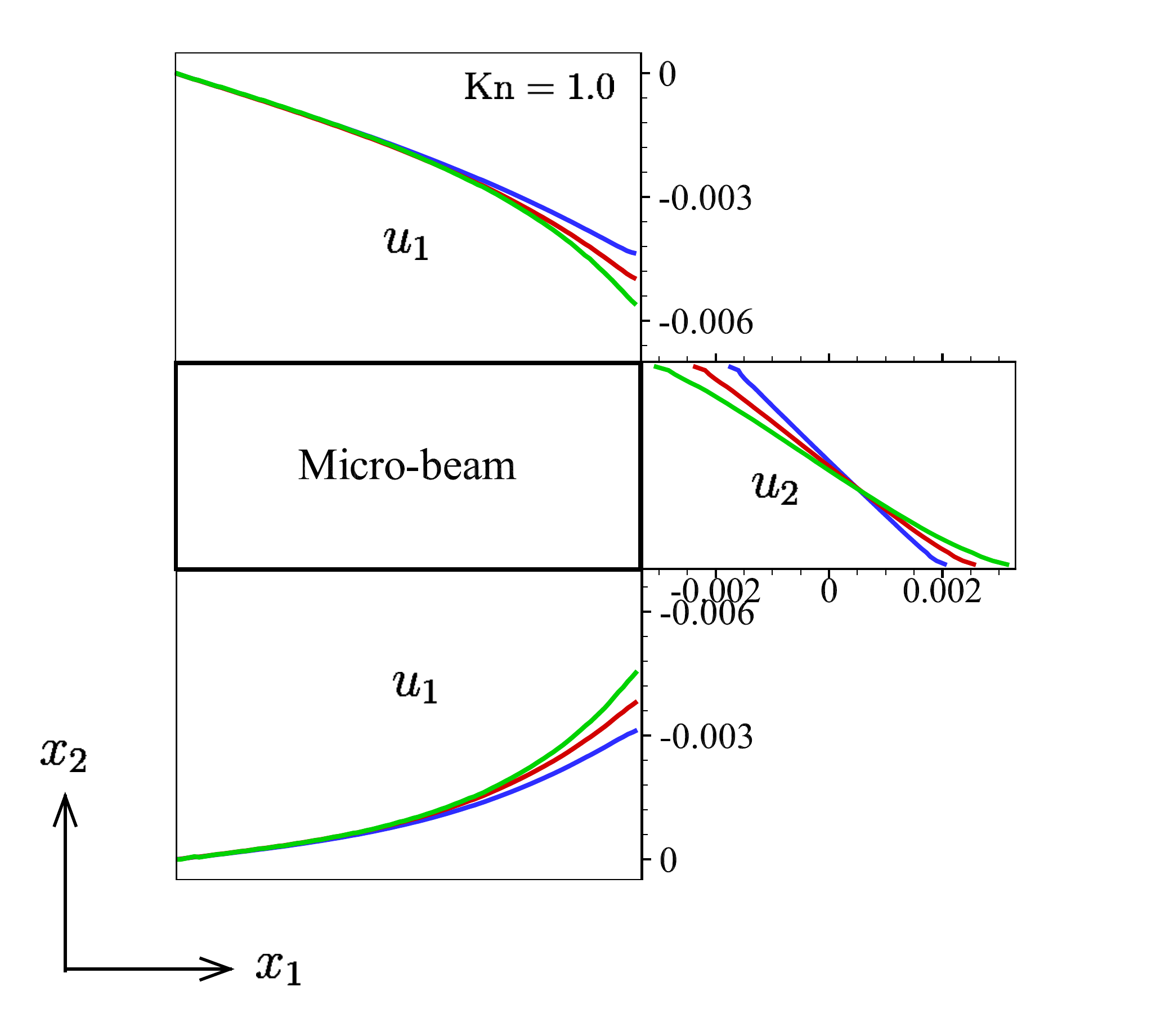}\label{fig:2DKnudsenForce_edge_Kn1_w:a}} \\
	\subfloat[]{\includegraphics[scale=0.3,clip=true]{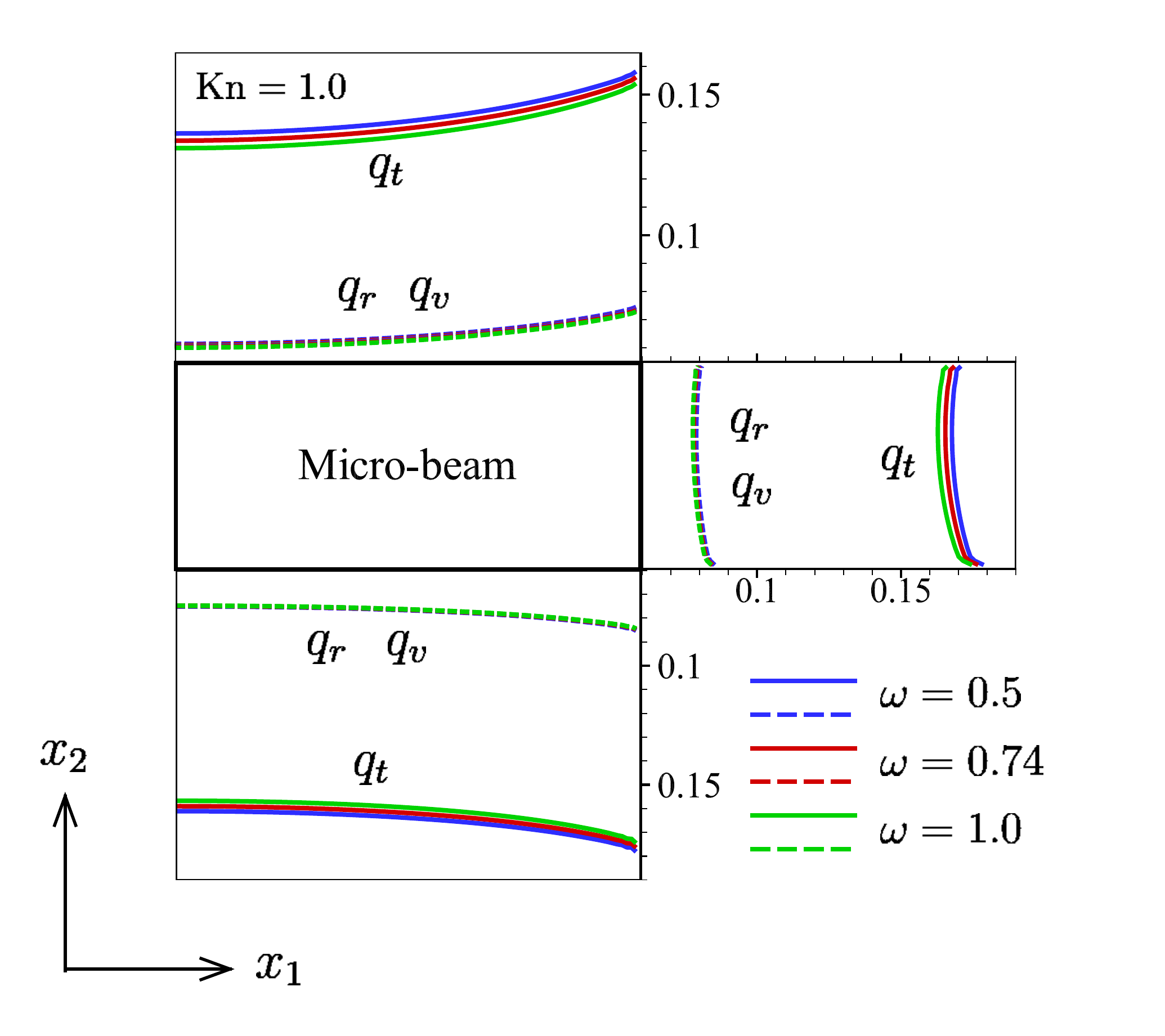}\label{fig:2DKnudsenForce_edge_Kn1_w:b}}
	\subfloat[]{\includegraphics[scale=0.3,clip=true]{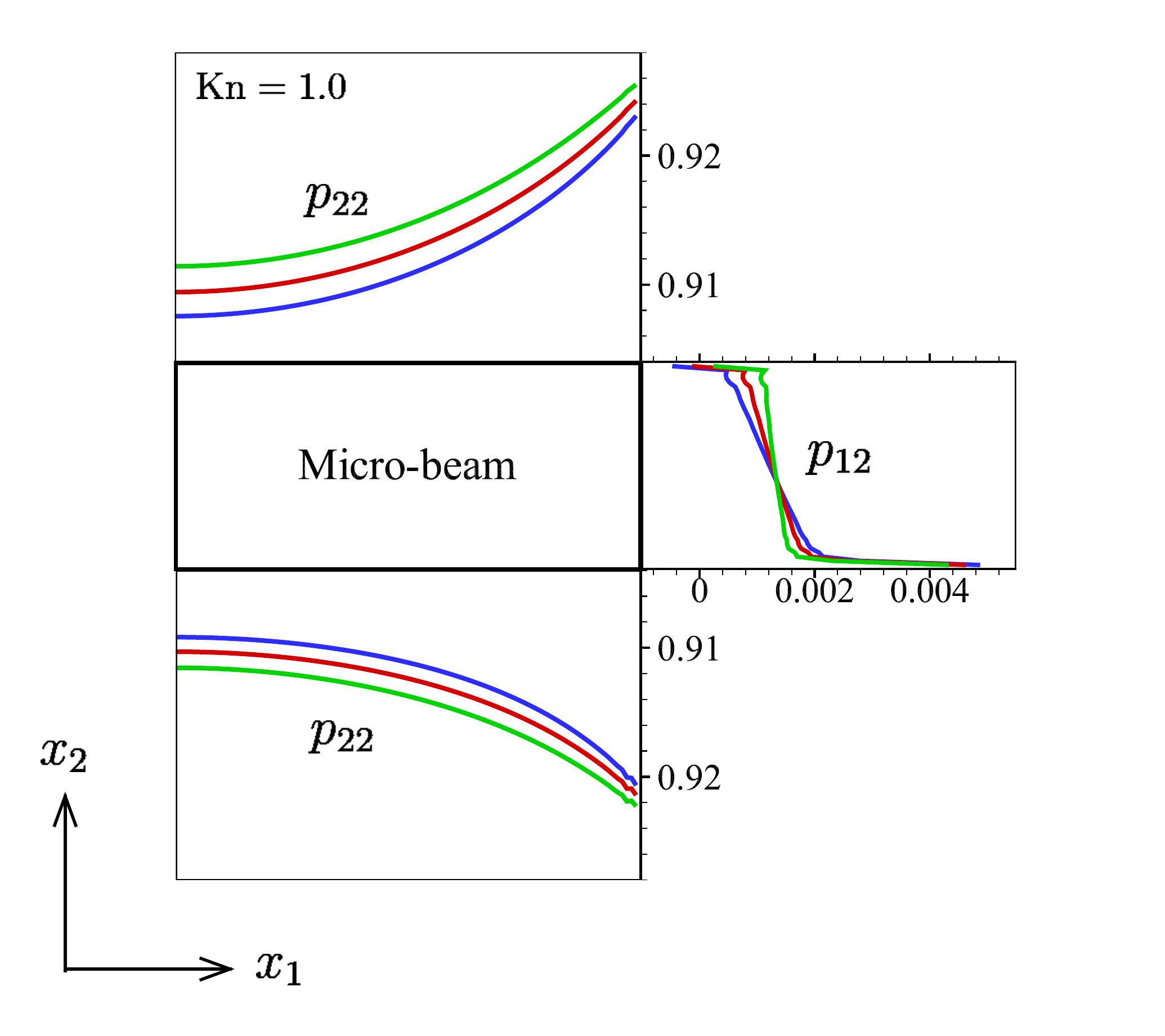}\label{fig:2DKnudsenForce_edge_Kn1_w:c}}
	\caption{The distribution of (a) velocity (normalized by $v_m$), (b) heat flux (normalized by $n_0k_BT_0v_m$) and (c) normal stress $p_{22}$ and shear stress $p_{12}$ (normalized by $n_0k_BT_0$) along the surface of the heated micro-beam solved by kinetic model equations when $\text{Kn}=1$. The viscosity index $\omega=0.5,~0.74,~1$ are represented by blue, red and greed lines, respectively.}
	\label{fig:2DKnudsenForce_edge_Kn1_w}
\end{figure}

The thermally induced flows redistribute the gas molecules and hence the pressure in the chamber. Therefore, it is expected to have a net force acting on the beam. The resultant force in horizontal direction is zero due to the symmetry about $x_1=0$. To investigate the vertical force acting on the beam, the normal pressure $p_{22}$ on the top and bottom surfaces and shear stress $p_{12}$ along the right surface of the beam are shown in figure \ref{fig:2DKnudsenForce_edge_Kn1_w:c}. The variation of normal pressure is found to be small along the surfaces, which is around 1\%. Therefore, although the normal pressure $p_{22}$ is about two orders of magnitude larger than the shear stress ${p_{12}}$, the resultant force of the normal pressure $F_n$ is of the same order or even smaller than the resultant shear force $F_s$. This is consistent with the fact that the origin of the Knudsen force is the thermally induced flows, which determine the order of magnitude of the shear force. Thus, the Knudsen force should be sensitive to the shear force. When $\text{Kn}=0.1$, $F_n=3.52\times 10^{-4}n_0k_BT_0L_0$ is much smaller than $F_s=1.32\times 10^{-3}n_0k_BT_0L_0$, and then the beam is subjected to a total Knudsen force $F=1.67\times 10^{-3}n_0k_BT_0L_0$ pointing to the positive $x_2$ direction. When $\text{Kn}=1$, the magnitudes of both $F_n$ and $F_s$ are larger than those when $\text{Kn}=0.1$ but in opposite directions. Competed by these two forces, the total force is $F=5.06\times 10^{-4}n_0k_BT_0L_0$, which points to the positive $x_2$ direction but is relatively small in its magnitude.

\subsubsection{Influence of intermolecular potential}

\begin{table}[t]
    \begin{center}
	  \begin{tabular}{ccccc}
        \hline
         $\text{Kn}$ & $\omega$ & \makecell[c]{$F_n\times10^{3}$ \\ $(n_0k_BT_0L_0)$} & \makecell[c]{$F_s\times10^{3}$ \\ $(n_0k_BT_0L_0)$} & \makecell[c]{$F\times10^{3}$ \\ $(n_0k_BT_0L_0)$} \\
		\specialrule{0em}{4pt}{4pt}
         \multirow{3}*{\quad 0.1 \quad}  & 0.5	& 0.646 & 1.29 & 1.93   \\
         					& 0.74	& 0.352 & 1.32 & 1.67   \\
         					& 1	& 0.222 & 1.38 & 1.60   \\
         \specialrule{0em}{3pt}{3pt}
		 \multirow{3}*{\quad 1 \quad}  & 0.5	& 0.274 & 2.63 & 2.91   \\
        					& 0.74	& -2.22  & 2.73 & 0.506 \\
        					& 1	& -4.70  & 2.80 & -1.90  \\
         \hline
      \end{tabular}
      \caption{
      	The Knudsen force calculated from kinetic model equation~\eqref{eq:kinetic_model_equation} for the viscosity index $\omega=0.5$, 0.74, and 1. $F_n$ is the resultant normal force from top and bottom surfaces of the beam, $F_s$ is the resultant shear force from the side surfaces, and the total force $F=F_n+F_s$, where the positive value indicates that the force points to the positive $x_2$ direction.
      }
    \label{tab:KnudsenForce_w}
    \end{center}
  \end{table}

The effect of intermolecular potential reflected in the viscosity index $\omega$ is also investigated. Similar to that in the thermal transpiration, the thermally induced velocity around the micro-beam changes significantly with $\omega$ as shown in figure \ref{fig:2DKnudsenForce_edge_Kn1_w:a}. For instance, when $\text{Kn}=1$, the maximum magnitude of velocity at the corner is increased by 1.74 times when $\omega$ changes from 0.5 to 1. However, figures \ref{fig:2DKnudsenForce_edge_Kn1_w:b} and \ref{fig:2DKnudsenForce_edge_Kn1_w:c} show that the heat flux and stress are not affected that much: the maximum difference is around 3.8\% in heat flux and less than 0.5\% in normal pressure. 

Both the magnitude and orientation of the resultant force acting on the micro-beam is found to be very sensitive to the viscosity index. Table \ref{tab:KnudsenForce_w} lists the normal, shear and total force for different $\omega$. As $\omega$ increases, the normal force tends to be stronger in negative $x_2$ direction. However, the shear force acting on the side surfaces increases slightly in the positive $x_2$ direction. When $\text{Kn}=1$, the opposite trends reverse the direction of the total force. It also implies that a zero net force exist at certain value of $\omega$, which happens to be around 0.74 (the value for variable soft sphere model of nitrogen) in this configuration.

\section{Conclusions}\label{sec:conclusion}

A kinetic model for molecular gas with internal DoF has been proposed. Compared with the previous works on the model equations, there are two features in our kinetic model: (i) realization of molecular velocity-dependent collision time, and consistent with the Boltzmann equation for monatomic gas when the translational-internal energy exchange is extremely slow; (ii) recovery of thermal relaxation processes and rates, and all  transport coefficients. Thus, this kinetic model has the ability to describe the influence of intermolecular potentials.

The accuracy of our model has been demonstrated by comparing with DSMC simulations for one-dimensional Fourier flow, Couette flow, creep flow driven by the Maxwell demon and normal shock wave. Then, the thermal transpiration and Knudsen force acting on micro-beam, which would need extreme long simulation time in DSMC, are investigated. It is found that the intermolecular potential, reflected through the viscosity index, has a big impact on the flow velocity and the Knudsen force exerted on the beam. This discovery is useful in the design of  micro-electromechanical systems for microstructure actuation and gas sensing \citep{Strongrich2015APL, Strongrich2017JMEMS}.


With the multiscale numerical method~\citep{SuArXiv2019,Zhu2021JCP} which is able to find the steady-state solution within dozens of iterations, the present kinetic model is expected to find applications in various areas with rarefied molecular gas dynamics, especially for high-temperature problems, such as shock wave that needs accurate velocity-dependent collision time in the kinetic model,  as well as for micro flows, where the deterministic numerical method is needed to resolve the small signals. 



\section*{Declaration of interests} 
The authors report no conflict of interest.

\bibliographystyle{jfm}
\bibliography{bibnew}

\end{document}